\documentclass[reqno,11pt]{amsart}
\usepackage{graphicx}
\usepackage{slashed}
\usepackage{amscd}
\usepackage{tensor}
\usepackage{amssymb}
\usepackage{esint}
\usepackage{enumitem}
\usepackage{accents}
\usepackage[usenames,dvipsnames]{pstricks}
\usepackage{pst-grad} 
\usepackage{pst-plot} 
\usepackage[mathscr]{eucal}
\numberwithin{equation}{section}
\allowdisplaybreaks[1]

\title[Baryogenesis for Causal Fermion Systems]{A Mechanism of Baryogenesis \\ for Causal Fermion Systems}

\author[F.\ Finster]{Felix Finster}
\address{Fakult\"at f\"ur Mathematik \\ Universit\"at Regensburg \\ D-93040 Regensburg \\ Germany}
\email{finster@ur.de, maximilian.jokel@ur.de}
\author[M.\ Jokel]{Maximilian Jokel}
\author[C.F.\ Paganini]{Claudio F. Paganini\\ \\ November 2021}
\address{Fakult\"at f\"ur Mathematik \\ Universit\"at Regensburg \\ D-93040 Regensburg \\ Germany}
\address{Max Planck Institute for Gravitational Physics (Albert Einstein Institute), Am M\"uh\-len\-berg 1, D-14476 Potsdam, Germany}
\email{claudio.paganini@ur.de}

\newtheorem{Def}{Definition}[section]
\newtheorem{Thm}[Def]{Theorem}
\newtheorem{Prp}[Def]{Proposition}
\newtheorem{Lemma}[Def]{Lemma}

\newtheorem{Example}[Def]{Example}

\newcommand{\Thanks}{\vspace*{.5em} \noindent \thanks}
\newcommand{\beq}{\begin{equation}}
\newcommand{\eeq}{\end{equation}}
\newcommand{\Proof}{\begin{proof}}
\newcommand{\QED}{\end{proof} \noindent}
\newcommand{\QEDrem}{\ \hfill $\Diamond$}

\newcommand{\la}{\langle}
\newcommand{\ra}{\rangle}
\newcommand{\bra}{\mathopen{<}}
\newcommand{\ket}{\mathclose{>}}
\newcommand{\Sl}{\mbox{$\prec \!\!$ \nolinebreak}}
\newcommand{\Sr}{\mbox{\nolinebreak $\succ$}}

\newcommand{\C}{\mathbb{C}}
\newcommand{\R}{\mathbb{R}}
\newcommand{\1}{\mbox{\rm 1 \hspace{-1.05 em} 1}}

\newcommand{\N}{\mathbb{N}}
\newcommand{\Pdd}{\mbox{$\partial$ \hspace{-1.2 em} $/$}}

\DeclareMathOperator{\Tr}{Tr}
\DeclareMathOperator{\tr}{tr}
\renewcommand{\O}{{\mathscr{O}}}
\renewcommand{\L}{{\mathcal{L}}}
\newcommand{\Sact}{{\mathcal{S}}}

\newcommand\B{{\mathscr{B}}}

\newcommand{\Cisc}{C^\infty_{\text{\rm{sc}}}}

\newcommand{\Dir}{{\mathcal{D}}}
\DeclareMathOperator{\supp}{supp}
\renewcommand{\H}{\mathscr{H}}

\newcommand{\Lin}{\text{\rm{L}}}
\newcommand{\F}{{\mathscr{F}}}

\newcommand{\D}{\mathscr{D}}

\DeclareMathOperator{\im}{Im}

\newcommand{\scrL}{\mycal L}
\newcommand{\scrM}{\mycal M}
\newcommand{\scrN}{\mycal N}

\newcommand{\itemD}{\item[{\raisebox{0.125em}{\tiny $\blacktriangleright$}}]}
\newcommand{\x}{\mathbf{x}}
\newcommand{\y}{\mathbf{y}}

\newcommand{\bitem}{\begin{itemize}[leftmargin=2em]}
\newcommand{\eitem}{\end{itemize}}
\DeclareFontFamily{OT1}{rsfso}{}
\DeclareFontShape{OT1}{rsfso}{m}{n}{ <-7> rsfso5 <7-10> rsfso7 <10-> rsfso10}{}
\DeclareMathAlphabet{\mycal}{OT1}{rsfso}{m}{n}

\newcommand{\s}{\mathfrak{s}}
\DeclareMathOperator{\Texp}{Texp}

\begin{document}

\maketitle

\begin{abstract}
It is shown that the theory of causal fermion systems gives rise to a novel mechanism of
baryogenesis. This mechanism is worked out computationally in globally hyperbolic
spacetimes in a way which enables the quantitative study in concrete cosmological situations.
\end{abstract}

\tableofcontents

\section{Introduction}
The present universe contains more matter than antimatter.
This particle/anti-particle asymmetry is so large that it cannot be explained by the
Standard Model of particle physics (for details see for example the PhD thesis~\cite[p.~60]{lucente2016implication} or the papers~\cite{huet1995electroweak,gavela1994standard, gavela1994standard1,gavela1994standard2} and references therein).
It is one of the outstanding problems of modern physics to explain how the magnitude of the predominance of matter comes about. Moreover, one needs to identify and quantify the underlying physical effects.
One scenario is that there was no predominance of matter right after the big bang,
and that the matter/anti-matter asymmetry was generated dynamically.
This scenario is usually referred to as {\em{baryogenesis}} (for a survey see for example~\cite{riotto} or~\cite{cline2006baryogenesis}). There are modifications of the theme referred to as {\em{leptogenesis}} where the dominant reaction inducing the asymmetry occurs in the leptonic sector of the particle model, which
in turn gives rise to an asymmetry also in the baryonic sector~\cite{kuzmin1985anomalous}. 
Another mechanism of baryogenesis in the guise of {\em{fermiogenesis}} has been studied recently in~\cite{maleknejad2020dark}.
Attempts to explain baryogenesis typically involve extensions of the Standard Model and/or grand unified theories.
 
 In this paper we propose a new mechanism for the dynamic generation of the matter/antimatter asymmetry
 in our universe. Our mechanism falls into the category of {\em{fermiogenesis}}, with the asymmetry occurring
in the same way for leptons and quarks, thereby guaranteeing for the matter content to be neutral
with respect to all charges. It is one of the core features of our mechanism that we do not need to
 extend the particle content of the Standard Model
 (except for the fact that the theory contains non-interacting right-handed neutrinos). Instead, we obtain the desired asymmetry by considering a mechanism based on the mathematical structures of the {\em{theory of causal fermion systems}}, a novel approach to unify General Relativity and Quantum Theory.
Our mechanism is quite different from those based on standard quantum field theory and particle physics
arguments, to such an extent that the connection to these more familiar mechanisms is not obvious.
\subsection{The Theory of Causal Fermion Systems}
Given that the theory of causal fermion systems is quite new, 
in this paper we include an outline of the basic concepts. In addition, the reader may find it helpful
to also consult introductory texts such as the reviews~\cite{dice2014, review, dice2018}, the textbook~\cite{cfs} or the website~\cite{cfsweblink}.
In a causal fermion system, spacetime and all objects therein are described by a measure~$\rho$ on a set~$\F$ of linear operators acting on a Hilbert space~$(\H, \la .|. \ra_\H)$. The physical equations are formulated via the so-called {\em{causal action principle}},
a nonlinear variational principle where an action~$\Sact$ is minimized under variations of the measure~$\rho$.
The causal action principle allows for the description of the gauge fields of the Standard Model
as well as the field equations of General Relativity.
Moreover, it includes non-interacting right-handed neutrinos.
It explains the fact that we observe three generations of fermions. Furthermore, it
gives a reason for the relative weakness of gravity as Newton's constant arises
is a quadratic function of the regularization length
(whereas the coupling constants of the gauge fields do not scale with the regularization length). Furthermore,
in the recent papers~\cite{fockbosonic, fockfermionic} the Quantum Field Theory description
of the interactions of the Standard Model has been recovered.

One key aspect of the set $\F$, that allows for the successful merger of the Standard Model and General Relativity in the theory of causal fermion systems, is that it constitutes an operator
manifold~\cite{banach}, thereby merging the basic mathematical structures on which Quantum Theory is built
(operators) with those on which General Relativity is built (manifolds). 
A causal variational principle
gives rise to a {\em{spacetime}}~$M$, being a distinguished subset of~$\F$.
The vectors in the Hilbert space~$\H$ are represented by spinorial wave functions in spacetime,
the so-called {\em{physical wave functions}}.
These structures generalize the usual notions of wave functions in classical
spacetimes and of spacetime itself.
The connection between the classical structures and the structures of a causal fermion system
is obtained with the help of the so-called {\em{local correlation map}}, which 
to every classical spacetime point associates a spacetime point operators in~$\F$.
This map contains information on the local densities and correlations of the physical wave function
at every spacetime point (for the detailed construction see Section~\ref{secdirac} below).
In this way, the physical wave functions encode the
geometric structure and the matter content of the spacetime described by the causal fermion system.
The causal action principle can be understood as a variational principle for all the physical wave functions.
Intuitively speaking, the causal action principle aims at bringing the ensemble of all physical wave functions into
an ``optimal'' configuration, as made precise by the specific form of the causal Lagrangian
(for details see Section~\ref{seccapreduced}). As a result of this minimization process, the
physical wave functions satisfy dynamical equations in spacetime, the so-called
{\em{dynamical wave equation}}. In this paper, we do not need to enter the analysis of
the dynamical wave equation (for details see~\cite{dirac}). But it is important to keep the following
points in mind:
\bitem
\item[(1)] In many examples and certain limiting cases, the dynamical wave equation
coincides with the Dirac equation.
\item[(2)] In general, the dynamical wave equation differs from the Dirac equation
by small corrections.
\eitem
Our mechanism of fermiogenesis is based on a specific correction to the Dirac equation
obtained by analyzing the nonlinear dynamics as described by the causal action principle.
\subsection{Conceptual Basis  of our Mechanism of Fermiogenesis}
We now outline the conceptual ideas underlying our mechanism of fermiogenesis.
In typical examples of causal fermion systems which describe Min\-kowski space
or a globally hyperbolic Lorentzian spacetime, the {\em{Dirac sea}} picture is built in,
meaning that the physical wave functions satisfy the Dirac equation (in agreement with
the above statement~(1)), and they include all the negative-frequency solutions
of the Dirac equation, with an ultraviolet regularization (for a detailed explanation of the Dirac sea picture see for example~\cite{srev}).
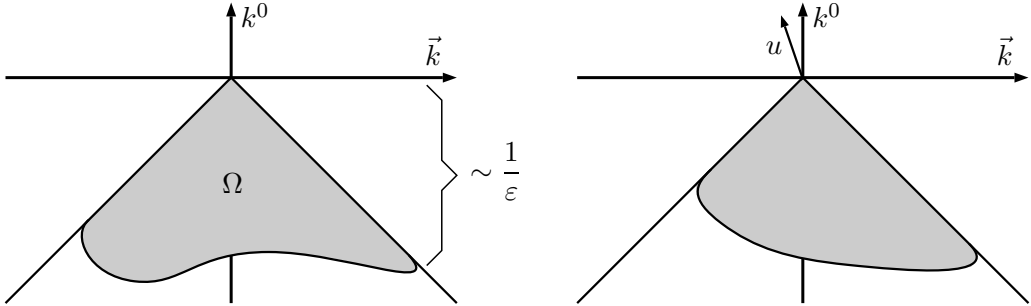
\begin{figure}
\psscalebox{1.0 1.0} 
{
\begin{pspicture}(0,27.809563)(13.660338,31.8699)
\definecolor{colour0}{rgb}{0.8,0.8,0.8}
\psline[linecolor=black, linewidth=0.04, arrowsize=0.05291667cm 2.0,arrowlength=1.4,arrowinset=0.0]{->}(10.620605,27.830168)(10.620605,31.830168)
\psline[linecolor=colour0, linewidth=0.01, fillstyle=solid,fillcolor=colour0](10.625606,30.81017)(12.915606,28.520168)(12.915606,28.420168)(12.855605,28.320168)(12.665606,28.290169)(12.025605,28.280169)(11.165606,28.360168)(10.745605,28.40017)(10.375606,28.492392)(10.015606,28.600168)(9.615605,28.840168)(9.375606,29.050169)(9.215606,29.270168)(9.2056055,29.420168)(10.175606,30.380169)
\psline[linecolor=black, linewidth=0.04, arrowsize=0.05291667cm 2.0,arrowlength=1.4,arrowinset=0.0]{->}(3.0206056,27.830168)(3.0206056,31.830168)
\pspolygon[linecolor=colour0, linewidth=0.01, fillstyle=solid,fillcolor=colour0](3.0656054,30.760168)(5.4556055,28.40017)(5.4756055,28.31017)(5.443383,28.25128)(5.2156053,28.260168)(4.8056054,28.340168)(4.3556056,28.440168)(3.8456054,28.500168)(3.3256054,28.505169)(2.9556055,28.460169)(2.5556054,28.330168)(2.3156054,28.23017)(2.1056054,28.15017)(1.9456055,28.120169)(1.7144943,28.115725)(1.5311611,28.159058)(1.2711611,28.304613)(1.1167166,28.459057)(1.0367166,28.670168)(1.0467166,28.870169)(3.0356054,30.820168)
\psline[linecolor=black, linewidth=0.04, arrowsize=0.05291667cm 2.0,arrowlength=1.4,arrowinset=0.0]{->}(0.02060547,30.830168)(6.0206056,30.830168)
\psline[linecolor=black, linewidth=0.03](3.0206056,30.830168)(6.0206056,27.830168)
\psline[linecolor=black, linewidth=0.03](3.0206056,30.830168)(0.02060547,27.830168)
\psbezier[linecolor=black, linewidth=0.03](1.1294943,28.939056)(0.89060545,28.747946)(1.1093776,28.192331)(1.7406055,28.12016845703125)(2.3718333,28.048004)(2.5206056,28.510168)(3.5206056,28.510168)(4.5206056,28.510168)(5.8618336,27.928005)(5.3806057,28.47017)
\psline[linecolor=black, linewidth=0.04, arrowsize=0.05291667cm 2.0,arrowlength=1.4,arrowinset=0.0]{->}(7.6206055,30.830168)(13.620605,30.830168)
\psline[linecolor=black, linewidth=0.03](10.620605,30.830168)(7.6206055,27.830168)
\psline[linecolor=black, linewidth=0.03](10.620605,30.830168)(13.620605,27.830168)
\psline[linecolor=black, linewidth=0.02](5.6206055,30.73017)(5.8206053,30.530169)(5.8206053,29.73017)(6.0206056,29.530169)(5.8206053,29.330168)(5.8206053,28.530169)(5.6206055,28.330168)
\psbezier[linecolor=black, linewidth=0.03](9.399494,29.607946)(9.182828,29.403502)(9.099378,29.222332)(9.540606,28.89016845703125)(9.981833,28.558004)(10.440605,28.420168)(11.210606,28.340168)(11.980605,28.260168)(13.311833,28.138004)(12.8306055,28.620169)
\psline[linecolor=black, linewidth=0.03, arrowsize=0.05291667cm 2.0,arrowlength=1.4,arrowinset=0.0]{->}(10.610605,30.840168)(10.3306055,31.660168)
\rput[bl](3.15,31.5){$k^0$}
\rput[bl](10.75,31.5){$k^0$}
\rput[bl](5.6,31){$\vec{k}$}
\rput[bl](13.2,31){$\vec{k}$}
\rput[bl](6.2,29.2){$\displaystyle \sim \frac{1}{\varepsilon}$}
\rput[bl](10.13,31.15){$u$}
\rput[bl](2.9,29.3){$\Omega$}
\end{pspicture}
}
\caption{Examples of regularized Dirac seas in Minkowski space.}
\label{figseareg}
\end{figure}%
The key point is that the regularization depends on the spacetime point.
In particular, the number of states needed for building up the Dirac sea may change
in time on cosmological scales. Since the number of physical wave functions remains unchanged, this
means that, starting from a system with a completely filled Dirac sea, at a later time
there may be too many or too few physical wave functions to fill up the sea.
If this is the case, either we have physical wave functions ``left over'' which must
occupy particle states, or else there will remain ``holes'' in the Dirac sea describing anti-particles. 
Therefore, in the effective description of the causal fermion system in familiar
language, the system will evolve from a vacuum spacetime to a spacetime with a matter/anti-matter asymmetry.
In this way, we obtain a dynamical mechanism of fermiogenesis. Note that this mechanism depends in an essential way on the unified description of gravity and matter provided by the theory of Causal Fermion Systems. This is exactly the kind of novel angle at classic problems which one would expect for a successful unification of General Relativity and Quantum Theory. 
Before going on, we note that the overall sign of our mechanism is of no relevance,
because in the effective matter/anti-matter description, what we call matter and anti-matter
is merely a convention.
We also point out that our effect of fermiogenesis does not merely describe local fluctuations
of particles versus anti-particles, but a global shift in one direction (towards particles or anti-particles),
as quantified by integrals over Cauchy surfaces.

\subsection{The Dirac Dynamics of the Regularization}
We now explain our mechanism of fermiogenesis in some more detail. 
In order to describe the vacuum in {\em{Minkowski space}}, we work with a hard cutoff, meaning that we
occupy all the states with momenta~$k \in \Omega$, where~$\Omega$ is a subset of the lower mass
shell which extends up to an energy scale of the order of the Planck energy
(or, equivalent, of the order~$\varepsilon^{-1}$, where~$\varepsilon$ is the regularization length);
see the left of Figure~\ref{figseareg}
(as the rest mass is much smaller than~$\varepsilon^{-1}$, the hyperbola looks like a cone).
A more concrete choice, shown on the right of Figure~\ref{figseareg},
is to occupy all the states with
\beq \label{ku}
k^2 = m^2 \qquad \text{and} \qquad -1 < k u < 0 \:,
\eeq
where~$u$ is a future-directed timelike vector whose length~$\sqrt{u^2} \sim \varepsilon$
is the regularization scale
(here~$k^2$ and~$ku$ denote the Minkowski inner
product with signature~$(+ - -\, -)$; for more details see~\eqref{Preghom} and~\eqref{Pregu}
in Section~\ref{secudef}).
Clearly, the regularization breaks Lorentz invariance, as becomes apparent from the fact that the
vector~$u$ distinguishes a time direction.
Nevertheless, it is not clear whether this violation of Lorentz invariance due to the regularization
has any effects on the dynamics of the system which could be captured by
classical tests of Lorentz invariance (for more details see the discussion below).

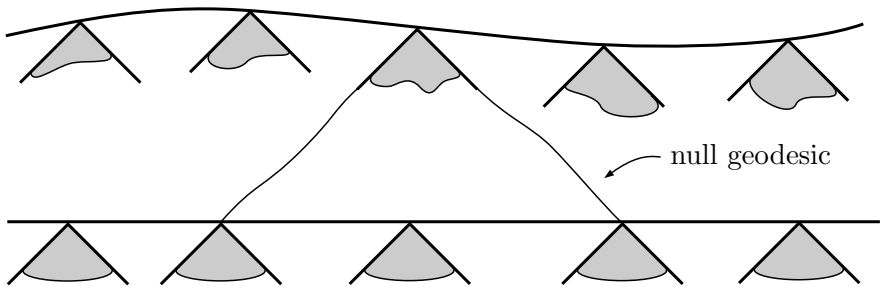
\begin{figure}
\psscalebox{1.0 1.0} 
{
\begin{pspicture}(0,25.422829)(11.644209,29.142332)
\definecolor{colour0}{rgb}{0.8,0.8,0.8}
\pspolygon[linecolor=colour0, linewidth=0.02, fillstyle=solid,fillcolor=colour0](5.489209,28.837252)(4.874209,28.21725)(4.879209,28.157251)(4.944209,28.10725)(5.064209,28.08225)(5.174209,28.10725)(5.269209,28.157251)(5.364209,28.167252)(5.459209,28.14225)(5.529209,28.08225)(5.609209,28.007252)(5.6792088,28.007252)(5.764209,28.09225)(5.834209,28.16225)(5.924209,28.18725)(6.019209,28.20725)(6.029209,28.292252)
\pspolygon[linecolor=colour0, linewidth=0.02, fillstyle=solid,fillcolor=colour0](0.83920896,26.229752)(0.23920898,25.639751)(0.279209,25.594751)(0.45420897,25.534752)(0.614209,25.514751)(0.83920896,25.504751)(1.024209,25.50975)(1.159209,25.524752)(1.334209,25.56475)(1.409209,25.604752)(1.419209,25.659752)
\pspolygon[linecolor=colour0, linewidth=0.02, fillstyle=solid,fillcolor=colour0](2.869209,26.22475)(2.274209,25.63475)(2.319209,25.594751)(2.444209,25.549751)(2.589209,25.524752)(2.8092089,25.50975)(2.999209,25.50975)(3.1842089,25.51975)(3.359209,25.56475)(3.444209,25.60975)(3.449209,25.66975)
\pspolygon[linecolor=colour0, linewidth=0.02, fillstyle=solid,fillcolor=colour0](5.386709,26.23475)(4.786709,25.64475)(4.826709,25.59975)(4.941709,25.55475)(5.121709,25.51975)(5.291709,25.50975)(5.456709,25.50975)(5.611709,25.514751)(5.771709,25.54475)(5.876709,25.569752)(5.956709,25.614752)(5.966709,25.66975)
\pspolygon[linecolor=colour0, linewidth=0.02, fillstyle=solid,fillcolor=colour0](8.209209,26.229752)(7.619209,25.65475)(7.634209,25.614752)(7.684209,25.57975)(7.789209,25.549751)(7.889209,25.534752)(8.029209,25.50975)(8.239209,25.50975)(8.389209,25.514751)(8.539209,25.52975)(8.674209,25.559752)(8.769209,25.604752)(8.804209,25.65475)
\pspolygon[linecolor=colour0, linewidth=0.02, fillstyle=solid,fillcolor=colour0](10.544209,26.229752)(9.969209,25.664751)(9.984209,25.62475)(10.029209,25.594751)(10.109209,25.569752)(10.209209,25.54475)(10.354209,25.52975)(10.484209,25.51975)(10.674209,25.51975)(10.824209,25.534752)(10.979209,25.55475)(11.094209,25.594751)(11.159209,25.659752)(10.569209,26.24975)
\pspolygon[linecolor=colour0, linewidth=0.02, fillstyle=solid,fillcolor=colour0](10.414209,28.674751)(9.914209,28.174751)(9.924209,28.094751)(9.969209,28.01975)(10.114209,27.89475)(10.279209,27.81475)(10.409209,27.784752)(10.499209,27.809752)(10.534209,27.88475)(10.584209,27.924751)(10.694209,27.93975)(10.844209,27.95475)(10.989209,27.959751)(11.039209,27.989752)(11.039209,28.059752)
\pspolygon[linecolor=colour0, linewidth=0.02, fillstyle=solid,fillcolor=colour0](7.961709,28.60975)(7.466709,28.084751)(7.476709,28.049751)(7.546709,28.024752)(7.666709,27.989752)(7.771709,27.95475)(7.861709,27.909752)(7.921709,27.844751)(7.951709,27.77975)(8.051709,27.72475)(8.201709,27.69975)(8.396709,27.684752)(8.521709,27.69975)(8.621709,27.739752)(8.676709,27.79475)(8.681709,27.89475)
\pspolygon[linecolor=colour0, linewidth=0.02, fillstyle=solid,fillcolor=colour0](3.266709,29.05475)(2.716709,28.50975)(2.711709,28.459751)(2.741709,28.399752)(2.776709,28.364752)(2.841709,28.344751)(2.936709,28.32975)(3.026709,28.32975)(3.116709,28.34975)(3.186709,28.39475)(3.241709,28.444752)(3.306709,28.479752)(3.416709,28.49975)(3.551709,28.504751)(3.666709,28.49975)(3.7467089,28.51975)(3.771709,28.559752)
\pspolygon[linecolor=colour0, linewidth=0.02, fillstyle=solid,fillcolor=colour0](1.004209,28.914751)(0.34920898,28.264751)(0.369209,28.22475)(0.464209,28.219751)(0.549209,28.254751)(0.65420896,28.309752)(0.77420896,28.364752)(0.874209,28.409752)(1.054209,28.41975)(1.2392089,28.424751)(1.3642089,28.43975)(1.404209,28.469751)(1.404209,28.51975)
\psline[linecolor=black, linewidth=0.04](0.034208983,26.27975)(11.634209,26.27975)
\psbezier[linecolor=black, linewidth=0.04](0.014208985,28.754036)(1.0281054,28.972237)(1.961896,29.16397)(3.0858757,29.0997509765625)(4.209855,29.035532)(6.1739388,28.76795)(7.566709,28.659752)(8.959479,28.551552)(10.670042,28.644037)(11.414209,28.91118)
\psline[linecolor=black, linewidth=0.04](2.464209,28.26975)(3.264209,29.069752)(4.064209,28.26975)
\psline[linecolor=black, linewidth=0.04](4.684209,28.039751)(5.484209,28.83975)(6.284209,28.039751)
\psline[linecolor=black, linewidth=0.04](0.20420898,28.129751)(1.004209,28.92975)(1.804209,28.129751)
\psline[linecolor=black, linewidth=0.04](9.614209,27.889751)(10.414209,28.68975)(11.214209,27.889751)
\psrotate(10.556653, 25.643196){0.075457245}{\psellipse[linecolor=black, linewidth=0.02, dimen=outer](10.556653,25.643196)(0.6022222,0.145)}
\pspolygon[linecolor=colour0, linewidth=0.02, fillstyle=solid,fillcolor=colour0](10.094987,25.78764)(9.979987,25.668196)(11.138876,25.66875)(11.009987,25.79764)
\psline[linecolor=black, linewidth=0.04](9.762209,25.46375)(10.562209,26.26375)(11.362209,25.46375)
\psrotate(0.83587563, 25.626417){0.075457245}{\psellipse[linecolor=black, linewidth=0.02, dimen=outer](0.83587563,25.62642)(0.6022222,0.145)}
\pspolygon[linecolor=colour0, linewidth=0.02, fillstyle=solid,fillcolor=colour0](0.374209,25.770863)(0.25920898,25.651417)(1.4180979,25.651974)(1.289209,25.780863)
\psline[linecolor=black, linewidth=0.04](0.041431207,25.446974)(0.8414312,26.246973)(1.6414312,25.446974)
\psrotate(8.206764, 25.631084){0.075457245}{\psellipse[linecolor=black, linewidth=0.02, dimen=outer](8.206764,25.631084)(0.6022222,0.145)}
\pspolygon[linecolor=colour0, linewidth=0.02, fillstyle=solid,fillcolor=colour0](7.7450976,25.775528)(7.630098,25.656084)(8.788987,25.65664)(8.660098,25.785528)
\psline[linecolor=black, linewidth=0.04](7.41232,25.45164)(8.21232,26.25164)(9.0123205,25.45164)
\psrotate(2.8697646, 25.628084){0.075457245}{\psellipse[linecolor=black, linewidth=0.02, dimen=outer](2.8697643,25.628084)(0.6022222,0.145)}
\pspolygon[linecolor=colour0, linewidth=0.02, fillstyle=solid,fillcolor=colour0](2.408098,25.77253)(2.293098,25.653084)(3.4519868,25.65364)(3.323098,25.782528)
\psline[linecolor=black, linewidth=0.04](2.07532,25.448639)(2.8753202,26.24864)(3.6753201,25.448639)
\psrotate(5.3795424, 25.631418){0.075457245}{\psellipse[linecolor=black, linewidth=0.02, dimen=outer](5.3795424,25.631418)(0.6022222,0.145)}
\pspolygon[linecolor=colour0, linewidth=0.02, fillstyle=solid,fillcolor=colour0](4.917876,25.775862)(4.8028755,25.656418)(5.9617643,25.656973)(5.8328757,25.785862)
\psline[linecolor=black, linewidth=0.04](4.585098,25.451973)(5.385098,26.251974)(6.1850977,25.451973)
\psline[linecolor=black, linewidth=0.04](7.164209,27.809752)(7.964209,28.60975)(8.764209,27.809752)
\psbezier[linecolor=black, linewidth=0.02](6.269209,28.06475)(6.704651,27.638966)(6.9894323,27.558437)(7.274209,27.2697509765625)(7.5589857,26.981066)(7.9336495,26.52653)(8.189209,26.27975)
\psbezier[linecolor=black, linewidth=0.02](4.689209,28.049751)(4.360473,27.751177)(4.4022326,27.680891)(3.914209,27.1997509765625)(3.4261854,26.71861)(3.340139,26.803518)(2.884209,26.289751)
\psbezier[linecolor=black, linewidth=0.02](4.959209,28.31225)(4.705848,28.099043)(5.020064,28.06617)(5.119209,28.0822509765625)(5.218354,28.098331)(5.36495,28.241209)(5.519209,28.07225)(5.673468,27.903294)(5.721093,28.053095)(5.819209,28.132252)(5.9173245,28.211407)(6.1460047,28.152262)(5.999209,28.327251)
\psbezier[linecolor=black, linewidth=0.02](0.344209,28.27725)(0.35617533,28.090395)(0.62701833,28.310284)(0.859209,28.3822509765625)(1.0913997,28.454218)(1.674051,28.322025)(1.284209,28.64225)
\psbezier[linecolor=black, linewidth=0.02](2.789209,28.58225)(2.4961753,28.355394)(3.0420184,28.215282)(3.219209,28.4072509765625)(3.3963997,28.599218)(4.0490513,28.357025)(3.659209,28.677252)
\psbezier[linecolor=black, linewidth=0.02](9.944209,28.227251)(9.741176,28.035395)(10.417018,27.615284)(10.519209,27.8422509765625)(10.6214,28.069218)(11.339051,27.767025)(10.924209,28.177252)
\psbezier[linecolor=black, linewidth=0.02](7.524209,28.157251)(7.2511754,27.980394)(7.8320184,28.030283)(7.924209,27.8122509765625)(8.016399,27.59422)(8.984051,27.637024)(8.594209,27.95725)
\psbezier[linecolor=black, linewidth=0.02, arrowsize=0.05291667cm 2.0,arrowlength=1.4,arrowinset=0.0]{->}(8.719209,27.139751)(8.524209,27.16975)(8.339209,27.12475)(7.969209,26.8597509765625)
\rput[bl](8.85,26.97){null geodesic}
\end{pspicture}
}
\caption{The regularization with Dirac dynamics.}
\label{figcone1}
\end{figure}

If we are in {\em{curved spacetime}}, the regularization may change on large scales.
Nevertheless, choosing a local reference frame, one can still work with the Fourier transform,
making it possible to describe the regularization in the neighborhood of any spacetime point
again as in Figure~\ref{figseareg} (but with the set~$\Omega$ or the vector~$u$ depending
on the spacetime point). This concept will be made precise in this paper.
Anticipating these results, we can visualize the regularization as shown in Figures~\ref{figcone1}
and~\ref{figcone2} by plotting a regularized cone at each spacetime point.
An important point to keep in mind is that the dynamics of the regularization
depends on and is determined by the dynamics of the physical wave functions.
The simplest and most familiar dynamics is the one described by the Dirac equation.
Already this {\em{Dirac dynamics}} of the wave functions
gives rise to a non-trivial and uniquely determined dynamics of the regularization.
More specifically, as first observed in~\cite{dgc} and worked out in detail in~\cite{reghadamard},
the dynamics of the regularization is described by transport equations along the light cone.
As a result, even if we begin initially with a regularization by a vector~$u$
(as shown on the right of Figure~\ref{figseareg}), the regularization at later times will no longer
have this simple structure. Indeed, at each spacetime point it will typically have the more general
form as shown on the left of Figure~\ref{figseareg}. This dynamical behavior is
illustrated in Figure~\ref{figcone1}.
From the point of view of fermiogenesis, the Dirac dynamics is an important starting point
for our analysis. 
As will be shown in detail in Section~\ref{secnobaryo}, the Dirac dynamics does {\em{not}} give rise to 
fermiogenesis, in agreement with the standard picture
that it allows only for pair creation while preserving the matter/anti-matter symmetry.
\subsection{Modifications of the Dirac Dynamics on Large Scales}
What makes our analysis interesting is that the Dirac dynamics is not the final answer.
Indeed, the dynamics of the regularization is determined by the causal action principle.
The causal action principle gives agreement with the Dirac dynamics only in a certain limiting
case (the so-called continuum limit). 
In this limiting case, the effective description of the dynamics is given by classical field theory.
In particular, one recovers the Maxwell and Einstein equations, while the underlying regularization
imprints only on the effective coupling constants.

\begin{figure}
\psscalebox{1.0 1.0} 
{
\begin{pspicture}(0,25.422829)(11.644209,29.142332)
\definecolor{colour0}{rgb}{0.8,0.8,0.8}
\pspolygon[linecolor=colour0, linewidth=0.02, fillstyle=solid,fillcolor=colour0](9.876709,27.389751)(9.251709,26.67975)(9.271709,26.629751)(9.396709,26.584751)(9.631709,26.569752)(9.836709,26.584751)(10.071709,26.62475)(10.281709,26.684752)(10.421709,26.74975)(10.471709,26.79475)(10.466709,26.87475)
\pspolygon[linecolor=colour0, linewidth=0.02, fillstyle=solid,fillcolor=colour0](7.009209,27.434752)(6.434209,26.879751)(6.459209,26.809752)(6.604209,26.73475)(6.859209,26.684752)(7.104209,26.664751)(7.359209,26.67975)(7.544209,26.719751)(7.614209,26.774752)(7.624209,26.83975)
\pspolygon[linecolor=colour0, linewidth=0.02, fillstyle=solid,fillcolor=colour0](4.301709,27.57975)(3.766709,27.06475)(3.801709,26.989752)(3.896709,26.92975)(4.076709,26.864752)(4.361709,26.82975)(4.576709,26.82975)(4.751709,26.85975)(4.8667088,26.909752)(4.911709,26.96475)
\pspolygon[linecolor=colour0, linewidth=0.02, fillstyle=solid,fillcolor=colour0](2.539209,27.44975)(2.009209,26.899752)(2.029209,26.844751)(2.154209,26.784752)(2.339209,26.739752)(2.534209,26.729752)(2.694209,26.73475)(2.884209,26.754751)(3.034209,26.799751)(3.114209,26.854752)(3.119209,26.93975)
\pspolygon[linecolor=colour0, linewidth=0.02, fillstyle=solid,fillcolor=colour0](10.394209,28.682251)(9.839209,28.08225)(9.864209,28.03725)(9.909209,27.96725)(10.024209,27.932251)(10.174209,27.90225)(10.314209,27.89225)(10.529209,27.89225)(10.699209,27.917252)(10.869209,27.95725)(10.979209,28.01725)(11.019209,28.07225)(11.014209,28.12725)
\pspolygon[linecolor=colour0, linewidth=0.02, fillstyle=solid,fillcolor=colour0](7.949209,28.597252)(7.264209,27.862251)(7.294209,27.81225)(7.454209,27.75225)(7.709209,27.727251)(7.954209,27.73225)(8.149209,27.75225)(8.309209,27.77725)(8.479209,27.827251)(8.604209,27.87725)(8.644209,27.96725)
\pspolygon[linecolor=colour0, linewidth=0.02, fillstyle=solid,fillcolor=colour0](5.486709,28.82225)(4.766709,28.112251)(4.786709,28.06225)(4.881709,28.007252)(5.056709,27.977251)(5.236709,27.98225)(5.456709,27.987251)(5.656709,28.01225)(5.896709,28.07225)(6.006709,28.11725)(6.071709,28.157251)(6.101709,28.212252)
\pspolygon[linecolor=colour0, linewidth=0.02, fillstyle=solid,fillcolor=colour0](3.284209,29.06225)(2.579209,28.42225)(2.584209,28.362251)(2.694209,28.29725)(2.864209,28.24225)(3.074209,28.20725)(3.264209,28.19725)(3.479209,28.192251)(3.669209,28.20725)(3.814209,28.237251)(3.904209,28.27725)(3.944209,28.352251)
\pspolygon[linecolor=colour0, linewidth=0.02, fillstyle=solid,fillcolor=colour0](0.82920897,27.54725)(0.35420898,27.02725)(0.394209,26.977251)(0.514209,26.927252)(0.689209,26.87725)(0.83920896,26.85725)(0.989209,26.84225)(1.1192089,26.837252)(1.264209,26.84225)(1.399209,26.862251)(1.459209,26.89225)(1.484209,26.94725)
\pspolygon[linecolor=colour0, linewidth=0.02, fillstyle=solid,fillcolor=colour0](0.979209,28.92225)(0.539209,28.417252)(0.57920897,28.347252)(0.71920896,28.26225)(0.859209,28.212252)(1.039209,28.16225)(1.204209,28.132252)(1.3692089,28.122252)(1.549209,28.122252)(1.679209,28.157251)(1.724209,28.20725)(1.699209,28.292252)
\pspolygon[linecolor=colour0, linewidth=0.02, fillstyle=solid,fillcolor=colour0](0.83920896,26.229752)(0.23920898,25.639751)(0.279209,25.594751)(0.45420897,25.534752)(0.614209,25.514751)(0.83920896,25.504751)(1.024209,25.50975)(1.159209,25.524752)(1.334209,25.56475)(1.409209,25.604752)(1.419209,25.659752)
\pspolygon[linecolor=colour0, linewidth=0.02, fillstyle=solid,fillcolor=colour0](2.869209,26.22475)(2.274209,25.63475)(2.319209,25.594751)(2.444209,25.549751)(2.589209,25.524752)(2.8092089,25.50975)(2.999209,25.50975)(3.1842089,25.51975)(3.359209,25.56475)(3.444209,25.60975)(3.449209,25.66975)
\pspolygon[linecolor=colour0, linewidth=0.02, fillstyle=solid,fillcolor=colour0](5.7733755,26.241417)(5.1733756,25.651417)(5.2133756,25.606417)(5.328376,25.561419)(5.5083756,25.526417)(5.6783757,25.516418)(5.8433757,25.516418)(5.9983754,25.521418)(6.1583757,25.551418)(6.2633758,25.576418)(6.3433757,25.621418)(6.3533754,25.676418)
\pspolygon[linecolor=colour0, linewidth=0.02, fillstyle=solid,fillcolor=colour0](8.235876,26.229752)(7.6458755,25.65475)(7.660876,25.614752)(7.7108755,25.57975)(7.8158755,25.549751)(7.9158754,25.534752)(8.055876,25.50975)(8.265876,25.50975)(8.415875,25.514751)(8.565876,25.52975)(8.700875,25.559752)(8.795876,25.604752)(8.830875,25.65475)
\pspolygon[linecolor=colour0, linewidth=0.02, fillstyle=solid,fillcolor=colour0](10.544209,26.229752)(9.969209,25.664751)(9.984209,25.62475)(10.029209,25.594751)(10.109209,25.569752)(10.209209,25.54475)(10.354209,25.52975)(10.484209,25.51975)(10.674209,25.51975)(10.824209,25.534752)(10.979209,25.55475)(11.094209,25.594751)(11.159209,25.659752)(10.569209,26.24975)
\psline[linecolor=black, linewidth=0.04](0.034208983,26.27975)(11.634209,26.27975)
\psbezier[linecolor=black, linewidth=0.04](0.014208985,28.754036)(1.0281054,28.972237)(1.961896,29.16397)(3.0858757,29.0997509765625)(4.209855,29.035532)(6.1739388,28.76795)(7.566709,28.659752)(8.959479,28.551552)(10.670042,28.644037)(11.414209,28.91118)
\psrotate(10.556653, 25.643196){0.075457245}{\psellipse[linecolor=black, linewidth=0.02, dimen=outer](10.556653,25.643196)(0.6022222,0.145)}
\pspolygon[linecolor=colour0, linewidth=0.02, fillstyle=solid,fillcolor=colour0](10.094987,25.78764)(9.979987,25.668196)(11.138876,25.66875)(11.009987,25.79764)
\psline[linecolor=black, linewidth=0.04](9.762209,25.46375)(10.562209,26.26375)(11.362209,25.46375)
\psrotate(0.83587563, 25.626417){0.075457245}{\psellipse[linecolor=black, linewidth=0.02, dimen=outer](0.83587563,25.62642)(0.6022222,0.145)}
\pspolygon[linecolor=colour0, linewidth=0.02, fillstyle=solid,fillcolor=colour0](0.374209,25.770863)(0.25920898,25.651417)(1.4180979,25.651974)(1.289209,25.780863)
\psline[linecolor=black, linewidth=0.04](0.041431207,25.446974)(0.8414312,26.246973)(1.6414312,25.446974)
\psrotate(8.233431, 25.631084){0.075457245}{\psellipse[linecolor=black, linewidth=0.02, dimen=outer](8.233432,25.631084)(0.6022222,0.145)}
\pspolygon[linecolor=colour0, linewidth=0.02, fillstyle=solid,fillcolor=colour0](7.7717648,25.775528)(7.6567645,25.656084)(8.815654,25.65664)(8.686765,25.785528)
\psline[linecolor=black, linewidth=0.04](7.438987,25.45164)(8.238987,26.25164)(9.038987,25.45164)
\psrotate(2.8697646, 25.628084){0.075457245}{\psellipse[linecolor=black, linewidth=0.02, dimen=outer](2.8697643,25.628084)(0.6022222,0.145)}
\pspolygon[linecolor=colour0, linewidth=0.02, fillstyle=solid,fillcolor=colour0](2.408098,25.77253)(2.293098,25.653084)(3.4519868,25.65364)(3.323098,25.782528)
\psline[linecolor=black, linewidth=0.04](2.07532,25.448639)(2.8753202,26.24864)(3.6753201,25.448639)
\psrotate(5.766209, 25.638084){0.075457245}{\psellipse[linecolor=black, linewidth=0.02, dimen=outer](5.766209,25.638084)(0.6022222,0.145)}
\pspolygon[linecolor=colour0, linewidth=0.02, fillstyle=solid,fillcolor=colour0](5.3045425,25.782528)(5.1895423,25.663084)(6.348431,25.66364)(6.2195425,25.792528)
\psline[linecolor=black, linewidth=0.04](4.9717646,25.45864)(5.7717648,26.25864)(6.5717645,25.45864)
\psbezier[linecolor=black, linewidth=0.04](0.04754232,27.480703)(1.0614387,27.698904)(1.6785626,27.387304)(3.149209,27.536417643229175)(4.6198554,27.685532)(6.147272,27.534618)(7.5400424,27.426418)(8.932813,27.318218)(10.5233755,27.450703)(11.367542,27.557846)
\psbezier[linecolor=black, linewidth=0.02](7.814209,26.656418)(7.9608755,26.476418)(8.040875,26.42975)(8.220876,26.276417643229166)
\psbezier[linecolor=black, linewidth=0.02](6.214209,26.663084)(6.0408754,26.469751)(5.834209,26.356417)(5.774209,26.269750976562467)
\psbezier[linecolor=black, linewidth=0.02](4.6908755,28.049751)(4.384209,27.74975)(4.5330977,27.845306)(4.3042088,27.600862087673608)
\psbezier[linecolor=black, linewidth=0.02](6.274209,28.043085)(6.6675425,27.716417)(6.7808757,27.663084)(6.994209,27.476417643229166)
\psrotate(1.139209, 28.309752){-10.053358}{\psellipse[linecolor=black, linewidth=0.02, dimen=outer](1.139209,28.309752)(0.6175,0.1825)}
\psrotate(0.919209, 26.96725){-5.4578414}{\psellipse[linecolor=black, linewidth=0.02, dimen=outer](0.919209,26.96725)(0.5775,0.14)}
\psrotate(3.256709, 28.34975){-3.0083165}{\psellipse[linecolor=black, linewidth=0.02, dimen=outer](3.256709,28.34975)(0.69,0.1775)}
\psrotate(2.571709, 26.881){0.5431756}{\psellipse[linecolor=black, linewidth=0.02, dimen=outer](2.571709,26.881)(0.57,0.17375)}
\psrotate(4.336709, 26.99975){-4.4698625}{\psellipse[linecolor=black, linewidth=0.02, dimen=outer](4.336709,26.99975)(0.58,0.1875)}
\psrotate(5.434209, 28.14225){5.368858}{\psellipse[linecolor=black, linewidth=0.02, dimen=outer](5.434209,28.14225)(0.6825,0.175)}
\psrotate(7.954209, 27.88725){3.8625417}{\psellipse[linecolor=black, linewidth=0.02, dimen=outer](7.954209,27.88725)(0.6975,0.175)}
\psrotate(7.036709, 26.817251){-2.8533757}{\psellipse[linecolor=black, linewidth=0.02, dimen=outer](7.036709,26.817251)(0.61,0.17)}
\psrotate(9.865459, 26.74225){7.8682613}{\psellipse[linecolor=black, linewidth=0.02, dimen=outer](9.865459,26.74225)(0.63375,0.175)}
\psrotate(10.439209, 28.059752){2.1876602}{\psellipse[linecolor=black, linewidth=0.02, dimen=outer](10.439209,28.059752)(0.605,0.1925)}
\pspolygon[linecolor=colour0, linewidth=0.02, fillstyle=solid,fillcolor=colour0](0.686709,28.577251)(0.561709,28.43725)(1.696709,28.29725)(1.491709,28.462252)
\psline[linecolor=black, linewidth=0.04](0.23370515,28.073889)(0.9755049,28.928133)(1.8297491,28.186333)
\pspolygon[linecolor=colour0, linewidth=0.02, fillstyle=solid,fillcolor=colour0](2.6842089,28.532251)(2.579209,28.432251)(3.9242089,28.35725)(3.734209,28.552252)
\psline[linecolor=black, linewidth=0.04](2.4464543,28.307407)(3.2828474,29.069277)(4.044718,28.232883)
\pspolygon[linecolor=colour0, linewidth=0.02, fillstyle=solid,fillcolor=colour0](4.921709,28.272251)(4.786709,28.13725)(6.081709,28.23225)(5.921709,28.37725)
\psline[linecolor=black, linewidth=0.04](4.684209,28.039751)(5.484209,28.83975)(6.284209,28.039751)
\pspolygon[linecolor=colour0, linewidth=0.02, fillstyle=solid,fillcolor=colour0](10.044209,28.29725)(9.859209,28.09225)(10.999209,28.13725)(10.819209,28.28725)
\psline[linecolor=black, linewidth=0.04](9.636903,27.846247)(10.39198,28.688778)(11.23451,27.9337)
\pspolygon[linecolor=colour0, linewidth=0.02, fillstyle=solid,fillcolor=colour0](9.399209,26.86725)(9.249209,26.692251)(10.484209,26.847252)(10.349209,26.96725)
\psline[linecolor=black, linewidth=0.04](9.123211,26.558794)(9.875836,27.403515)(10.720558,26.650888)
\pspolygon[linecolor=colour0, linewidth=0.02, fillstyle=solid,fillcolor=colour0](6.639209,27.077251)(6.444209,26.882252)(7.624209,26.827251)(7.459209,26.987251)
\psline[linecolor=black, linewidth=0.04](6.2123075,26.666714)(7.0222096,27.456688)(7.8121834,26.646786)
\pspolygon[linecolor=colour0, linewidth=0.02, fillstyle=solid,fillcolor=colour0](3.966709,27.25225)(3.811709,27.067251)(4.921709,26.977251)(4.721709,27.15225)
\psline[linecolor=black, linewidth=0.04](3.494428,26.79567)(4.299991,27.590069)(5.094389,26.784506)
\pspolygon[linecolor=colour0, linewidth=0.02, fillstyle=solid,fillcolor=colour0](2.179209,27.069752)(2.014209,26.909752)(3.139209,26.914751)(2.944209,27.08975)
\psline[linecolor=black, linewidth=0.04](1.7722532,26.647558)(2.5424097,27.476328)(3.371179,26.706171)
\pspolygon[linecolor=colour0, linewidth=0.02, fillstyle=solid,fillcolor=colour0](0.504209,27.17975)(0.369209,27.034752)(1.484209,26.934752)(1.264209,27.119751)
\psline[linecolor=black, linewidth=0.04](0.0642027,26.711218)(0.8245642,27.548983)(1.6623293,26.78862)
\pspolygon[linecolor=colour0, linewidth=0.02, fillstyle=solid,fillcolor=colour0](7.484209,28.097252)(7.269209,27.872252)(8.649209,27.952251)(8.459209,28.112251)
\psline[linecolor=black, linewidth=0.04](7.1792855,27.780619)(7.949458,28.609373)(8.778213,27.839201)
\end{pspicture}
}
\caption{The regularization with locally rigid dynamics.}
\label{figcone2}
\end{figure}
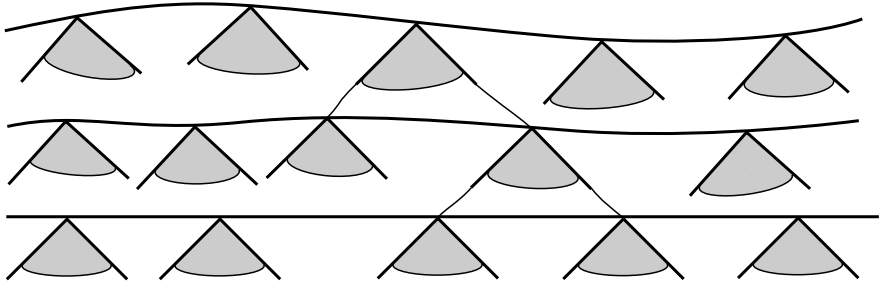

Going beyond the effective description in the continuum limit, the causal action principle
describes corrections to the Dirac equation, thereby also giving rise to fermiogenesis.
In contrast to the description in the continuum limit, these corrections will depend on the
regularization.
In order to quantify these effects, in the recent PhD thesis~\cite{jokeldr} it has been studied which types
of regularizations are compatible with the Euler-Lagrange (EL) equations of the causal action principle.
The results of this thesis suggest that the regularization is rather ``rigid'' in the sense that
changing the local form of the regularization typically violates the EL equations,
except for changes which describe Lorentz transformations and/or scalings of the system.
We here implement these findings by demanding that at each spacetime point the
regularization should have the form as shown on the right of Figure~\ref{figseareg}.
Such regularizations can be described by a vector field~$u(x)$, the so-called {\em{regularizing
vector field}}. We refer to the resulting dynamics of the regularization as the {\em{locally rigid dynamics}}. 
It is illustrated in Figure~\ref{figcone2}.
As we shall see, the locally rigid dynamics does give rise to fermiogenesis.
We shall also analyze the resulting effect quantitatively.

Before going on, we point out that the locally rigid dynamics
signifies a major departure from the Dirac dynamics.
It is motivated by the theory of causal fermion systems and, so far,
has been studied only in this context.
Since the goal of this paper is to work out the applications to fermiogenesis,
here we shall not derive the locally rigid dynamics from first principles.
Instead, we take it as an ad-hoc assumption and analyze its implications.
Moreover, for conceptual clarity we here focus on
fermiogenesis, but disregard other effects such as the
possibility of dynamically changing coupling constants, similar as suggested for the Dirac
dynamics in~\cite{dgc}.

Working out fermiogenesis for a locally rigid dynamics makes it necessary to address the
following two questions:
\bitem
\item[\rm{(a)}] Given the regularizing vector field~$u$, how is the Dirac dynamics to be modified?
How large is the resulting effect of fermiogenesis?
\item[\rm{(b)}] What is the dynamics of the regularizing vector field~$u$?
\eitem
Here we focus mainly on question~(a). In this way, we demonstrate that fermiogenesis does occur for
causal fermion systems, and we work out this effect quantitatively, given a regularizing vector field~$u$.
Question~(b) is considerably harder and
not yet fully understood. One main difficulty is that fermiogenesis might have a
back reaction on the regularizing vector field, an effect that we cannot address here.
Disregarding such back reactions, in Appendix~\ref{appA} two slightly different dynamical
equations are proposed. These dynamical equations should not be considered as the final answer,
but instead as a suitable starting point for
a more detailed modeling of the dynamical equations for~$u$.

We now explain a bit more in detail how we analyze question~(a).
Modifying the Dirac dynamics is a rather subtle issue because we need to preserve
a number of conservation laws. For the Dirac dynamics, these conservation laws
correspond to current conservation for all wave functions. When going over to the dynamics described by the
causal action principle, there are corresponding conservation laws formulated in terms
of so-called surface layer integrals (see~\cite{noether, osi}).
These conservation laws pose strong constraints for how the dynamics can
be modified. Making use of these constraints, given a global foliation~$(\scrN_t)_{t \in \R}$ of spacetime,
we obtain a canonical modification of the Dirac dynamics to a locally rigid dynamics.
This is accomplished by forming adiabatic projections to spectral subspaces of
a spatial operator~$\tilde{A}(t)$ constructed from the regularizing vector field~$u|_{\scrN_t}$ at time~$t$.
To leading order, the resulting rate of fermiogenesis~$B(t)$ is given by the simple formula
\beq \label{approxintro}
B(t) = - \tr \big( \tilde{E}_{-m}(t) \: \dot{\tilde{A}}(t) \big) \:,
\eeq
where~$\tilde{E}_{-m}(t)\, dm$ is the spectral measure of the interacting Hamiltonian
at energy~$-m$ corresponding to the cusp of the lower mass shell
(for detail see Section~\ref{secapprox}). This formula shows that fermiogenesis arises if the
eigenfunctions of~$\tilde{A}(t)$ cross the ``sea level'' as described by the eigenvalue $-m$ of the Hamiltonian.
\subsection{General Remarks}
The locally rigid dynamics resulting from the adiabatic projections
has the property that it is a small modification of the Dirac
dynamics relevant only on cosmological scales.
Even more, the Dirac dynamics agrees with the locally rigid dynamics in
spatially homogeneous FLRW universes, simply because in this case a regularization which is initially spherically symmetric and homogeneous will remain so under the Dirac dynamics, thereby giving agreement with the
locally rigid dynamics for a spatially homogeneous regularizing vector field.
These findings are in agreement with the fact that fermiogenesis is not observed in today's universe.
Moreover, these considerations show deviations of the metric from FLRW,
either through spatial fluctuations or bubble formation, are essential for our mechanism
of fermiogenesis. Working out the mechanism quantitatively for various relevant spacetime configurations
goes beyond the scope of this paper. Here we merely
discuss some aspects relevant to these future applications as well as a possible physical interpretation of the
here-introduced mechanism in the outlook section of this paper.
In particular, we will discuss the link to the blue print for a modified history of the universe introduced in~\cite{paganini2020proposal} on the basis of preliminary ideas leading to the present work.

We close the introduction with a few remarks:
\bitem
\itemD We point out that the regularization vector field gives a {\em{preferred observer}} for the dynamics of the fermionic projector. It is not clear to what extent this breaking of Lorentz invariance carries over to the dynamics of the effective fields. However, in the context of the matter/anti-matter asymmetry this is not an issue, as on universal scale we do in fact have a preferred rest frame defined by a vanishing of the dipole moment in the CMB power spectrum. 
\itemD Let us discuss how and to which extent our mechanism of fermiogenesis
satisfies {\em{Sakharov's criteria}}~\cite{Sakharov:1967dj}.
We first note that, under both $C$ and $CP$ inversions, states of the Dirac sea are
interchanged with particle states and vice versa. In this way, the causal fermion system
entails an asymmetry naturally satisfying Sakharov's second condition. The focus of the present work will be to show that in a suitably general setting, Sakharov's first criterion of baryon number non-conservation, can be satisfied, although in the form of fermion number non-conservation across all fermionic sectors (i.e.\ equally for all leptons and quarks\footnote{This guarantees for the resulting matter content to be neutral with respect to all charges.}). 
We finally comment on Sakharov's last condition, non-thermality. It is not entirely clear how it is satisfied in the 
context of the here presented mechanism, as the mechanism does not really entail any sort of reaction rates that would need to be out of equilibrium. Instead, the third criterion is replaced by the requirement that~\eqref{approxintro} be large enough in the relevant spacetime configuration.
Independent of the details of the mechanism, we know that all particles (in all fermionic sectors and in
all generations) are created in the Fermi ground state (i.e.\ the minimal energy configuration allowed by the Pauli exclusion principle). As discussed in~\cite{paganini2020proposal} this might have interesting consequences on reheating and the subsequent kinetics of dark matter. \\[-1em]
\itemD As a final observation, we want to remark that in the development of our new mechanism for fermiogenesis one inevitably comes to introduce {\em{non-Riemannian-measures}} in the effective description. Modified gravity theories with non-Rie\-mannian-measures were originally introduced in \cite{guendelman1996principle, earlynrmt1, earlynrmt2} with the aim to remove the dependence of the gravitational field equations on an additive constant to the Lagrangian in the action. In recent work \cite{nrmtinflation1, nrmtinflation2} it was shown that these modifications in fact can give rise to an asymptotically de Sitter universe with an inflationary early phase. For a review on non-Riemannian-measure-theories see \cite{nrmtreview}. Note, that it is quite intuitive from the construction of the local correlation map that Causal Fermion Systems would be compatible with non-Riemannian-measure theories. What is surprising is that to obtain fermiogenesis these theories are in fact inevitable in the effective description. 
\eitem
\subsection{Organization of the Paper}
The paper is organized as follows. In Section~\ref{sec:prel} we introduce the general concepts the reader needs to be familiar with. In Section~\ref{sec:cfs} we introduce the theory of causal fermion systems with a focus on those aspects that are essential for the present paper. In Section~\ref{secdiraccosmo} we introduce the relevant conservation laws of causal fermion systems and how they relate to integral quantities in the effective spacetime description. In Section~\ref{secrigid} we introduce the locally rigid regularization and the associated dynamics. Then in Section~\ref{sec:baryo} we introduce the tools to detect fermiogenesis, show the absence of fermiogenesis in the context of the Dirac dynamics and the presence thereof under the locally rigid dynamics. In Section~\ref{secperturb} we give a first order approximation of the rate of fermiogenesis under the locally rigid dynamics. In Section~\ref{sec:outlook} we discuss future research and possible physical interpretations of the fermiogenesis mechanism introduced in this paper. 
Finally, the appendices provide additional material and clarify certain aspects.
In Appendix~\ref{appA} we suggest dynamical equations for the regularizing vector field.
Appendix~\ref{appB} explains in which sense our results are robust to the choice of
foliations. In Appendix~\ref{appgauge} it is explained why the operator used in the detection of fermiogenesis
must be carefully adjusted to the local geometry.

\section{Preliminaries}\label{sec:prel}
\subsection{The Dirac Equation in Globally Hyperbolic Spacetimes}
let~$(\scrM, g)$ be a smooth, globally hyperbolic, time-oriented
Lorentzian spin manifold of dimension four.
For the signature of the metric we use the convention~$(+ ,-, -, -)$.
We denote the corresponding spinor bundle by~$S\scrM$. Its fibers~$S_x\scrM$ (with~$x \in \scrM$)
are endowed
with an inner product~$\Sl .|. \Sr_x$ of signature~$(2,2)$, the so-called {\em{spin inner product}}.
Clifford multiplication is described by a mapping~$\gamma$
which satisfies the anti-commutation relations, i.e.\
\[ 
\gamma \::\: T_x\scrM \rightarrow \Lin(S_x\scrM) \qquad
\text{and} \qquad \gamma(u) \,\gamma(v) + \gamma(v) \,\gamma(u) = 2 \, g(u,v)\,\1_{S_x\scrM} \]
 for all~$u, v \in T_x \scrM$ (and~$\1_{S_x\scrM}$ is the identity operator on the spinor space).
We also write Clifford multiplication in components with the Dirac matrices~$\gamma^j$.
The metric connections on the tangent bundle and the spinor bundle are denoted by~$\nabla$.
The sections of the spinor bundle are also referred to as wave functions.

We denote the smooth sections of the spinor bundle by~$C^\infty(\scrM, S\scrM)$.
The Dirac operator~$\Dir$ is defined by
\[ \Dir := i \gamma^j \nabla_j \::\: C^\infty(\scrM, S\scrM) \rightarrow C^\infty(\scrM, S\scrM)\:. \]
Given a real parameter~$m \in \R$ (the {\em{mass}}), the Dirac equation reads
\beq \label{Direq}
(\Dir - m) \,\psi = 0 \:.
\eeq

In the globally hyperbolic setting considered here
there is a global smooth foliation denoted by~$(\scrN_t)_{t \in \R}$
(for the proof see~\cite{bernal+sanchez}).
The parameter~$t$ is a global time function.
This time function can also be chosen as the time of local coordinates. Then the
metric takes the form
\beq \label{lapseshift}
ds^2 = N(x)^2\: dt^2 - 2 \sum_{\alpha}^3 \beta_\alpha(x) \: dt\: dx^\alpha
- \sum_{\alpha, \beta=1}^3 g_{\alpha \beta}(x)\: dx^\alpha\: dx^\beta \:.
\eeq
The function~$N$ and the spatial vector field~$\beta$ are referred to as the {\em{lapse function}}
and the {\em{shift vector}}, respectively
(for details see for example~\cite[Section~VI.3]{choquet}). Moreover, $g_{\alpha \beta}$ is the induced
Riemannian metric on the Cauchy surface~$\scrN_t$.

The Cauchy problem for the Dirac equation is well-posed.
In particular, taking smooth and compactly supported initial data,
one gets a unique global smooth solution. Moreover, due to finite propagation speed,
the restriction of this solution to any other Cauchy surface also has compact support.
We thus obtain solutions in the class~$\Cisc(\scrM, S\scrM)$ of smooth sections
with spatially compact support. On such solutions, one has the scalar product
\beq \label{print}
(\psi | \phi)_m = \int_\scrN \Sl \psi \,|\, \gamma(\nu)\, \phi \Sr_x\: d\mu_\scrN(x) \:,
\eeq
where~$\scrN$ denotes any Cauchy surface and~$\nu$ its future-directed normal
(due to current conservation, the scalar product is
in fact independent of the choice of~$\scrN$; for details see~\cite[Section~2]{finite}).
Forming the completion of the obtained scalar product space gives the Hilbert space~$(\H_m, (.|.)_m)$.

The fact that the Cauchy problem is well-posed makes it possible to introduce the
{\em{retarded}} and {\em{advanced Green's operators}}~$s_m^\wedge$ and~$s_m^\vee$ as
linear mappings (see for example~\cite{dimock3, baer+ginoux})
\[ s_m^\wedge, s_m^\vee \::\: C^\infty_0(\scrM, S\scrM) \rightarrow \Cisc(\scrM, S\scrM)\:. \]
They satisfy the defining equation of the Green's operator
\[ 
(\Dir - m) \left( s_m^{\wedge, \vee} \phi \right) = \phi \qquad \text{for all~$\phi \in C^\infty_0(\scrM, S\scrM)$}\:, \]
where~$C^\infty_0(\scrM, S\scrM)$ denotes the smooth sections with compact support.
Moreover, they are uniquely determined by the condition that the support of~$s_m^\wedge \phi$
(or~$s_m^\vee \phi$) lies in the future (respectively the past) of~$\supp \phi$.
The {\em{causal fundamental solution}}~$k_m$ is introduced by
\[ 
k_m := \frac{1}{2 \pi i} \left( s_m^\vee - s_m^\wedge \right) \::\: C^\infty_0(\scrM, S\scrM) \rightarrow \Cisc(\scrM, S\scrM) \cap \H_m \:. \]
It maps to a dense subspace of the Dirac solution space~$\H_m$.

Next, on the compactly supported wave functions (not necessarily Dirac solutions),
one can introduce a Lorentz invariant inner product by integrating the spin inner product over spacetime,
\begin{align*} 
\bra \psi|\phi \ket \,&:\: C^\infty(\scrM, S\scrM) \times C^\infty_0(\scrM, S\scrM) \rightarrow \C\:, \\
\bra \psi|\phi \ket &:= \int_\scrM \Sl \psi | \phi \Sr_x\: d\mu_\scrM
\end{align*}
(it clearly suffices to assume that one of the wave function has compact support).
This indefinite inner product is related to the scalar product via the causal fundamental solution
by the relation (see for example~\cite[Proposition~3.1]{finite})
\beq \label{pairing}
2 \pi\: (\psi \,|\, k_m \,\phi)_m = \bra \psi | \phi \ket \qquad
\text{for all~$\psi \in \H_m$ and~$\phi \in C^\infty_0(\scrM, S\scrM)$}\:.
\eeq

\subsection{The Unregularized Kernel of the Fermionic Projector} \label{seckfp}
The {\em{fermionic projector}} is obtained by choosing a closed subspace~$\H \subset \H_m$
and by composing~$k_m$ with the projection operator onto this subspace,
\beq \label{Pdef}
P := -\pi_\H \,k_m \::\: C^\infty_0(\scrM, S\scrM) \rightarrow \H \:.
\eeq
This operator can be represented by a
bi-distribution~$P \in \D'(\scrM \times \scrM)$, i.e.\ (for details see for example~\cite[Theorem~3.12]{finite})
\beq \label{Pkern}
\bra \phi | P \psi \ket = P \big( \overline{\phi} \otimes \psi \big) \qquad
\text{for all~$\phi, \psi \in C^\infty_0(\scrM, S\scrM)$}
\eeq
(for notational simplicity, we denote the distribution with the same symbol as the
corresponding operator).
This makes it possible to express~$P$ as an integral operator with distributional kernel,
\[ (P \psi)(x) = \int_\scrM P(x,y)\: \psi(y)\: d\mu_\scrM(y) \:. \]
The bi-distribution~$P(x,y)$ is referred to as the {\em{unregularized kernel of the fermionic projector}}.
The computation
\beq
P \big( \overline{\phi} \otimes \psi \big) \overset{\eqref{Pkern}}{=}
\bra \phi | P \psi \ket \overset{\eqref{Pdef}}{=} \bra \phi \,|\, \pi_\H \,k_m \psi \ket 
\overset{\eqref{pairing}}{=} 2 \pi\: (k_m \phi \,|\, \pi_\H \,k_m \psi )_m  \label{Psymm}
\eeq
shows that this kernel is symmetric in the sense that
\[ P(x,y)^* = P(y,x) \:. \]

Before going on, we briefly explain the underlying physical picture.
The vectors in the subspace~$\H \subset \H_m$ have the interpretation
as those Dirac wave functions which are realized in the physical system under
consideration. If one describes the vacuum in Minkowski space, one chooses~$\H$ as the subspace of all
negative-energy solutions of the Dirac equation. The resulting fermionic projector describes
the completely filled Dirac sea. Likewise, a system of non-interacting particles and/or anti-particles in Minkowski
space can be described by adding to~$\H$ solutions of positive energy and/or removing solutions
of negative energy.

In general curved spacetimes, there is no canonical splitting of the solution space~$\H_m$
into solutions of positive and negative energy. Consequently, there is no canonical choice
for the subspace~$\H$. Nevertheless, there is a distinguished class of subspaces
for which the bi-distribution~$P$ is of {\em{Hadamard form}}.
In our setting, the Hadamard condition is tantamount to demanding that any~$x \in \scrM$
has a geodesically convex neighborhood in which the 
bi-distribution~$P(x,y)$ has the form (see~\cite{sahlmann2001microlocal} or~\cite[page~156]{hack})
\beq \label{hadamard1}
P(x,y) = \lim_{\varepsilon \searrow 0} \;i \Pdd_x \left( \frac{U(x,y)}{\sigma_\varepsilon(x,y)}
+ V(x,y)\: \log \sigma_\varepsilon(x,y) + W(x,y) \right) ,
\eeq
where
\beq \label{sigmadef}
\sigma_\varepsilon(x,y) := \Gamma(x,y) - i \varepsilon \,(y-x)^0 \:,
\eeq
and~$U$, $V$ and~$W$ are smooth functions on~$\scrM \times \scrM$
taking values in the~$4 \times 4$-matrices acting on the spinors
(we always denote spacetime indices by latin letters running from~$0, \ldots, 3$),
and~$\Gamma(x,y)$ is the geodesic distance squared (with the sign convention that~$\Gamma$ is positive in timelike and negative in spacelike directions).

\subsection{The Regularized Kernel with Dirac Dynamics} \label{secdirac}
We next introduce an ultraviolet regularization. The simplest method is to regularize
while preserving the Dirac equation (more general regularization which violate the
Dirac equation will be considered in Section~\ref{secrigid}).
To this end, given a regularization length~$\varepsilon>0$,
we introduce a {\em{regularization operator}}~$({\mathfrak{R}}_\varepsilon)_{\varepsilon>0}$
as a mapping from the Dirac solutions to continuous solutions
(for more details see~\cite[Definition~4.1]{finite})
\beq \label{Repsdirac}
{\mathfrak{R}}_\varepsilon \::\: \H_m \rightarrow C^0(\scrM, S\scrM) \cap \H_m \:.
\eeq
There are various ways to choose such regularization operators.
One method is to choose finite-dimensional
subspaces~$\H^{(\ell)} \subset \Cisc(\scrM, S\scrM) \cap \H_m$ which exhaust~$\H_m$ in the
sense that~$\H^{(0)} \subset \H^{(1)} \subset \cdots$ and~$\H_m = \overline{\cup_\ell \H^{(\ell)}}$.
Setting~$\ell(\varepsilon) = \max ( [0, 1/\varepsilon] \cap \N)$, we can introduce the
operators~${\mathfrak{R}}_\varepsilon$ as the orthogonal projection operators to~$\H^{(\ell(\varepsilon))}$.
An alternative method is to choose a Cauchy hypersurface~$\scrN$, to mollify the restriction~$\psi_m|_\scrN$
to the Cauchy surface on the length scale~$\varepsilon$, and to define~${\mathfrak{R}}_\varepsilon \psi_m$
as the solution of the Cauchy problem for the mollified initial data.

Having chosen a regularization operator, the {\em{regularized fermionic projector}}~$P^\varepsilon$ is defined by
\beq \label{regPdef}
P^\varepsilon := -{\mathfrak{R}}_\varepsilon\, \pi_\H\, {\mathfrak{R}}_\varepsilon^*
\,k_m \::\: C^\infty_0(\scrM, S\scrM) \rightarrow \H_m \:.
\eeq
This operator can be represented by the {\em{regularized kernel of the fermionic
projector}}~$P^\varepsilon \in \D'(\scrM \times \scrM)$, i.e.
\beq \label{Pdistr}
\bra \phi | P^\varepsilon \psi \ket = P^\varepsilon \big( \overline{\phi} \otimes \psi \big) \qquad
\text{for all~$\phi, \psi \in C^\infty_0(\scrM, S\scrM)$} \:.
\eeq
Similar to~\eqref{Psymm}, the computation
\[ P^\varepsilon \big( \overline{\phi} \otimes \psi \big) =
\bra \phi | P^\varepsilon \psi \ket = \bra \phi \,|\, {\mathfrak{R}}_\varepsilon\, \pi_\H\, {\mathfrak{R}}_\varepsilon^* \,k_m \psi \ket 
= 2 \pi\:(k_m \phi \,|\, {\mathfrak{R}}_\varepsilon\, \pi_\H\, {\mathfrak{R}}_\varepsilon^* \,k_m \psi )_m \]
shows that also the regularized kernel is symmetric, i.e.
\[ P^\varepsilon(x,y)^* = P^\varepsilon(y,x) \:. \]
Moreover, since~${\mathfrak{R}}_\varepsilon$ maps to solutions~\eqref{Repsdirac}, also the regularized
kernel satisfies the Dirac equation,
\[ (\Dir - m)\, P^\varepsilon(x,y) = 0 \:. \]
Therefore, also the dynamics of the regularization is described by the Dirac equation.
As worked out in detail in~\cite{reghadamard}, this means more concretely that
the regularized kernel has the so-called {\em{regularized Hadamard}} expansion
\begin{align}
P^\varepsilon(x,y) &= (\Dir + m)\, T^\varepsilon(x,y)
\;\bigg(1 + \O\Big( \frac{\varepsilon^2}{\Gamma} \Big) \bigg)\qquad \text{with} \label{Preghadamard} \\
T^\varepsilon(x,y)  &= \frac{X_{-1}(x,y)}{\Gamma_{-1}(x,y)} \\
&\quad\, + \sum_{n=0}^{\infty} X_n(x,y)\; \Gamma_{n}(x,y)^n\:\log \Gamma_{[n]}(x,y)
+ \sum_{n=0}^{\infty} Y_n(x,y)\: \Gamma_{n}(x,y)^n \:,
\end{align}
where
\begin{align*}
\Gamma_n(x,y) &= \Gamma(x,y) + i \varepsilon \,f(x,y) \\
X_n(x,y) &= A_n(x,y) + i \varepsilon \,a_{n}(x,y) \\
Y_n(x,y) &= B_n(x,y) + i \varepsilon \,b_{n}(x,y) 
\end{align*}
with smooth real-valued functions~$A_{-1}$, $A_n$ and~$B_n$ as well as
real-valued continuous functions~$f$, $a_n$ and~$b_n$ which are smooth away from the diagonal~$x=y$.
The dynamics of the regularization is described by the function~$f$ which is determined
by transport equations along null geodesics,
\beq \label{ftransport}
2\, \langle \nabla \Gamma, \nabla_x f(x,y) \rangle = 4\,f(x,y) \qquad \text{and} \qquad f(x,x)=0\:.
\eeq
These transport equations will be the starting point for our analysis of the dynamics of the
regularization in Section~\ref{secrigid}.

\subsection{Construction of a Corresponding Causal Fermion System} \label{seccfsconstruct}
We again let~$\H \subset \H_m$ be the closed subspace chosen at the beginning of Section~\ref{seckfp}.
Endowed with the induced scalar product
\[ \la .|. \ra_\H := (.|.)_m \big|_{\H \times \H} \:, \]
it is again a Hilbert space. Restricting the regularization operator~\eqref{Repsdirac} to~$\H$ and
evaluating at a spacetime point~$x$ gives the
{\em{regularized wave evaluation operator}}~$\Psi^\varepsilon(x)$,
\beq \label{Pepsdirac}
\Psi^\varepsilon(x) = {\mathfrak{R}}_\varepsilon(x) \::\: \H \rightarrow S_x \scrM \:.
\eeq
Taken its adjoint (with respect to the Hilbert space
scalar product~$\la .|. \ra_\H$ and the spin inner product~$\Sl .|. \Sr_x$) gives the mapping
\[ \big(\Psi^\varepsilon(x) \big)^* \::\: S_x \scrM \rightarrow \H  \:. \]
Multiplying~$\Psi^\varepsilon(x)$ by its adjoint gives the operator
\beq \label{Fepsprod}
F^\varepsilon(x) := - \big( \Psi^\varepsilon(x) \big)^* \,\Psi^\varepsilon(x) \::\: \H \rightarrow \H \:,
\eeq
referred to as the {\em{local correlation operator}} at
the spacetime point~$x$. The local correlation operator is also characterized by the relation
\[ 
(\psi \,|\, F^\varepsilon(x)\, \phi) = -\Sl ({\mathfrak{R}}_\varepsilon\psi)(x) | 
({\mathfrak{R}}_\varepsilon \phi)(x) \Sr_x \qquad \text{for all~$\psi, \phi
\in \H$} \:. \]
Taking into account that the inner product on the Dirac spinors at~$x$ has signature~$(2,2)$,
it is a symmetric operator on~$\H$
of rank at most four, which (counting multiplicities) has at most two positive and at most two negative eigenvalues.
Varying the spacetime point, we obtain a mapping
\[ F^\varepsilon \::\: \scrM \rightarrow \F \subset \Lin(\H)\:, \]
where~$\F$ denotes all symmetric operators of
rank at most four with at most two positive and at most two negative eigenvalues.

Taking the push-forward of the volume measure on~$\scrM$ gives a
measure~$d\rho:= (F^\varepsilon)_* \,d\mu_\scrM$ on~$\F$
(thus~$\rho(\Omega) := \mu_\scrM((F^\varepsilon)^{-1}(\Omega))$).
In this way,
one obtains a causal fermion system (for details see for example~\cite[Section~1]{nrstg}).
Here we proceed slightly differently and rescale the local operators such that they all have trace one,
\beq \label{rescale1}
G^\varepsilon(x) := \left\{ \begin{array}{cl}
\displaystyle \frac{F^\varepsilon(x)}{\tr \big(F^\varepsilon(x) \big)} & \text{if $\tr \big(F^\varepsilon(x) \big) \neq 0$}
\\[1em]
0 & \text{otherwise} \end{array} \right.
\eeq
(the reason for this rescaling will be explained in the paragraph after~\eqref{volconstraint} below).
Now we introduce the measure~$\rho$ by
\beq \label{rescale2}
d\rho := (G^\varepsilon)_*  \,d\mu_\scrM \:.
\eeq
In this way, we obtain a causal fermion system of spin dimension two
with fixed local trace, as will be defined abstractly at the beginning of the next section.

We finally remark that the ansatz~\eqref{rescale2} can be generalized by multiplying the
volume measure on~$\scrM$ by a positive weight function~$\sigma \in C^\infty(\scrM, \R^+)$,
\beq \label{rescale3}
d\rho := \: (G^\varepsilon)_*  \,\big( \sigma\: d\mu_\scrM\big)  \:.
\eeq
This generalization will be needed in order to satisfy the Euler-Lagrange equations
in cosmological situations (see Section~\ref{secdiraccosmo}  below). In this way, one gets a direct connection to
physical theories involving {\em{non-Riemannian measures}} (see~\cite{nrmtreview} and references therein).

\section{A Few Basics on Causal Fermion Systems} \label{sec:cfs}
This section provides the necessary abstract background on causal fermion systems.
\subsection{Causal Fermion Systems with Fixed Local Trace} \label{seccap}
We now give the general definition of a causal fermion system with fixed local trace.
\begin{Def} \label{defcfs} {\em{ 
Given a separable complex Hilbert space~$\H$ with scalar product~$\la .|. \ra_\H$
and a parameter~$n \in \N$ (the {\em{spin dimension}}), we let~$\F \subset \Lin(\H)$ be the set of all
symmetric operators~$A$ on~$\H$ of finite rank which have trace one,
\beq \label{fixedtrace}
\tr A = 1 \:,
\eeq
and which (counting multiplicities) have
at most~$n$ positive and at most~$n$ negative eigenvalues. On~$\F$ we are given
a positive measure~$\rho$ (defined on a $\sigma$-algebra of subsets of~$\F$).
We refer to~$(\H, \F, \rho)$ as a {\em{causal fermion system with fixed local trace}}.
}}
\end{Def} \noindent
On~$\F$ we consider the topology induced by the operator norm
\[ 
\|A\| := \sup \big\{ \|A u \|_\H \text{ with } \| u \|_\H = 1 \big\} \:. \]
{\em{Spacetime}}~$M$ is defined to be the support of this measure,
\[ 
M := \supp \rho \subset \F \:. \]
It is a topological space (again with the topology induced by the operator norm).
The fact that the spacetime points are operators gives rise to
many additional structures which are {\em{inherent}} in the sense
that they only use information already encoded in the causal fermion system.
In the next sections (Section~\ref{seccapreduced}--\ref{secHextend})
we shall introduce those structures needed later on; a more complete
treatment can be found in~\cite[Section~1.1]{cfs}.

\subsection{The Reduced Causal Action Principle} \label{seccapreduced}
In order to single out the physically admissible
causal fermion systems, one must formulate physical equations. To this end, we impose that
the measure~$\rho$ should be a minimizer of the causal action principle,
which we now introduce. For any~$x, y \in \F$, the product~$x y$ is an operator of rank at most~$2n$. 
However, in general it is no longer symmetric because~$(xy)^* = yx$,
and this is different from~$xy$ unless~$x$ and~$y$ commute.
As a consequence, the eigenvalues of the operator~$xy$ are in general complex.
We denote the nontrivial eigenvalues counting algebraic multiplicities
by~$\lambda^{xy}_1, \ldots, \lambda^{xy}_{2n} \in \C$
(more specifically,
denoting the rank of~$xy$ by~$k \leq 2n$, we choose~$\lambda^{xy}_1, \ldots, \lambda^{xy}_{k}$ as all
the non-zero eigenvalues and set~$\lambda^{xy}_{k+1}, \ldots, \lambda^{xy}_{2n}=0$).
Given a parameter~$\kappa>0$ (which will be kept fixed throughout this paper),
we introduce the $\kappa$-Lagrangian and the causal action by
\begin{align}
\text{$\kappa$-\em{Lagrangian:}} && \L(x,y) &= \frac{1}{4n} \sum_{i,j=1}^{2n} \Big( \big|\lambda^{xy}_i \big|
- \big|\lambda^{xy}_j \big| \Big)^2 + \kappa\: \bigg( \sum_{j=1}^{2n} \big|\lambda^{xy}_j \big| \bigg)^2
\label{Lagrange} \\
\text{\em{causal action:}} && \Sact(\rho) &= \iint_{\F \times \F} \L(x,y)\: d\rho(x)\, d\rho(y) \:. \label{Sdef}
\end{align}
The {\em{reduced causal action principle}} is to minimize~$\Sact$ by varying the measure~$\rho$
under the
\beq
\text{\em{volume constraint}} \qquad \rho(\F) = \text{const} \label{volconstraint} \:.
\eeq

This variational principle is obtained from the general causal action principle as introduced in~\cite[\S1.1.1]{cfs}
as follows. Using that minimizing measures are supported on operators of constant trace
(see~\cite[Proposition~1.4.1]{cfs}), we may fix the trace of the operators. Moreover, 
by rescaling all the operators according to~$x \rightarrow \lambda x$ with~$\lambda \in \R$,
we may assume without loss of generality that this trace is equal to one~\eqref{fixedtrace}.
Next, the $\kappa$-Lagrangian arises when treating the so-called boundedness constraint
with a Lagrange multiplier term. Here we slightly simplified the setting
by combining this Lagrange multiplier term with the Lagrangian right from the beginning.

This variational principle is mathematically well-posed if~$\H$ is finite-dimensional.
For the existence theory and the analysis of general properties of minimizing measures
we refer to~\cite{discrete, continuum, lagrange}.
In the existence theory one varies in the class of regular Borel measures
(with respect to the topology on~$\Lin(\H)$ induced by the operator norm),
and the minimizing measure is again in this class. With this in mind, here we always assume
that~$\rho$ is a {\em{regular Borel measure}}.

We now illustrate the abstract setting of causal fermion systems
by explaining how the causal action and the constraints
can be computed for the causal fermion system
constructed in Section~\ref{seccfsconstruct}. We first note that, for the push-forward measure~\eqref{rescale2},
the $\rho$-integrals in the reduced causal action can be rewritten as
$\mu_\scrM$-integrals over the Lorentzian spacetime,
\[ \Sact= \iint_{\scrM \times \scrM} \L \big(G^\varepsilon(x), G^\varepsilon(y) \big)\: d\mu_\scrM(x)\, d\mu_\scrM(y) \:. \]
Moreover, using that the $\kappa$-Lagrangian~\eqref{Lagrange} is homogeneous of degree two in both arguments,
the rescaling in~\eqref{rescale1} and~\eqref{rescale2} give rise to
\beq \label{Srescale}
\Sact= \iint_{\scrM \times \scrM} \frac{\L \big(F^\varepsilon(x), F^\varepsilon(y) \big)}{\big( \tr F^\varepsilon(x) \big)^2 \, \big(\tr F^\varepsilon(y) \big)^2}\;
\: d\mu_\scrM(x)\, d\mu_\scrM(y)
\eeq
(similarly, for the measure~\eqref{rescale3} one inserts the weight functions~$\sigma(x)$ and~$\sigma(y)$
into the integrand; for simplicity we here omit the resulting formulas).
The causal action can be expressed even in terms of the regularized kernel of the fermionic projector,
as the following consideration shows. We return to the setting of
the Dirac dynamics with regularization introduced in Section~\ref{secdirac}.
Working again with the regularized wave evaluation operator~\eqref{Pepsdirac},
the kernel of the regularized fermionic projector (as defined by~\eqref{regPdef} and~\eqref{Pdistr})
can be written as
\beq \label{PPsidirac}
P^\varepsilon(x,y) = -\Psi^\varepsilon(x)\, \Psi^\varepsilon(y)^*
\eeq
(for details see~\cite[Proposition~1.2.7]{cfs}). Comparing with~\eqref{Fepsprod}, one sees that
the trace of~$F^\varepsilon(x)$ in~\eqref{Srescale} can be
expressed in terms of the regularized kernel by
\[ \tr \big(F^\varepsilon(x) \big) = \Tr_{S_x\scrM} \big( P^\varepsilon(x,x) \big) \:. \]
Moreover, using that all the non-zero eigenvalues of an operator product as well as the
corresponding algebraic multiplicities
do not change when the factors of the operator product are cyclically commuted
(for details see~\cite[Section~1.1.3]{cfs}), one also sees that
\[ F^\varepsilon(x)\, F^\varepsilon(y) : \H \rightarrow \H \qquad \text{is isospectral to} \qquad
P^\varepsilon(x,y)\, P^\varepsilon(y,x) : S_x\scrM \rightarrow S_x\scrM \:. \]
Therefore, the eigenvalues~$\lambda^{xy}_1, \ldots, \lambda^{xy}_4$ in the
$\kappa$-La\-grangian~\eqref{Lagrange} coincide with the eigenvalues
of the {\em{closed chain}} $A_{xy} := P^\varepsilon(x,y)\, P^\varepsilon(y,x)$.
In this way, the computation of the $\kappa$-Lagrangian is reduced to computing the
eigenvalues of a $4 \times 4$-matrix.

Expressing the causal action principle in terms of the regularized kernel
also gives a different perspective on the nature of the causal action principle.
Namely, choosing an orthonormal basis~$(\psi_k)$ of the subspace~$\H \subset \H_m$
and inserting the completeness relation into~\eqref{PPsidirac}, the
regularized kernel can be written in terms of the regularized wave functions of
the occupied states of the system,
\beq \label{Pepsstate}
P^\varepsilon(x,y) = - \sum_k | ({\mathfrak{R}}_\varepsilon \psi_k)(x)\Sr \Sl
({\mathfrak{R}}_\varepsilon \psi_k)(y)| \:.
\eeq
In this way, the causal action principle becomes a variational principle for the ensemble of all
these wave functions. These wave functions are referred to as the
{\em{physical wave functions}}. The physical wave functions give a
representation of the abstract Hilbert space~$(\H, \la .|. \ra_\H)$ in terms of an ensemble of spinorial wave functions.
The analysis of the continuum limit as carried out in~\cite{cfs}
shows that forming suitable ensembles of Dirac solutions gives minimizers in
a suitable limiting case in which the regularization is removed.
This also shows that solutions of a Dirac equation give good candidates for approximate
minimizers. However, there is no reason to expect that Dirac solutions give rise to
exact minimizers. In other words, the dynamics as described by the causal action principle
goes beyond the Dirac dynamics.
It is the goal of the present paper to analyze the differences
of these dynamics in the context of fermiogenesis.

\subsection{The Euler-Lagrange Equations}
A minimizer of the reduced causal action
satisfies the following {\em{Euler-Lagrange (EL) equations}}:
For a suitable value of the parameter~$\s>0$,
the function~$\ell$ defined by
\beq \label{elldef}
\ell \::\: \F \rightarrow \R\:,\qquad \ell(x) := \int_M \L(x,y)\: d\rho(y) - \s
\eeq
is minimal and vanishes in spacetime,
\[ 
\ell|_M \equiv \inf_\F \ell = 0 \:. \]
The parameter~$\s$ can be viewed as the Lagrange parameter
corresponding to the volume constraint. By rescaling the measure,
one can give~$\s$ an arbitrary non-zero value. With this in mind, we keep the
parameter~$\s$ fixed throughout the paper.
For the derivation and further details on the EL equations we refer to~\cite[Section~2]{jet}
or~\cite[Chapter~7]{intro}.

\subsection{The Commutator Inner Product} \label{secHextend}
In the setting of causal fermion systems, integrals over hypersurfaces
are replaced by so-called {\em{surface layer integrals}}, which are
double integrals of the general form
\beq \label{osi}
\int_\Omega \bigg( \int_{M \setminus \Omega} (\cdots)\: \L(x,y)\: d\rho(y) \bigg)\, d\rho(x) \:,
\eeq
where $(\cdots)$ stands for suitable variational derivatives of the Lagrangian, and~$\Omega$
is a Borel subset of~$M$.
The connection can be understood most easily in the case when~$\L(x,y)$ vanishes
unless~$x$ and~$y$ are close together. In this case, we only get a contribution to~\eqref{osi}
if both~$x$ and~$y$ are close to the boundary of~$\Omega$.
A more detailed explanation of the idea of a surface layer integrals is given in~\cite[Section~2.3]{noether}.

Surface layer integrals were first introduced in~\cite{noether} in order to 
make a connection between symmetries and conservation laws for surface layer integrals.
Here we will make essential use of the conservation law corresponding to the
symmetry under unitary transformations on the Hilbert space~$\H$.
For a minimizing measure~$\rho$, it gives rise to a conservation law for an inner product on the physical wave functions of the form
\beq \label{OSIdyn}
\begin{split}
\la \psi | \phi \ra^t_\rho = -2i \,\bigg( \int_{\Omega^t} \!d\rho(x) \int_{M \setminus \Omega^t} \!\!\!\!\!\!\!d\rho(y) 
&- \int_{M \setminus \Omega^t} \!\!\!\!\!\!\!d\rho(x) \int_{\Omega^t} \!d\rho(y) \bigg)\\
&\times\:
\Sl \psi(x) \:|\: Q(x,y)\, \phi(y) \Sr_x \:,
\end{split}
\eeq
where~$\Omega_t$ is the past of a Cauchy surface~$\scrN_t$.
Here ``conservation law'' means that this inner product is independent of~$t$.
The inner product~\eqref{OSIdyn} is referred to as the {\em{commutator inner product}}
(the name comes from the fact that the unitary invariance can be expressed in terms of
commutators; see~\cite[Section~3]{dirac} for details).
The kernel~$Q(x,y)$ appearing in this formula is the first variational derivative of the Lagrangian,
where we identified the spinor spaces~$S_x \scrM$ with corresponding
spin spaces of the causal fermion system (see~\cite[\S1.4.1 and Proposition~1.6]{cfs}
or~\cite{dirac} for more details).
The wave functions~$\psi$ and~$\phi$ in~\eqref{OSIdyn}
are wave functions in Minkowski space. But, as already mentioned after~\eqref{Pepsstate}
in Section~\ref{seccapreduced}, these wave functions do not need to be solutions of the Dirac equation.

\section{Compatibility of the Conservation Laws} \label{secdiraccosmo}
This section is devoted to a careful analysis of the various conservation laws.
On the one hand, we have the conservation law for the commutator inner product~\eqref{OSIdyn}.
Coming from the causal action principle, this conservation law is fundamental in the sense
that it must hold even for the modified dynamics. On the other hand, there is the
scalar product~\eqref{print}, which is conserved for solutions of the Dirac dynamics, but might be
violated for the modified dynamics. 
In order to get a first connection between the Dirac dynamics and the modified dynamics as described by the
causal action principle, we need to analyze how~\eqref{print} and~\eqref{OSIdyn} are related to each other.
This also makes it necessary to analyze the effect of the scalings like~\eqref{rescale1}.
Our conclusion will be that, making use of this scaling freedom, we can {\em{arrange}}
that~\eqref{print} agrees with~\eqref{OSIdyn} up to a universal constant
(with a well-defined error term).
As a consequence, the scalar product~\eqref{print} will also be conserved for the
modified dynamics. This result is very convenient and useful, simply because the surface integral~\eqref{print}
is more familiar and easier to compute than the surface layer integral~\eqref{OSIdyn}.

In~\cite[Section~5]{noether} it was shown for non-interacting Dirac systems in {\em{Minkowski space}}
that, asymptotically for~$\varepsilon \searrow 0$,
the commutator inner product is indeed proportional to the current integral, i.e.
\beq \label{currcompute}
\la \psi | \phi \ra^t_\rho = \mathfrak{c}\: \int_{\R^3} \Sl \psi | \gamma^0 \phi \Sr(t,\vec{x})\: d^3x \:.
\eeq
Using the normalization conventions in~\cite{noether}, a scaling argument shows that the
proportionality factor has length dimension minus four. Since it is finite even without regularization,
it must scale like (more details on such scaling arguments are given in~\cite{action})
\[ \mathfrak{c} \simeq m^4 \]
Here and in what follows, $\simeq$ means proportionality with an irrelevant
numerical constant.
However, the normalization conventions in~\cite{noether} are not compatible with
our assumption of fixed local trace~\eqref{fixedtrace}. This makes it necessary to rescale
the measure. In order to clarify the notation, we denote the push-forward of the volume measure
of Minkowski space by~$F^\varepsilon$ by an additional subscript~$\scrM$,
\[ 
d\rho_\scrM := (F^\varepsilon)_* \,d\mu_\scrM \:. \]
We also add a subscript~$\scrM$ to other objects constructed from this scaling.

A direct computation (again using the normalization conventions in~\cite{noether}; see
also~\cite[Section~2.5]{cfs} and~\cite[Appendix~A]{jacobson}) yields
\begin{align*}
\tr \big( F^\varepsilon_\scrM(x) \big) &\simeq \frac{m}{\varepsilon^2} \\
\ell_\scrM(x) + \s_\scrM &\simeq \frac{(\varepsilon m)^p}{\varepsilon^8} + \frac{\kappa}{\varepsilon^8}\:.
\end{align*}
where the parameter~$p \geq 5$ depends on the unknown microstructure of spacetime.
In what follows, we shall not need the specific form of the above formula for~$\ell_\scrM$.
Therefore, it is preferable to write the function~$\ell_\scrM$ in the shorter form
\beq \label{adef}
\ell_\scrM +\s_\scrM \simeq \frac{a}{\varepsilon^8} \qquad
\text{with} \qquad a := (\varepsilon m)^p + \kappa\:.
\eeq
Now we rescale the measure by (see also~\cite[Appendix~A]{jacobson})
\beq \label{tilrho}
\rho(\Omega) := \sigma\: \rho_\scrM\Big( \frac{\Omega}{\lambda} \Big) \qquad \text{with~$\lambda, \sigma>0$}\:.
\eeq
(where~$\Omega/\lambda = \{ x/\lambda \:|\: x \in \Omega \}$ is the set obtained
by rescaling all the operators in~$\Omega$; note that~$\F$ is a subset of~$\Lin(\H)$ which is invariant under such rescalings).
We remark that the rescaled measure could also be written in the form~\eqref{rescale3} if we chose
\[ \sigma(x) = \sigma \qquad \text{and} \qquad G^\varepsilon(x) = \lambda\, F^\varepsilon(x)\:. \]
We denote the rescaled quantities without the index~$\scrM$. They take the form
\begin{align*}
\tr \big( F^\varepsilon(x) \big) &= \lambda\, \tr \big( F^\varepsilon_\scrM(x) \big)
\simeq \lambda\: \frac{m}{\varepsilon^2} \\
\ell(x) + \s &\simeq \frac{\lambda^4\: \sigma}{\varepsilon^8}\: a \\
\la \psi | \phi \ra^t_\rho
&\simeq m^4\: \lambda^4\: \sigma^2 \int_{\R^3} \Sl \psi(x) \,|\, \gamma^0\, \phi(x) \Sr_x\: d^3x \:.
\end{align*}
In order to arrange~\eqref{fixedtrace} and~\eqref{elldef}, we must choose
\[ \lambda \simeq \frac{\varepsilon^2}{m} \qquad \text{and} \qquad
\sigma \simeq \frac{m^4}{a} \:\s \:. \]
We thus obtain
\begin{align*}
\la \psi | \phi \ra^t_\rho
&\simeq \varepsilon^8\: \sigma^2
\int_{\R^3} \Sl \psi(x) \,|\, \gamma^0\, \phi(x) \Sr_x\: d^3x \\
&= ( \varepsilon m)^8\: \frac{\s^2}{a^2}
\int_{\R^3} \Sl \psi(x) \,|\, \gamma^0\, \phi(x) \Sr_x\: d^3x \:.
\end{align*}
We conclude that the factor~$\mathfrak{c}$ in~\eqref{currcompute} becomes
\beq \label{alphaeps}
\mathfrak{c} \simeq \varepsilon^8\: \sigma^2 \:.
\eeq

Before going on, we briefly explain this scaling behavior and explain how it can be understood
directly from ~\eqref{currcompute} and~\eqref{OSIdyn}. The parameter~$\sigma$ multiplies the
measure in~\eqref{tilrho} and therefore simply scales the volume by an overall constant.
This scaling has no physical significance; it also transforms minimizing measures again to
minimizers. This scaling freedom can be used to give the parameter~$\s$ in~\eqref{elldef}
a prescribed value (for example, we could arrange that~$\s=1$).
Alternatively, we can simply disregard the scaling freedom by choosing~$\sigma=1$.
The parameter~$\lambda$, on the other hand, changes the local trace.
It is determined by our convention~\eqref{fixedtrace}.
Having fixed the local trace in this way, the Lagrangian and its variational derivatives are dimensionless.
In particular, the integrand in~\eqref{OSIdyn} is dimensionless. Carrying out an eight-dimensional
integral, the surface layer integral in~\eqref{OSIdyn} has length dimension eight.
The Dirac current integral on the right side of~\eqref{currcompute}, on the other hand,
is dimensionless (because it is the usual scalar product of Dirac theory, and we normalize the
physical states to have $L^2$-norm one). Therefore, the factor~$\mathfrak{c}$ in~\eqref{currcompute} must have
length dimension eight, explaining~\eqref{alphaeps}.

We now move on to {\em{curved spacetime}} with the {\em{Dirac dynamics}}.
It is a general conclusion from the regularized Hadamard expansion~\eqref{Preghadamard}
that the regularization length~$\varepsilon$ is {\em{not fixed}}, but it has a dynamics on its own
as described by the transport equation~\eqref{ftransport}.
As a result, the parameter~$\varepsilon$ varies in spacetime on cosmological scales
(as is illustrated in Figure~\ref{figcone1}).
Since the Dirac dynamics is known to be
at least a very good approximation to the exact physical dynamics,
our strategy for deriving the locally rigid dynamics is to slightly modify the Dirac dynamics.
Proceeding in this way, the parameter~$\varepsilon$ will again vary in spacetime
(however, in a way which at this stage is unknown).
Consequently, the same will be true for the parameter~$a$ in~\eqref{adef}.
We assume that both~$\varepsilon$ and~$a$ vary only on cosmological scales.
This means that, when computing current integrals of wave functions in a laboratory, 
we can choose Gaussian coordinates and work again with the above formula~\eqref{currcompute},
but with the prefactor~$\mathfrak{c}$ replaced by a function~$\mathfrak{c}(x)$
which varies on cosmological scales. This leads us to the relation
\begin{gather}
\la \psi | \phi \ra^t_\rho = \int_{\scrN_t} \mathfrak{c}(x)\: \Sl \psi \,|\, \gamma(\nu)\, \phi \Sr(x)\: d\mu_{\scrN_t}(x)
+ \O \bigg( \frac{\ell_\text{lab}}{\ell_{\text{cosmo}}} \bigg)
\qquad \text{with}
\label{currcompute2} \\
\mathfrak{c}(x) \simeq \varepsilon(x)^8\: \sigma(x)^2 \qquad \text{and} \qquad
\sigma(x) = \frac{m^4}{a(x)} \:\s\:, \label{sigmaform}
\end{gather}
where~$\ell_\text{lab}$ is the size of the laboratory where~$\phi$ and~$\psi$ are measured, whereas~$\ell_\text{cosmo}$ is the cosmological
scale on which~$\varepsilon(x)$ varies.
Here~$a(x)$ stands for the function in~\eqref{adef} with~$\varepsilon$ replaced by~$\varepsilon(x)$.
Moreover, as in~\eqref{print}, we chose a Cauchy surface~$\scrN$ with future-directed normal~$\nu$.

This consideration has the following consequences. First, in order to satisfy the
EL equations, the weight function~$\sigma$ in the ansatz~\eqref{rescale3}
must be chosen as a non-constant function~\eqref{sigmaform} which varies on cosmological scales.
In other words, causal fermion systems necessarily involve non-Riemannian measures~\cite{nrmtreview}
and a deviation from the standard Dirac dynamics.
This is relevant on the conceptual level, as it makes our mechanism for baryogenesis consistent in the effective description - at least on the qualitative level. As shown in one of the eraliest papers on non-Riemannian measure theories~\cite{guendelman1996principle} a deviation from Riemannian measures leads to a matter stress energy tensor that is not divergence free, as one would expect in the context of our mechanism that effectively creates the matter anti-matter assymmetrie in a spacetime which evolves from vacuum initial data. 
Second and more importantly, one sees from~\eqref{currcompute2} that the conservation of the Dirac current
and the conservation law for the commutator inner product are not compatible.
This also shows that there is a discrepancy between the Dirac dynamics and the dynamics of the system
as described by the causal action principle. In order to cure this inconsistency, one can
absorb the prefactor~$\mathfrak{c}(x)$ in~\eqref{currcompute2} into the Dirac wave functions by
formulating the Dirac equation for a rescaled wave function,
\beq \label{tildepsi}
(\Dir - m)\, \tilde{\psi} = 0 \qquad \text{with} \qquad \tilde{\psi}(x) = \sqrt{\mathfrak{c}(x)}\: \psi(x) \:.
\eeq
Then~\eqref{currcompute2} becomes
\[ 
\la \psi | \phi \ra^t_\rho = \int_{\scrN_t} \Sl \tilde{\psi} \,|\, \gamma(\nu)\, \tilde{\phi} \Sr(x)\: d\mu_{\scrN_t}(x)
+ \O \bigg( \frac{\ell_\text{lab}}{\ell_{\text{cosmo}}} \bigg) \:. \]
Note that both sides of this equation are conserved (the left side as a
consequence of the conservation law for the commutator inner product,
and the right side due to current conservation for the Dirac equation in~\eqref{tildepsi}.
Therefore, the conservation laws are compatible. However, one should keep in mind
that, as a consequence of the scaling in~\eqref{tildepsi}, the original wave function~$\psi$
no longer satisfies the Dirac equation~\eqref{Direq} and violates current conservation.
This can be described more explicitly by an additional term in the Dirac equation
for the unrescaled wave function,
\beq \label{Dircosmo}
\Big(\Dir + \frac{i}{2}\: \gamma^j \,\big( \partial_j \log \mathfrak{c} \big) - m \Big) \psi = 0 \:.
\eeq
This additional term changes the amplitude of the Dirac waves. 
Being anti-symmetric, it leads to
a violation of Dirac current conservation, in such a way that~\eqref{currcompute2} holds.
We refer to~\eqref{Dircosmo} as the {\em{cosmological Dirac equation}}. It is interesting to note that studies in the cosmological setting~\cite{dynamicalmass} of Dirac equations with similar modifications which successfully reproduce inflationary scenarios have been motivated by considerations in non-Riemannian-measure theories.

At first sight, the change of the amplitude of the Dirac wave functions
as described by the cosmological Dirac equation
seems to imply that the coupling of the Dirac waves to other fields (like the electromagnetic or gravitational
field) may also change on cosmological scales.
However, this conclusion is not correct, because we must take into account that the
change of amplitude affects all the Dirac waves in the same way. As a consequence, it drops
out of the causal fermion system, as the following argument shows:
According to~\eqref{PPsidirac} and~\eqref{Fepsprod},
the rescaling~\eqref{tildepsi} means that
\[ \tilde{P}^\varepsilon(x,y) = \sqrt{\mathfrak{c}(x)}\, \sqrt{\mathfrak{c}(y)}\, P^\varepsilon(x,y) 
\qquad \text{and} \qquad \tilde{F}^\varepsilon(x) = \mathfrak{c}(x)\: F^\varepsilon(x) \:. \]
However, the rescaling drops out when
forming the operators~$G^\varepsilon(x)$ of fixed trace~\eqref{rescale1}.
Therefore, the causal fermion system remains unchanged. With this in mind, in what follows we may
disregard the rescaling and work just as well with the geometric Dirac equation~\eqref{Direq}.

A general result of the above considerations is that, if we modify the Dirac dynamics,
we can always arrange that the dynamics respects the conservation law
\beq \label{conserve}
\int_{\scrN_t} \Sl \psi \,|\, \gamma(\nu)\, \phi \Sr(x)\: d\mu_{\scrN_t}(x) = \text{const} \:.
\eeq
Thus, in what follows, we have two conserved scalar products: the commutator inner product~\eqref{OSIdyn}
and the $L^2$-scalar product~\eqref{conserve}. It is most convenient to identify these scalar products at any time
by a unitary transformation~$V(t)$. This allows us to work exclusively with the more familiar scalar $L^2$-scalar product. We do not need to specify the unitary mapping~$V(t)$, which depends on the unknown
microstructure of spacetime. Instead, we simply demand that the modified dynamics must be unitary
with respect to the scalar product~\eqref{conserve}.

\section{The Locally Rigid Dynamics of the Regularization} \label{secrigid}
Having clarified the role of the conservation laws, we can now derive the
locally rigid dynamics. Our guiding principles are the following:
\begin{itemize}[leftmargin=2em]
\item[{\rm{(i)}}] The Dirac dynamics should be modified only slightly.
\item[{\rm{(ii)}}] The regularization should have locally the form~\eqref{ku}
with a regularizing vector field~$u(x)$ (as shown in Figure~\ref{figcone2}).
\item[{\rm{(iii)}}] Current conservation should hold, in the sense that
for a chosen foliation~$(\scrN_t)_{t \in \R}$ by Cauchy surfaces,
the integral~\eqref{conserve} must be independent of time~$t$ for all physical wave functions~$\psi$ and~$\phi$.
\end{itemize}
The guiding principle~(ii) is inspired by the results in the PhD thesis~\cite{jokeldr}.
Additionally, it can be motivated from the wish that the regularization should
preserve as many local symmetries as possible.

We begin with a preliminary construction of the regularized kernel of the fermionic projector
described by a regularizing vector field~$u(x)$ (Section~\ref{secudef}). This method satisfies
the properties~(i) and~(ii), but it violates~(iii). In order to also arrange current conservation,
we describe the regularization by an operator~$\tilde{A}(t)$ (Section~\ref{secAintro})
and modify the Dirac dynamics by adiabatic projections to spectral subspaces of~$\tilde{A}(t)$
(Section~\ref{secadiabatic}). In this way, we arrange~(iii) while preserving~(i) and~(ii).
A-priori, our construction will depend on the choice of the foliation~$(\scrN_t)_{t \in \R}$.
However, as shown in detail in Appendix~\ref{appB}, the leading order effect
in an expansion in~$\varepsilon/\ell_{\text{macro}}$ turns out to be independent of the choice of foliation.

\subsection{The Regularizing Vector Field} \label{secudef}
In Section~\ref{secdirac}, the regularized kernel with Dirac dynamics was introduced.
However, while being a good approximation,
there is no compelling reason why the regularized kernel should satisfy a
Dirac equation. In order to go beyond the Dirac dynamics, one generalizes~\eqref{Repsdirac} by 
introducing a regularization operator which maps to continuous wave functions which
need not be Dirac solutions,
\[ 
{\mathfrak{R}}_\varepsilon \::\: \H_m \rightarrow C^0(\scrM, S\scrM) \:. \]
The corresponding regularized kernel of the fermionic projector is introduced
by taking the previous relations~\eqref{Pepsdirac} and~\eqref{PPsidirac} as the definition
of the {\em{regularized wave evaluation operator}} (for details see~\cite[Section~1.2.2]{cfs}),
\[ 
\Psi^\varepsilon(x) := {\mathfrak{R}}_\varepsilon(x) \::\: \H \rightarrow S_x \scrM \:. \]
Then the kernel of the regularized fermionic projector can be written as
\[ 
P^\varepsilon(x,y) := -\Psi^\varepsilon(x)\, \Psi^\varepsilon(y)^* \::\: S_y\scrM \rightarrow S_x\scrM \:. \]
We again point out that this kernel does not need to satisfy the Dirac equation.

We next specify the regularization for the {\em{Minkowski vacuum}}.
In the previous section, we arranged that the commutator inner product agrees with
the usual current integral (see~\eqref{conserve}). This implies that the physical wave functions
from which the regularized wave evaluation operator is built up are all orthonormalized with
respect to the usual scalar product on the wave functions, i.e.\
\[ ( \Psi^\varepsilon u \,|\, \Psi^\varepsilon v )_t = c\: \la u | v \ra_\H \]
(again with a universal constant~$c$; here the scalar product on the left given by~\eqref{print}
for~$\scrN$ chosen as the Cauchy surface of constant time).
In other words, denoting the Hilbert space with scalar product~$(.|.)_t$ at time~$t$ by~$\H_t
\simeq L^2(\R^3, \C^4)$, the wave evaluation operator is an isometric embedding,
\[ \Psi^\varepsilon : \H \hookrightarrow \H_t \qquad \text{isometry for all~$t$} \:. \]
Consequently, the regularized kernel is described by the projection operator
onto the image of the operator~$\Psi^\varepsilon$. In the homogeneous setting of
the regularized Minkowski vacuum, this projection operator reduces to a multiplication
by a characteristic function in momentum space.
In a more physical language, we must work with a {\em{hard cutoff
in momentum space}} (i.e.\ a regularization where, as on the left of Figure~\ref{figseareg},
we multiply in momentum space by the characteristic function of a set~$\Omega$).
Moreover, we assume that in the chosen reference frame, the regularization is
spherically symmetric. Thus we choose
\beq \label{Pvac}
P^\varepsilon_\text{vac}(x,y) = \int \frac{d^4k}{(2 \pi)^4}\: \hat{P}^\varepsilon_\text{vac}(k)\: e^{-ik (y-x)}
\eeq
with
\beq \label{Pregmink}
\hat{P}^\varepsilon_\text{vac}(k) = \big( k_j \gamma^j +m \big)\: \delta \big( k^2-m^2 \big)\: \Theta\big( -k^0 \big) \:
\Theta\big( 1 + \varepsilon k^0 \big) \:.
\eeq
Before going on, we again point out that this hard cutoff is enforced by the identification of
the conservation laws which we introduced in order to get into the position of working with the
usual $L^2$-scalar product~\eqref{conserve} (see the discussion at the end of Section~\ref{secdiraccosmo}).

In the next step, we return to curved spacetime but consider 
the regularized kernel locally in a Gaussian coordinate system.
We assume that the spacetime region described by this coordinate system is so small that
curvature effects can be neglected. Moreover, the regularized kernel describing the
Dirac sea can be assumed in good approximation to be homogeneous, making it possible to take its
Fourier transform,
\[ P^\varepsilon(x,y) \approx \int \frac{d^4k}{(2 \pi)^4}\: \hat{P}^\varepsilon(k)\: e^{-ik (y-x)}\:. \]
This homogeneous setting was analyzed in detail~\cite{jokeldr}. The obtained results give an
indication that the regularizations which satisfy the EL equations are rigid in the sense that
they differ from the kernel of the Minkowski vacuum~\eqref{Pvac} at most by a change of the regularization scale
and a Lorentz boost. This motivates the assumption that the Fourier kernel~$\hat{P}^\varepsilon$ should be of
the form
\begin{align}
\hat{P}^\varepsilon(k) &= \hat{P}^\text{vac}\big( k, u \big) \qquad \text{with} \label{Preghom} \\
\hat{P}^\text{vac}(k, u) &= \big( k_j \gamma^j +m \big)\: \delta \big( k^2-m^2 \big)\: \Theta\big( -k^0 \big)
\: \Theta\big( 1+k u \big) \:, \label{Pregu}
\end{align}
where~$u$ is a future-directed, timelike vector (note that the regularization scale is given by~$\sqrt{u^2}$).
From here on, we use this form of the regularization as an auxiliary hypothesis
and show that it has interesting implications.

In order to take into account that the regularization may vary on cosmological scales, we choose
a timelike and future-directed vector field~$u(x)$, referred to as the {\em{regularizing vector field}}.
The dynamics of the regularizing vector field can be determined by taking the average of the
solutions of the transport equation~\eqref{ftransport} over all null directions.
This will be explained further in the Appendices~\ref{appA} and~\ref{appB}.
In the main text of this paper, we focus on the question of how the regularized kernel
can be introduced for a given regularizing vector field.
In order to explain the difficulties, we begin by discussing a first naive idea
of defining the regularized kernel with a quasi-homogeneous ansatz
similar to the Wigner function, i.e.\ in local Gaussian coordinates
\beq \label{Pepsnaive}
P^\varepsilon(x,y) :=
\int \frac{d^4k}{(2 \pi)^4}\: \hat{P}^\text{vac}\bigg( k, u\Big( \frac{x+y}{2} \Big) \bigg)\: e^{-ik (y-x)}\:.
\eeq
This simple method has the shortcoming that the conservation law as set up in Section~\ref{secdiraccosmo}
is in general not respected.
In order to improve the situation, we need to modify the Dirac dynamics carefully
while preserving current conservation, in such a way
that, locally, the regularization has the desired form~\eqref{Preghom}.
The next sections are devoted to the derivation of this modified dynamics.

\subsection{A Spectral Description of the Regularization} \label{secAintro}
In order to avoid the problems of the naive ansatz~\eqref{Pepsnaive}, we want to describe the
regularization at any time~$t$ by a projection operator~$\tilde{E}(t)$.
To this end, we again choose a foliation~$(\scrN_t)_{t \in \R}$ of the globally hyperbolic
spacetime~$\scrM$ by Cauchy surfaces. It is convenient to realize the projection operator as
a spectral projection operator of an operator denoted by~$\tilde{A}(t)$,
\beq \label{Egen}
\tilde{E}(t) := \chi_I \big( \tilde{A}(t) \big)
\eeq
where~$I$ an interval, which in view of the freedom in rescaling~$\tilde{A}(t)$ we simply choose as
\[ I := \big( -1, 1 \big) \:. \]
In order to describe the regularization~\eqref{Pregmink} in the {\em{Minkowski vacuum}}, we can simply choose~$\tilde{A}=
\varepsilon H$
as a multiple of the Dirac Hamiltonian
\beq \label{Hvac}
H := - i (\gamma^0)^{-1} \gamma^\alpha \partial_\alpha + \gamma^0 m \:.
\eeq
Then~\eqref{Egen} describes a hard cutoff at frequency~$\omega=-\varepsilon^{-1}$.
The regularization~\eqref{Pregu}, on the other hand, can be arranged by choosing~$\tilde{A}=A(u)$ as the differential operator
\beq \label{Au}
A(u) = u^0\: H + i \sum_{\alpha=1}^3 u^\alpha\: \frac{\partial}{\partial x^\alpha} \:.
\eeq
Here~$H$ and~$-i \partial_\alpha$ are the energy and momentum operators. Therefore,
plane wave solutions of momentum~$k$ are eigenfunctions of~$A(u)$
corresponding to the eigenvalues~$ku$. With this in mind, the projection operator~\eqref{Egen} implements
the condition~$ku>-1$ in~\eqref{ku}, in a way which can be generalized to curved spacetime.

In {\em{curved spacetime}}, we need to choose~$\tilde{A}$ in such a way that it has the following properties:
\bitem
\item[(a)] Locally in a Gaussian coordinate system, it has the form~\eqref{Au}.
\item[(b)] It is symmetric with respect to the scalar product~\eqref{conserve} with~$\scrN=\scrN_t$.
\eitem
The first idea would be to choose~$\tilde{A}$ in generalization of~\eqref{Hvac} as the
Dirac Hamiltonian~$\tilde{H}$, which in curved spacetime is introduced by writing the Dirac equation
in the Hamiltonian form
\beq \label{tilHdef}
i \partial_t \psi = \tilde{H}\, \psi
\eeq
(where~$t$ is the time function corresponding to the foliation~$(\scrN_t)_{t \in \R}$).
However, this choice has two shortcomings: First, the Hamiltonian is
in general not symmetric with respect to the scalar product~\eqref{conserve}
(for details see~\cite[Section~3.6]{intro} or~\cite{arminjon2}). Second, the Hamiltonian~$\tilde{H}$
has the undesirable property that it depends on the lapse function.
In analogy to~\eqref{Au} we set
\beq \label{Atildef}
\tilde{A}(t) = \frac{1}{4} \: \Big\{ u^0 ,\; \big( H + H^* \big) \Big\} + 
\frac{1}{4} \sum_{\alpha=1}^3 \Big\{ u^\alpha,\: \big(i \,\nabla_\alpha - i \,\nabla_\alpha^* \big) \Big\} \:,
\eeq
where the star denotes the formal adjoint on the Hilbert space~$\H_t$.
Here the anti-commutators are needed in order to get a symmetric operator with respect to
the scalar product~\eqref{conserve}, where we choose the domain as the smooth and compactly supported
spinors,
\[ \D \big( \tilde{A}(t) \big) = C^\infty_0(\scrN_t, S\scrM) \:. \]
 Using Chernoff's method~\cite{chernoff73}, one sees that this operator is essentially selfadjoint.
 We denote the unique selfadjoint extension again by~$\tilde{A}(t)$.
 Then the corresponding projection operator~$\tilde{E}(t)$ can be defined by~\eqref{Egen}.

\subsection{The Dirac Dynamics Modified by Adiabatic Projections} \label{secadiabatic}
Now we can implement the guiding principles~(i)--(iii) stated at the beginning of this section.
Namely, in order to realize~(ii) while preserving the conservation law~(iii), we
modify the Dirac dynamics by repeated projections with the operator~$\tilde{E}(t).$\footnote{It is interesting to note that the locally rigid dynamics in~\eqref{Vtdef} with its continuous adiabatic projection 
bears some similarity with the dynamics of open isolated quantum systems in Fröhlich's ETH approach to Quantum Theory; see \cite{froehlich2019review} for a review of the approach, \cite{eth-cfs} for a detailed comparison with the theory of causal fermion systems and \cite{frohlich2021time} for a detailed discussion of the modified dynamics under the ETH approach. It is important to note, however, that the projections in Fröhlich's ETH approach to Quantum Theory are {\em{not}} adiabatic and hence, in contrast to the evolution given by \eqref{Vtdef}, their resulting dynamics are not unitary. Nevertheless, it should be well worth keeping Fröhlich's ETH approach in mind when further investigating the dynamics in~\eqref{Vtdef}. In particular in the light, that the transition from the inflationary period to our present day matter and radiation filled universe can be seen as a transition from unitary to stochastic evolution. Conceptually this fits neatly into the new story line of the universe developed in~\cite{paganini2020proposal} based on preliminary considerations to the present work.}
Thus
\beq \label{Vtdef}
V^t_{t_0} \, \psi_0 = \lim_{N \rightarrow \infty}
\tilde{E}_I(t) \,U^{t}_{t-\Delta t}\, 
\cdots
\tilde{E}_I\big(t_0+2 \Delta t\big) \,U^{t_0+ 2 \Delta t}_{t_0+\Delta t}\, \tilde{E}_I\big(t_0+\Delta t\big) \,U^{t_0+\Delta t}_{t_0}\, \psi(0) \:,
\eeq
where we set~$\Delta t := (t-t_0)/N$, and
\beq \label{Udef}
U^{t''}_{t'} : \H_{t'} \rightarrow \H_{t''}
\eeq
denotes the unitary Dirac dynamics
from time~$t'$ to~$t''$. This time evolution can be written infinitesimally as
\beq \label{dtV}
\frac{d}{dt} V^t_{t_0} = \big( \dot{\tilde{E}}_I(t) -i \tilde{E}_I(t) \,\tilde{H}(t) \big)\, V^t_{t_0} \:,
\eeq
where~$\tilde{H}(t)$ is again the Dirac Hamiltonian~\eqref{tilHdef}.
This equation can be solved with a time-ordered exponential,
\beq \label{VTexp}
V^t_{t_0} = \Texp \bigg( \int_0^t \big( \dot{\tilde{E}}_I(\tau) -i \tilde{E}_I(\tau) \,\tilde{H}(\tau) \big)\: d\tau \bigg) \:,
\eeq
defined via the Dyson series
\begin{align*}
&\Texp \bigg( \int_0^t \big( B(\tau)\, d\tau \bigg)
:= \1 + \int_0^t \big( B(\tau)\, d\tau
+ \int_0^t d\tau \int_0^\tau d\tau'\: B(\tau)\: B(\tau') + \cdots \:.
\end{align*}
The method of continuous projections in~\eqref{Vtdef} is also referred to as
{\em{adiabatic projections}} (see for example~\cite[Section~1.1]{teufel-adiabatic} or~\cite[Section~8.2]{avron}).
As can be verified directly from~\eqref{dtV}, these adiabatic projections
give rise to a unitary dynamics (for details in a somewhat different context
see for example~\cite[proof of Proposition~4.2]{norm}).
In this way, we have achieved our goal: At any time~$t$, the regularization
has the desired form as described by the
projection operator~$\tilde{E}(t)$. Moreover, the dynamics is compatible with the conservation law~\eqref{conserve}.
Therefore, the ordered exponential~\eqref{VTexp} seems the right ansatz for modifying the Dirac dynamics.
The time evolution~$V^t_{t_0}$ is the desired {\em{locally rigid dynamics}}.

\section{Detecting Fermiogenesis}\label{sec:baryo}
The goal of this section is to give a concise procedure for detecting fermiogenesis.
To this end, we again choose a foliation~$(\scrN_t)_{t \in \R}$ in a globally hyperbolic spacetime. We denote the
projection operator to the occupied states at time~$t$ by
\begin{align*} 
& \tilde{\Pi}(t) \::\: \H_t \rightarrow \H_t \\
& (\tilde{\Pi}(t)\, \psi)(\x) = -\int_{\scrN_t} \tilde{P}^\varepsilon\big( (t,\x), y \big)\; \gamma(\nu)(y) \: \psi(y)\: d\mu_{\scrN_t}(y) \:.
\end{align*}
Its time evolution is given either by the Dirac dynamics or by the modified dynamics with adiabatic projections, i.e.\
\begin{align}
\tilde{\Pi}(t) &= U^t_{t_0} \,\tilde{\Pi}(t_0)\, (U^t_{t_0})^{-1} && \text{(Dirac dynamics)} \label{Utime} \\
\tilde{\Pi}(t) &= V^t_{t_0} \,\tilde{\Pi}(t_0)\, (V^t_{t_0})^{-1} && \text{(locally rigid dynamics)} \label{Vtime}
\end{align}
(where~$U^t_{t_0}$ is unitary operator describing the Dirac time evolution~\eqref{Udef},
whereas~$V^t_{t_0}$ is the modified time evolution operator given by~\eqref{Vtdef} or~\eqref{VTexp}).
In view of the limited energy range accessible to current experiments,
for determining whether fermiogenesis occurs we may restrict attention to an energy scale~$\Lambda$,
which we assume to be in the range
\beq \label{Lamrange}
m, \ell_\text{macro}^{-1} \ll \Lambda \ll \varepsilon^{-1} \:.
\eeq
Next, we choose an operator~$\eta_\Lambda$ acting on the spatial Hilbert space~$\H_t$ such that the trace
\beq \label{trLam}
\tr \big( \eta_\Lambda\: \tilde{\Pi}(t) \big)
\eeq
tells us about the number of particles in the energy range~$(-\Lambda, \Lambda)$.
The rate of fermiogenesis~$B(t)$ (as measured in the global time function of the foliation
by Cauchy surfaces~$(\scrN_t)_{t \in \R}$), should then be given by the time change of this quantity,
\beq \label{fermiogenesis}
B(t) := \frac{d}{dt} \tr \big( \eta_\Lambda\: \tilde{\Pi}(t) \big) \:.
\eeq
Since the choice of the energy scale~$\Lambda$ is arbitrary, the function~$B(t)$ should
not depend on the choice of~$\Lambda$, as long as it is in the range~\eqref{Lamrange}.
As we shall see, the choice of the operator~$\eta_\Lambda$ is not straightforward.
Therefore, we first consider the situation for the Dirac dynamics~\eqref{Utime},
in which case we shall work out that, as expected naively, there is indeed {\em{no}} fermiogenesis
(Section~\ref{secnobaryo}). Then we move on to the locally rigid dynamics~\eqref{Vtime}
and verify that in general there will be fermiogenesis (Section~\ref{secbaryomod}).
We also derive explicit formulas for~$B(t)$.

\subsection{Absence of Fermiogenesis for the Dirac Dynamics} \label{secnobaryo}
Before coming to the locally rigid dynamics, we need to verify that the Dirac dynamics does {\em{not}}
give rise to fermiogenesis. This statement corresponds to the usual conception that the Dirac dynamics
only allows for the generation of particle/anti-particle pairs. 
The difficulty in making this statement precise is that the particles detected by~\eqref{trLam}
also include the states of the Dirac sea. Before we can make sense of fermiogenesis via~\eqref{fermiogenesis},
we need to make sure that the contribution by the sea states to~\eqref{trLam} is constant,
because only then the time derivative~\eqref{fermiogenesis} tells us exclusively about fermiogenesis.
In curved spacetime, this makes it necessary to carefully adjust the operator~$\eta_\Lambda$
to the local geometry. The necessity of these adjustments is illustrated in Appendix~\ref{appgauge}
by a simple counter example. Our method is to subtract counter terms which depend on the
geometry in a neighborhood of the Cauchy surface~$\eta_t$. It is most convenient to
determine these counter terms from the Hadamard expansion of~$P(x,y)$.
Since the operator~$\eta_\Lambda$ already involves a cutoff on the energy scale~$\Lambda$,
we can leave out the regularization on the scale~$\varepsilon$ and work with the unregularized
kernel. We will show that, having determined~$\eta_\Lambda$ in this way, the function~$B(t)$ vanishes,
even taking into account the smooth contributions in~\eqref{hadamard1} (see Theorem~\ref{thmnobaryo} below).
Before entering the details, we remark that this method bears some similarity with the point splitting
renormalization of the energy-momentum tensor in curved spacetime as
used for example in~\cite{fewster+verch2}.
Our method can be regarded as an adaptation of this procedure to the problem of fermiogenesis.

For the detailed construction, we truncate the Hadamard expansion~\eqref{hadamard1}.
Thus, given~$N \geq 2$, we define the bi-distribution~$P^N(x,y)$ by
\[ 
\tilde{P}^N(x,y) := \lim_{\varepsilon \searrow 0} \bigg( X_{-2} \: \Dir_x \Big(
\frac{1}{\sigma_\varepsilon} \Big) + \frac{X_{-1}}{\sigma_\varepsilon} 
+  \log \sigma_\varepsilon \sum_{n=0}^{N} X_n \:\sigma_\varepsilon^n + 
\sum_{n=1}^{N} Y_n \:\sigma_\varepsilon^n \bigg) \:, \]
where~$\sigma_\varepsilon$ is again the function~\eqref{sigmadef}.
Here the parameter~$N>0$ gives the order of the expansion (it will be specified
in Theorem~\ref{thmnobaryo} below).
Here the functions~$X_n, Y_n : S_y\scrM \rightarrow S_x\scrM$
are smooth matrix-valued functions which can be calculated explicitly
by solving transport equations along the geodesic joining~$x$ and~$y$
(for computational details see~\cite{baer+ginoux}, \cite[Section~5 and Appendix~A]{lqg} or~\cite{reghadamard}).
The truncated kernel is again symmetric (for details see~\cite{moretti}).
It satisfies the Dirac equation with an error term,
\beq \label{Diracapprox}
(\Dir_x - m)\: \tilde{P}^N(x,y) = \tilde{e}(x,y) \:,
\eeq
and this error term is small near the diagonal~$x=y$ (for details see the proof of Theorem~\ref{thmnobaryo} below).

An important point for what follows is that all the terms of the Hadamard expansion
depend only on the local geometry in a neighborhood of the Cauchy surface~$\scrN$.
They describe what an observer on the Cauchy surface would consider to be a ``vacuum state.''
With this in mind, for detecting fermiogenesis we modify~$\tilde{\Pi}(t)$ by subtracting the truncated Hadamard distribution,
\beq \label{tilPireg} \begin{split}
& \tilde{\Pi}^N(t) \::\: \H_t \rightarrow \H_t \\
& (\tilde{\Pi}^N(t)\, \psi)(\x) = -\int_{\scrN_t} \Big( \tilde{P}^\varepsilon\big( (t,\x), y \big)
- \tilde{P}^N\big( (t,\x),y \big) \Big)\; \gamma\big(\nu(y)\big) \: \psi(y)\: d\mu_{\scrN_t}(y) \:.
\end{split}
\eeq
We also modify~\eqref{fermiogenesis} accordingly by setting
\beq \label{baryoN}
B^N(t) := \frac{d}{dt} \tr \big( \eta_\Lambda\: \tilde{\Pi}^N(t) \big) \:.
\eeq
We choose~$\eta_\Lambda$ on~$\scrN_t$ as the integral operator
\beq \label{etaLdef}
(\eta_\Lambda \psi)(x) := \int_{\scrN_t} \eta_\Lambda(x,y) \: \psi(y)\: d\mu_{\scrN_t}(y) \:,
\eeq
where~$\eta_\Lambda(x,y)$ is the integral kernel
\beq \label{etaLkernel}
\eta_\Lambda(x,y) := \eta \Big( \frac{d(x,y)}{\Lambda} \Big) \: {\mathcal{J}}^y_x \:.
\eeq
Here~$d(.,.)$ is the Riemannian distance function on~$\scrN_t$, and~$\mathcal{J}^y_x : S_y \scrM \rightarrow S_x \scrM$
is the parallel transport corresponding to the spinorial Levi-Civita connection on~$S\scrM$
along the shortest geodesic in~$\scrN_t$ joining the points~$x$ and~$y$.
Moreover, $\eta \in C^\infty_0([-1,1], \R^+_0)$ is a compactly supported test function. We always choose~$\Lambda$
so large that~$\eta_\Lambda(x,.)$ is supported in a convex neighborhood of~$x$ in~$\scrN_t$.

\begin{Thm} \label{thmnobaryo} For the Dirac dynamics and choosing~$N \geq 2$,
\beq \label{nobaryo}
B^N(t) = \O \Big( \frac{1}{\Lambda} \Big) \:,
\eeq
uniformly in~$\varepsilon$.
\end{Thm}

Before coming to the proof, we explain how the error term in~\eqref{nobaryo} is to be understood.
Clearly, when choosing~$\Lambda$ large, one must keep in mind the condition~$\Lambda \ll \varepsilon^{-1}$
in~\eqref{Lamrange} must be satisfied. The statement on the uniformity in~$\varepsilon$ in Theorem~\ref{thmnobaryo}
makes it possible to take the simultaneous limit~$\Lambda \rightarrow \infty$ and~$\varepsilon \searrow 0$
while respecting~\eqref{Lamrange}. The relevant length scales for the decay in~\eqref{nobaryo}
is the energy scale~$\ell_\text{macro}^{-1}$ of macroscopic physics. Thus one could write the error term
in a scale invariant way as
\[ \O \Big( \frac{\ell_\text{macro}}{\Lambda} \Big) \:. \]

\Proof[Proof of Theorem~\ref{thmnobaryo}]
The computation of the time derivative~\eqref{fermiogenesis} is a bit subtle, because
also the integration measure~$d\mu_\scrN$ is time dependent.
For this reason, it is preferable to first
consider the time derivative of the probability integral~\eqref{print}.
For convenience, we work
in a coordinate system with vanishing shift vector. Thus, given the foliation~$(\scrN_t)_{t \in \R}$, we choose the coordinates on the leaves such that the vector field~$\partial_t$ is orthogonal.
Then, compared to~\eqref{lapseshift}, the metric takes the form
\beq \label{Nchart}
ds^2 = N(x)^2\: dt^2 - \sum_{\alpha, \beta=1}^3 g_{\alpha \beta}(x)\: dx^\alpha\: dx^\beta \:.
\eeq

Let~$\psi, \phi \in \Cisc(\scrM, S\scrM)$ be two wave functions (not necessarily Dirac solutions). 
We write the difference of the probability integrals at times~$t_1$ and~$t_0$ as
\[ (\psi | \phi)_t \big|_{t_0}^{t_1} = 
-i \int_\scrM \chi_{[t_0, t_1]}(t) \Big( \Sl \psi \,|\, (\Dir - m) \phi \Sr_x - \Sl (\Dir - m) \psi \,|\, \phi \Sr_x
\Big)\: d\mu_\scrM(x) \:. \]
This formula is immediately verified using integration by parts; it makes manifest that for solutions of the
Dirac equation, the probability integral is conserved in time. Using Fubini, we write the spacetime integral as
\[ (\psi | \phi)_t \big|_{t_0}^{t_1} = -i \int_{t_0}^{t_1} dt \int_{\scrN_t} 
\Big( \Sl \psi \,|\, (\Dir - m) \phi \Sr_x - \Sl (\Dir - m) \psi \,|\, \phi \Sr_x \Big)\: N(x)\: d\mu_{\scrN_t}(x) \:. \]
where~$N$ is the lapse function~\eqref{Nchart}.
Now we can differentiate w.r.t.\ time to obtain
\beq \label{dtscal}
\frac{d}{dt} (\psi | \phi)_t = -i \int_{\scrN_t} 
\Big( \Sl \psi \,|\, (\Dir - m) \phi \Sr_x - \Sl (\Dir - m) \psi \,|\, \phi \Sr_x \Big)\: N(x)\: d\mu_{\scrN_t}(x) \:.
\eeq
The main purpose of this formula is that it makes precise how the current integral changes if~$\psi$ or~$\phi$
do {\em{not}} satisfy the Dirac equation.

In order to use~\eqref{dtscal} for the computation of~\eqref{baryoN}, we write the difference of the
symmetric kernels in~\eqref{tilPireg} symbolically with bra-ket notation as
\[ \tilde{P}^\varepsilon(x,y) - \tilde{P}^N(x,y) = \sum_a c_a\: |\psi_a(x) \Sr \Sl \psi_a(y)| \]
(with some wave functions~$\phi_a$ and coefficients~$c_a$).
Then the trace in~\eqref{baryoN} can be written as
\[ \tr \big( \eta_\Lambda\: \tilde{\Pi}^N(t) \big) = 
\sum_a c_a \:\big( \psi_a \:\big|\: \eta_\Lambda\: \psi_a \big)_t \:. \]
Now we can apply~\eqref{dtscal} to obtain
\begin{align*}
&\frac{d}{dt} \tr \big( \eta_\Lambda\: \tilde{\Pi}^N(t) \big) \\
&= -i \int_{\scrN_t} \sum_a c_a \:
\Big( \Sl \psi_a \,|\, (\Dir - m) \,\eta_\Lambda\: \psi_a \Sr_x - \Sl (\Dir - m) \,\psi_a \,|\, \eta_\Lambda\: \psi_a \Sr_x \Big)\: N(x)\: d\mu_{\scrN_t}(x) \:.
\end{align*}
Now we can put in the inhomogeneous Dirac equation~\eqref{Diracapprox} to obtain
\beq \label{baryoreg}
\begin{split}
\frac{d}{dt} \tr \big( \eta_\Lambda\: \tilde{\Pi}^N(t) \big) 
= -i \int_{\scrN_t} \Tr_{S_x\scrM} \Big( &[\Dir, \eta_\Lambda] \,(\tilde{P}^\varepsilon-\tilde{P}^N) \big) (x,x) \\
&-\big( \eta_\Lambda\:(\tilde{e} - \tilde{e}^*) \big)(x,x)  \Big) \, N(x)\: d\mu_{\scrN_t}(x) \:.
\end{split}
\eeq

Having chosen~$N \geq 2$, the error term~$\tilde{e}$ decays at least linearly near the diagonal, i.e.
\[ \big\| \tilde{e}(t,\x; t,\y) \big\| = \O \big( (\x-\y)^2 \big) \:. \]
As a consequence, the contributions involving~$\tilde{e}$ and~$\tilde{e}^*$ to fermiogenesis
are of the order~$\O(\Lambda^{-1})$.

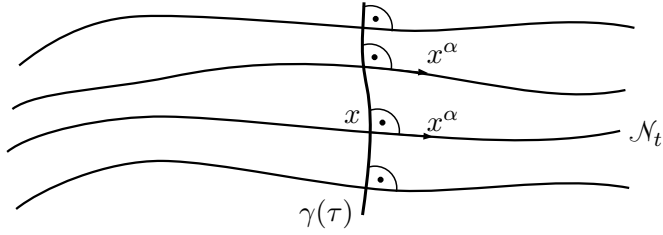
\begin{figure}
\psscalebox{1.0 1.0} 
{
\begin{pspicture}(0,28.36396)(8.372659,31.217607)
\psbezier[linecolor=black, linewidth=0.03](0.019448243,29.216824)(0.34750894,29.482914)(1.353032,29.696392)(2.1113622,29.68682373046875)(2.8696923,29.677256)(4.0355654,29.532913)(4.9771633,29.466824)(5.9187613,29.400734)(7.189084,29.277254)(8.169448,29.496824)
\psbezier[linecolor=black, linewidth=0.03](0.08944824,29.786823)(0.41750893,30.027328)(1.343032,30.04028)(2.1813622,30.211631422776364)(3.0196924,30.382982)(3.8055654,30.44079)(4.9971633,30.326824)(6.188761,30.212858)(7.259084,29.841444)(8.239449,30.0399)
\psbezier[linecolor=black, linewidth=0.03](0.17944825,30.366823)(0.50750893,30.632914)(1.023032,30.966393)(1.7113622,31.01682373046875)(2.3996923,31.067255)(4.2255654,30.902914)(5.1671634,30.836824)(6.1087613,30.770735)(7.399084,31.077255)(8.359448,30.866823)
\psbezier[linecolor=black, linewidth=0.03](0.13944824,28.456823)(0.46750894,28.722914)(1.333032,29.106392)(2.0913622,29.09682373046875)(2.8496923,29.087255)(4.1555653,28.772913)(5.097163,28.706823)(6.0387616,28.640734)(7.329084,28.917255)(8.289448,28.736824)
\psbezier[linecolor=black, linewidth=0.04](4.764448,31.206823)(4.7432013,30.664282)(4.7115154,30.536798)(4.804448,30.06682373046875)(4.8973813,29.59685)(4.8236012,28.924152)(4.744448,28.376823)
\psarc[linecolor=black, linewidth=0.02, dimen=outer](4.909448,28.736824){0.29}{-10.0}{110.0}
\pscircle[linecolor=black, linewidth=0.02, fillstyle=solid,fillcolor=black, dimen=outer](4.9194484,30.986824){0.03}
\pscircle[linecolor=black, linewidth=0.02, fillstyle=solid,fillcolor=black, dimen=outer](4.904448,30.476824){0.03}
\pscircle[linecolor=black, linewidth=0.02, fillstyle=solid,fillcolor=black, dimen=outer](5.009448,29.606823){0.03}
\pscircle[linecolor=black, linewidth=0.02, fillstyle=solid,fillcolor=black, dimen=outer](4.9844484,28.846825){0.03}
\psline[linecolor=black, linewidth=0.02, arrowsize=0.05291667cm 2.0,arrowlength=1.4,arrowinset=0.0]{->}(4.864448,29.476824)(5.6844482,29.416824)
\psline[linecolor=black, linewidth=0.02, arrowsize=0.05291667cm 2.0,arrowlength=1.4,arrowinset=0.0]{->}(4.7744484,30.346825)(5.614448,30.266823)
\psarc[linecolor=black, linewidth=0.02, dimen=outer](4.9444485,29.496824){0.29}{-10.0}{110.0}
\psarc[linecolor=black, linewidth=0.02, dimen=outer](4.8394485,30.361824){0.29}{-10.0}{110.0}
\psarc[linecolor=black, linewidth=0.02, dimen=outer](4.849448,30.871824){0.29}{-10.0}{110.0}
\rput[bl](8.3,29.3){$\scrN_t$}
\rput[bl](4.5,29.6){$x$}
\rput[bl](5.6,29.5){$x^\alpha$}
\rput[bl](5.6,30.4){$x^\alpha$}
\rput[bl](3.9,28.2){$\gamma(\tau)$}
\end{pspicture}
}
\caption{The tubular chart near~$x$.}
\label{figchart}
\end{figure}

The remaining task is to estimate the commutator~$[\Dir, \eta_\Lambda]$. To this end, it is most convenient to work in another adapted coordinate system
near~$x \in \scrN_t$ which is constructed as follows. We let~$\gamma(\tau)$ with~$\tau \in (t-\delta, t+\delta)$
and~$\delta>0$ be a curve with~$\gamma(t)=x$
and~$\gamma(\tau) \in \scrN_\tau$ for all~$\tau$. Moreover, its tangent vector~$\dot{\gamma}(\tau)$
should be orthogonal to~$\scrN_\tau$ for all~$\tau$ (see Figure~\ref{figchart}).
This curve should be thought of as the curve~$(\tau, \x)$ for fixed spatial coordinates in our
chart with metric~\eqref{Nchart}.
Now we choose a tubular chart around~$\gamma$ such that on each Cauchy surface~$\scrN_\tau$ it is
a Gaussian chart around~$\gamma(\tau)$. Denoting the resulting coordinates by~$(\tau, \y)$,
the metric takes the form
\[ ds^2 = \alpha(\tau, \y)\: d\tau^2 - 2 \beta_\alpha(\tau, \y)\: d\tau\, dy^\alpha - 
\sum_{\alpha, \beta=1}^3 g_{\alpha \beta}(\tau, \y)\: dy^\alpha\, dy^\beta \:, \]
where~$\alpha(x)$ and~$\beta(x)$ are the {\em{lapse function}} and the~{\em{shift vector}}
(for details see again~\cite[Section~VI.3]{choquet}). Moreover, the curve~$\gamma(\tau)$ has again
the coordinates~$(\tau, \x)$.
Along this curve,
\[ 
\beta_\alpha(\tau, \x) = 0 \qquad \text{and} \qquad
g_{\alpha \beta}(\tau, \x) = \delta_{\alpha \beta} \:,\quad \partial_\gamma g_{\alpha \beta}(\tau, \x)= 0 \:. \]
We now compute,
\begin{align}
\Dir &= i \gamma^j \partial_j + \B \notag \:, \\
& \!\!\!\!\! \big( [\Dir, \eta_\Lambda] \psi \big)(x) = \Dir_x \int_{\scrN_t} \eta_\Lambda(x,y) \: \psi(y)\: d\mu_{\scrN_t}(y)
- \int_{\scrN_t} \eta_\Lambda(x,y) \: (\Dir_y \psi(y))\: d\mu_{\scrN_t}(y) \notag \\
&= \int_{\scrN_t} \big( \B(x) \,\eta_\Lambda(x,y) - \eta_\Lambda(x,y)\: \B(y) \big) \: \psi(y)\: d\mu_{\scrN_t}(y) \label{t0} \\
&\quad\: + i \gamma^0(\tau, \x)\, \frac{\partial}{\partial \tau} \int \eta_\Lambda(\tau, \x; \tau, \y) \: \psi(\tau, \y)\: \sqrt{\det g_{\alpha \beta}(\tau, \y)}\: d^3\y \label{t1} \\
&\quad\: - i \int \eta_\Lambda(\tau, \x; \tau, \y) \: 
\Big( \gamma^0(\tau, \y)\, \partial_\tau \psi(\tau, \y) \Big)\: \sqrt{\det g_{\alpha \beta}(\tau, \y)}\: d^3\y \label{t2} \\
&\quad\: +i \sum_{\alpha=1}^3 \gamma^\alpha(\tau, \x)\, \frac{\partial}{\partial x^\alpha} \int \eta_\Lambda(\tau, \x; \tau, \y) \: \psi(\tau, \y)\: \sqrt{\det g_{\alpha \beta}(\tau, \y)}\: d^3\y \label{t3} \\
&\quad\: - i \sum_{\alpha=1}^3 \int \eta_\Lambda(\tau, \x; \tau, \y) \: 
\Big( \gamma^\alpha(\tau, \y)\, \partial_{\y^\alpha} 
\psi(\tau, \y) \Big)\: \sqrt{\det g_{\alpha \beta}(\tau, \y)}\: d^3\y \:. \label{t4} 
\end{align}
The term~\eqref{t0} is of the order~$\O(\Lambda^{-1})$. Likewise, in~\eqref{t1} the $\tau$-derivatives
of the metric are of the order~$\O(\Lambda^{-1})$. The remaining contribution can be combined with~\eqref{t2}
to
\[ i \int \eta_\Lambda(\tau, \x; \tau, \y) \: 
\Big( \big(\gamma^0(\tau, \x) - \gamma^0(\tau, \y) \big)\, \partial_\tau \psi(\tau, \y) \Big)\: \sqrt{\det g_{\alpha \beta}(\tau, \y)}\: d^3\y \:, \]
which is again of the order~$\O(\Lambda^{-1})$. In~\eqref{t3}, the $\x$-derivative of~$\eta_\Lambda$ can
be written as minus the $\y$-derivative, up to an error of the order~$\O(\Lambda^{-1})$.
Integrating by parts, we can combine the resulting contributions with~\eqref{t4} to obtain
\beq \label{dyterms}
i \sum_{\alpha=1}^3 \int \eta_\Lambda(\tau, \x; \tau, \y) \: 
\Big( \big( \gamma^\alpha(\tau, \x) - \gamma^\alpha(\tau, \y) \big)\, \partial_{\y^\alpha} 
\psi(\tau, \y) \Big)\: \sqrt{\det g_{\alpha \beta}(\tau, \y)}\: d^3\y \:,
\eeq
again up to an error of the order~$\O(\Lambda^{-1})$.
Finally, also the contribution~\eqref{dyterms} is of the order~$\O(\Lambda^{-1})$,
concluding the proof.
\QED

\subsection{Fermiogenesis for the Locally Rigid Dynamics} \label{secbaryomod}
In order to detect fermiogenesis for the locally rigid dynamics~\eqref{Vtime}, we again 
subtract the truncated Hadamard expansion according to~\eqref{tilPireg} 
and consider the trace~\eqref{baryoN}, where~$\eta_\Lambda$ is again the
integral operator~\eqref{etaLdef} with kernel~\eqref{etaLkernel}.
We thus again obtain~\eqref{baryoreg}, but now with the error term~$\tilde{e}$ given by
\begin{align*}
\tilde{e}(x,y) := -(\Dir_x - m)\: \big( \tilde{P}^\varepsilon(x,y) - \tilde{P}^N(x,y) \big) \:.
\end{align*}
The commutator in~\eqref{baryoreg} can be treated exactly as in the proof of Theorem~\ref{thmnobaryo}.
The same is true for the error term~$(\Dir_x - m)\: \tilde{P}^N(x,y)$. But we clearly need to take
into account the fact that~$P^\varepsilon(x,y)$ no longer satisfies the Dirac equation. Indeed,
using~\eqref{Vtime} and~\eqref{dtV} we obtain
\begin{align*}
\tilde{e}(x,y) &= -(\Dir_x - m)\: \tilde{P}^\varepsilon(x,y) = -\gamma^0 (i \partial_t - \tilde{H} \big) \tilde{P}^\varepsilon(x,y) \\
&= -\gamma^0 \big( i \dot{\tilde{E}}_I(t) + \tilde{E}_I(t) \,\tilde{H}(t) - \tilde{H}(t) \big) \tilde{P}^\varepsilon(x,y) \\
&= -\gamma^0 \big( i \dot{\tilde{E}}_I(t) + [ \tilde{E}_I(t), \,\tilde{H}(t) ] \big) \tilde{P}^\varepsilon(x,y) \:.
\end{align*}
We conclude that
\begin{align}
B(t) &= \frac{d}{dt} \tr \big( \eta_\Lambda\: \tilde{\Pi}^N(t) \big) \label{bt}
= i \int_{\scrN_t} \Tr_{S_x\scrM} \Big( \big( \eta_\Lambda\:(\tilde{e} - \tilde{e}^*) \big)(x,x)  \Big) \, N(x)\: d\mu_{\scrN_t}(x) \\
&= -2 \,\im \int_{\scrN_t} \Tr_{S_x\scrM} \Big( \big( \eta_\Lambda\:\tilde{e}  \big)(x,x)  \Big) \, N(x)\: d\mu_{\scrN_t}(x)
\notag \\
&= 2 \,\im \int_{\scrN_t} \Tr_{S_x\scrM} \Big( \big( \eta_\Lambda\: \gamma^0 
\big( i \dot{\tilde{E}}_I(t) + [ \tilde{E}_I(t), \,\tilde{H}(t) ] \big) \tilde{P} \big)(x,x)  \Big) \, N(x)\: d\mu_{\scrN_t}(x) \:.
\notag
\end{align}

In order to clarify the structure of this formula, we now proceed by analyzing its ingredients
in a perturbation expansion. Our goal is to derive an approximate formula
for fermiogenesis which is manifestly independent of~$\Lambda$
(see~\eqref{approx} in Section~\ref{secapprox}).

\section{Perturbative Description} \label{secperturb}
\subsection{Perturbation Expansion of the Spectral Measure} \label{secperturbE}
In order to obtain explicit results, we must employ perturbative methods.
To this end, we need to assume that~$\tilde{A}(t)$ has a perturbation expansion of the form
\beq \label{Apert}
\tilde{A}(t) = A + \Delta A(t) \qquad \text{with} \qquad \Delta A(t) := \sum_{p=1}^\infty \lambda^p\, A^{(p)}(t) \:,
\eeq
where~$A$ is again the spatial operator in Minkowski space~\eqref{Au}
(we treat the series as a formal power series, disregarding issues of convergence).
Likewise, the resolvent has the expansion
\begin{align}
\tilde{R}_\omega(t) &:= \big( \tilde{A} - \omega \big)^{-1} 
= \big( A + \Delta A - \omega \big)^{-1} 
= \Big( (A-\omega) \big( 1 + R_\omega \,\Delta A(t) \big) \Big)^{-1} \notag \\
&= \big( 1 + R_\omega \,\Delta A(t) \big)^{-1} R_\omega 
= \sum_{p=0}^\infty \big( - R_\omega \,\Delta A(t) \big)^p \,R_\omega \:, \label{Rpert}
\end{align}
where we used the Neumann series and set~$R_\omega = (A - \omega)^{-1}$.
Substituting the power ansatz~\eqref{Apert} gives a corresponding perturbation
expansion of the resolvent. For example, to second order we obtain
\[ 
\tilde{R}_\omega = R_\omega - \lambda \,R_\omega A^{(1)} R_\omega
+ \lambda^2 \,R_\omega A^{(1)} R_\omega A^{(1)} R_\omega
- \lambda^2 \,R_\omega A^{(2)} R_\omega + \O\big( \lambda^3 \big) \:. \]

Next, we want to compute the resulting spectral measure~$\tilde{E}$.
Our method is to use Stone's formula (see for example~\cite[Theorem~VII.13]{reed+simon}).
For technical simplicity, we assume that the spectra of both~$A$ and~$\tilde{A}(t)$ are
absolutely continuous. Then Stone's formula takes the simple form
\[ d\tilde{E}_\omega(t) = \tilde{E}_\omega(t)\: d\omega \qquad \text{with} \qquad
\tilde{E}_\omega(t) = \frac{1}{2 \pi i} \;s\!-\!\lim_{\varepsilon \searrow 0} \Big( \tilde{R}_{\omega+i \varepsilon}(t) - 
\tilde{R}_{\omega-i \varepsilon}(t) \Big) \:. \]
Now one can substitute the perturbation expansion for the resolvent~\eqref{Rpert}
to obtain corresponding formulas for the perturbation expansion of the spectral measure~$\tilde{E}$.

\subsection{Expansion of the Ordered Exponential}
The ordered exponential~\eqref{VTexp} can be expanded in a Dyson series
\begin{align*}
V^t_{t_0} &= \int_0^t \big( \dot{\tilde{E}}_I(\tau) -i \tilde{E}_I(\tau) \,\tilde{H}(\tau) \big)\: d\tau \\
&\quad\; + \int_0^t d\tau \int_0^{\tau} d\tau' \:\big( \dot{\tilde{E}}_I(\tau) -i \tilde{E}_I(\tau) \,\tilde{H}(\tau) \big)
\big( \dot{\tilde{E}}_I(\tau') -i \tilde{E}_I(\tau') \,\tilde{H}(\tau') \big) + \cdots \:.
\end{align*}
Now one can insert the above perturbation expansion for the spectral measures
to obtain explicit formulas for the modified dynamics including the adiabatic projections.

The drawback of this method is that the effect of the adiabatic projections
can be described only nonlinearly.
In other words, we cannot expect that expanding~$V^t_{t_0}$ to first or second order
will give us a good approximation of the exact modified dynamics.
Instead, one should take into account that adiabatic projections ensure that the states
preserve the orthonormality of all physical wave functions.
As a consequence, changing one physical wave function (for example of the energy on the
Planck scale) also gives rise to a collective change of all the other physical wave functions (in particular, of
wave functions of the energy on the Compton scale). This mutual interaction of all the physical
wave functions cannot be described perturbatively. But it can be described by a spectral flow,
as we now explain.

\subsection{A Simple Approximate Formula for Fermiogenesis} \label{secapprox}
Let us return to the formula for the modified dynamics~\eqref{Vtime},
\[ \tilde{\Pi}(t) = V^t_{t_0} \,\tilde{\Pi}(t_0)\, (V^t_{t_0})^{-1} \:. \]
Having shown that the Dirac dynamics does not give rise to fermiogenesis
(see Section~\ref{secnobaryo}), it seems a good approximation to take into account only
the adiabatic projections, i.e.
\[ V^t_{t_0} \approx \lim_{N \rightarrow \infty}
\tilde{E}_I(t) \,\cdots
\tilde{E}_I\big(t_0+2 \Delta t\big) \,\tilde{E}_I\big(t_0+\Delta t\big) \, \psi(0) \:. \]
Next, we assume that, initially, the operator~$\tilde{\Pi}$ projects approximately
onto the negative spectral subspace of the operator~$\tilde{A}(t)$. Thus, taking into account
the ultraviolet cutoff,
\[ \tilde{\Pi}(t_0) \approx \int_{-\frac{1}{\varepsilon}}^{-m} \tilde{E}_\omega(t_0)\: d\omega \:. \]
After these approximations, the adiabatic projection simply describes a spectral flow of the operator~$\tilde{A}(t)$, i.e.\
\[ \tilde{\Pi}(t) \approx \int_{-\frac{1}{\varepsilon}}^{-m} \tilde{E}_\omega(t)\: d\omega \:. \]

As a further simplification, let us detect fermiogenesis with~\eqref{fermiogenesis}, where we
construct~$\eta_\Lambda$ again with the help of the spectral calculus of~$\tilde{A}(t)$, i.e.\
\[ \eta = \int_{-\infty}^\infty \eta \Big( \frac{\omega}{\Lambda} \Big) \: \tilde{E}_\omega(t)\: d\omega \]
with a test function~$\eta \in C^\infty_0((-1,1), \R^+_0)$. We thus obtain
\begin{align*}
B(t) &= \frac{d}{dt} \tr \big( \eta_\Lambda\: \tilde{\Pi}(t) \big)
= \int_{-\frac{1}{\varepsilon}}^{-m} \tr \big( \eta_\Lambda(t)\: \dot{\tilde{E}}_\omega(t) \big) \:d\omega \:.
\end{align*}
Using the perturbation expansion of Section~\ref{secperturbE}, we get
\begin{align*}
B(t) &= -\frac{1}{2 \pi i} \;s\!-\!\lim_{\delta \searrow 0} \int_{-\infty}^\infty d\omega'\:
\eta \Big( \frac{\omega'}{\Lambda} \Big) \int_{-\frac{1}{\varepsilon}}^{-m} d\omega\:
\tr \big( \tilde{E}_{\omega'} \: \tilde{R}_{\omega+i s \delta} \dot{\tilde{A}} \tilde{R}_{\omega+i s \delta} \big) 
\bigg|_{s=-1}^{s=1} \\
&= -\frac{1}{2 \pi i} \;s\!-\!\lim_{\delta \searrow 0} 
\int_{-\infty}^\infty d\omega'\:
\eta \Big( \frac{\omega'}{\Lambda} \Big) \int_{-\frac{1}{\varepsilon}}^{-m} d\omega\:
\tr \big( \tilde{E}_{\omega'} \: \dot{\tilde{A}} \big) \: \frac{1}{(\omega'- \omega- i s \delta)^2} \bigg|_{s=-1}^{s=1} \\
&= \frac{1}{2 \pi i} \;s\!-\!\lim_{\delta \searrow 0} \int_{-\infty}^\infty d\omega'\:
\eta \Big( \frac{\omega'}{\Lambda} \Big) \int_{-\frac{1}{\varepsilon}}^{-m} d\omega\:
\tr \big( \tilde{E}_{\omega'} \: \dot{\tilde{A}} \big) \: \frac{\partial}{\partial \omega'} \Big( \frac{1}{\omega'- \omega- i s \delta} \Big) \bigg|_{s=-1}^{s=1} \\
&= \frac{1}{2 \pi i} \int_{-\infty}^\infty d\omega'\:
\eta \Big( \frac{\omega'}{\Lambda} \Big) \int_{-\frac{1}{\varepsilon}}^{-m} d\omega\:
\tr \big( \tilde{E}_{\omega'} \: \dot{\tilde{A}} \big) \: \frac{\partial}{\partial \omega'} \,2 \pi i\: \delta(\omega'- \omega) \\
&= - \eta(-m/\Lambda)\: \tr \big( \tilde{E}_{-m} \: \dot{\tilde{A}}  \big) \:.
\end{align*}
For large~$\Lambda$, we can set approximately~$\eta(-m/\Lambda) \approx \eta(0)$.
Normalizing~$\eta$ by~$\eta(0)=1$, we end up with the simple formula
\beq \label{approx}
B(t) = - \tr \big( \tilde{E}_{-m}(t) \: \dot{\tilde{A}}(t) \big) \:.
\eeq
As desired, this formula is independent of~$\Lambda$.
It gives a simple approximate formula for our mechanism of fermiogenesis,
which in future studies will serve as the starting point for a quantitative analysis in specific geometries.

\section{Conclusion and Outlook} \label{sec:outlook}
In the present paper we introduced the abstract formalism for a new mechanism of fermiogenesis. In this section we discuss some of the scenarios of interest as well as conceptual points, such as to clarify how one might think of the mechanism presented here. Furthermore, we discus how this new mechanism could lead to observable predictions.

The primary interests for future investigations will be in the cosmological setting.
As already mentioned in the introduction and demonstrated in detail in Section~\ref{secnobaryo},
the presented mechanism does not give rise to fermiogenesis in spatially homogeneous spacetimes such as the FLRW universe. This is reassuring in the sense that we do not need
to be concerned about late time fermiogenesis. The reason we emphasize this is that,
in view of the fact that the present mechanism does not involve any kind of high energy processes, there is no a-priori reason why the creation of matter should stop at late times. However, there is no observational evidence that the universe exhibits a continuous increase of the matter/anti-matter asymmetry. 
Hence the fact that in FLRW spacetimes the locally rigid dynamics and the Dirac dynamics coincide and that as a consequence of that there is no fermiogenesis in these spacetimes is an important consistency check for the here-presented mechanism.

The first cosmological scenario to investigate will be the case of {\em{small metric perturbations on de Sitter space}}. The perturbations break homogeneity and thereby introduce a
deviation of dynamics from Dirac dynamics. The advantage of this situation is that it allows for a perturbative calculation. However, there is a chance that, due to the infinitesimal nature, the effect might not be
large enough.

If this turns out to be the case, an alternative setting for investigations would be the case of a {\em{bubble with large regularization length}} forming inside an "inflationary" ambient spacetime
with a much shorter regularization length. In this scenario, the gravitational entropy density associated with the cosmological horizon, would drop across the bubble wall. In contrast to the discussion in~\cite{paganini2020proposal}, where the consequences of a potential mechanism of fermiogenesis based on causal fermion systems
was sketched, we here consider the entropy density given by\footnote{In contrast to the discussions in \cite{paganini2020proposal} where the author looked at the total gravitational entropy, the so-defined gravitational entropy density decreases in both scenarios, $\Lambda_{\text{high}}\rightarrow \Lambda_{\text{low}}$ with fixed regularization length (and thereby fixed gravitational constant), and $\varepsilon_{\text{low}}\rightarrow\varepsilon_{\text{high}}$ with fixed cosmological constant, i.e.\ fixed effective volume of the interacting region. Therefore, the difference with respect to the change in entropy in the two scenarios discussed in \cite{paganini2020proposal} is probably irrelevant, as the discontinuous decrease in gravitational entropy 
density would occur in both cases.}
\begin{equation}
    s_C=\frac{S_C}{V_C} = \frac{\sqrt{3\Lambda}}{4\hbar G} \:.
\end{equation}
It is important to note that a decrease in entropy density is a-priori nothing unusual. The FLRW universe is a good example for a system with an decreasing entropy density with the gravitational entropy density from the cosmological horizon as the asymptotic limit once all matter and radiation has dispersed. At the bubble wall, however, the entropy density would drop non-dispersively. Using the mechanism of fermiogenesis proposed in the present paper,
this drop in gravitational entropy density will be compensated by the entropy related to the matter.
Hence, locally for the fermiogenesis mechanism, we should expect the relation
\begin{equation}
    s_C(\varepsilon_{\text{low}})= s_C(\varepsilon_{\text{high}}) + s_{\text{matter}}
\end{equation}
to hold across the bubble wall. This would imply that in contrast to the scenarios considered in~\cite{susskind2021three} our cosmological scenario is compatible with the second law of thermodynamics~\footnote{If we look at the two scenarios in \cite{paganini2020proposal}: $\Lambda_{\text{high}} \rightarrow \Lambda_{\text{low}}$ and $\varepsilon_{\text{low}}\rightarrow\varepsilon_{\text{high}}$  they are equivalent from the perspective of entropy density the latter would lead to a violation of (local) energy conservation in the effective description (if we add the cosmological constant to the energy budget of the universe that is not the case in former scenario).  }.   The fermiogenesis mechanism then transfers entropy from gravitational degrees of freedom to matter degrees of freedom. This suggests that we associate the states that occupy the negative energy states with the gravitational sector of the effective theory. This is a crucial insight uncovering the fashion in which the theory of Causal Fermion Systems unites matter and gravitation. 
In this picture, it makes sense that a ``change in the sea 
level'' converts gravitational entropy into matter entropy.
The unification of gravity and matter in the fundamental description therefore provides a mechanism by which they can be converted into each other in the effective description.
Finally, the states which cross the sea level are free to disperse across the (cosmological) horizon which constitutes the boundary of the interacting region until the system returns to a steady state with only gravitational entropy.

For either of these scenarios, it should be interesting to investigate the observational consequences of all fermionic matter being created in the ground state. The relevant questions which need to be addressed in order to determine cosmological consequences have been outlined in~\cite{paganini2020proposal}. Of particular interest are the consequences of a fermionic dark sector in which an excess of matter is created in the Fermi ground state and not thermalized in the evolution of the universe.

It will be subject of further research to study how this this new mechanism compares to Hawking radiation, the only known mechanism in semi-classical physics where gravitational entropy can be converted into matter entropy. There is an interesting qualitative distinction to begin with. Namely, the fact that the matter created in Hawking radiation is in a thermal state and occurs for all Quantum Field Theories, while the fermiogenesis mechanism described here only affects fermions and leads to particle creation in a Fermi ground state across all three generations.

On a more technical level, an interesting question to ask is under which conditions and in which effective spacetime configurations the expressions~\eqref{approx} (or respectively~\eqref{bt}) have a fixed sign. This translates to the question under which conditions the change in "sea level" is monotone and hence in which scenarios one gets a proper matter/anti-matter asymmetry based on this mechanism.

Finally a detailed study of the relationship between non-Riemannian-measure theories and Causal Fermion Systems seems promising. The latter might provide additional motivation why the former are indeed of physical importance. Furthermore it might provide constraints on which modifications of the measure are admissible. In return the former might serve as a model for modifications to gravity in the effective description of the latter, thus providing an avenue towards numerical predictions. 

%
%
%
%
%

\appendix
\section{The Dynamics of the Regularizing Vector Field} \label{appA}
In this and the next appendix, we shall make proposals for the dynamics of the regularizing vector field~$u$
introduced in Section~\ref{secudef}. Our method is to begin from the Dirac dynamics of the
regularization as described by the transport equation~\eqref{ftransport}, and to deduce~$u$ by taking
``spatial averages'' of the resulting regularization. Before entering the details, we point out that
this method might be too naive, because we do not take into account the possible back reaction of the
matter created in the fermiogenesis process on the regularization.
With this in mind, the analysis in these appendices is intended only as a first suggestions on how the dynamics
of~$u$ could look like. A more detailed model for the dynamics
which incorporates all relevant physical effects remains to be developed.

As shown in Figures~\ref{figcone1} and~\ref{figcone2}, the main difference between the
Dirac dynamics and the locally rigid dynamics is that in the latter case the boundary of the
regularized Dirac sea has locally the form of a round sphere, possibly in a boosted reference frame.
In the adiabatic projection method in Section~\ref{secadiabatic}, this is implemented by
continuously projecting onto the states of the locally rigid regularization.
Conceptually, this means that the system adjusts instantaneously 
and goes back to the locally rigid regularization.
This situation is considered in Appendix~\ref{appA1}.
It might be more sensible physically to assume that the system does not adjust instantaneously,
but that this process certain finite time, referred to as the {\em{relaxation time}}.
This case is considered in Appendix~\ref{appA2}.

At present, the relaxation time~$\Delta t$ is an unknown function
of the parameters~$\varepsilon$ and~$m$ of the dimension of length, i.e.\
\[ \Delta t = \varepsilon\: (\varepsilon m)^q \]
with an unknown real parameter~$q$. The relaxation time will be analyzed 
in detail in the forthcoming paper~\cite{secvar}.
In simple terms, the causal action principle tries to adjust the Dirac sea configuration such that the causal action becomes minimal. The strength of this tendency to even out non-optimal configurations can be quantified 
in terms of second variations of the causal action as studied in~\cite{jokeldr}. The stronger this tendency, the smaller the relaxation time.

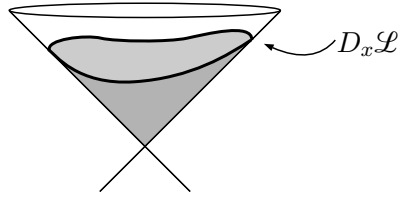
\begin{figure}
\begin{center}
\psscalebox{1.0 1.0} 
{
\begin{pspicture}(0,-2.4875)(4.997568,-0.0825)
\definecolor{colour0}{rgb}{0.7019608,0.7019608,0.7019608}
\definecolor{colour1}{rgb}{0.8,0.8,0.8}
\pspolygon[linecolor=colour0, linewidth=0.02, fillstyle=solid,fillcolor=colour0](1.345,-0.8825)(2.445,-1.9925)(3.805,-0.6125)(3.545,-0.7925)(3.225,-0.9425)(2.875,-1.0625)(2.615,-1.1125)(2.225,-1.1625)(1.925,-1.1425)(1.705,-1.0825)(1.505,-0.9925)
\pspolygon[linecolor=colour1, linewidth=0.02, fillstyle=solid,fillcolor=colour1](1.185,-0.7325)(1.405,-0.9125)(1.635,-1.0225)(1.935,-1.1025)(2.305,-1.1225)(2.645,-1.0825)(2.975,-1.0025)(3.365,-0.8625)(3.685,-0.6925)(3.855,-0.5525)(3.775,-0.4625)(3.515,-0.4725)(3.225,-0.5125)(2.945,-0.5825)(2.445,-0.6225)(2.045,-0.6025)(1.775,-0.5525)(1.555,-0.5425)(1.255,-0.6125)
\psline[linecolor=black, linewidth=0.02, fillstyle=solid,fillcolor=black](0.65,-0.1975)(3.05,-2.5975)
\psellipse[linecolor=black, linewidth=0.02, dimen=outer](2.4475,-0.1875)(1.8125,0.105)
\psframe[linecolor=white, linewidth=0.02, dimen=outer](0.04,-4.4475)(0.0,-4.4875)
\psline[linecolor=black, linewidth=0.02, fillstyle=solid,fillcolor=black](1.85,-2.5975)(4.25,-0.1975)
\psbezier[linecolor=black, linewidth=0.04](1.165,-0.7075)(1.2543558,-0.7723676)(1.54607,-1.1887468)(2.31,-1.1375)(3.07393,-1.0862533)(3.5639343,-0.7761285)(3.86,-0.5825)
\psbezier[linecolor=black, linewidth=0.04](1.18,-0.7225)(1.1250777,-0.5700321)(1.6626096,-0.50969714)(1.865,-0.5575)(2.0673904,-0.6053029)(2.7650778,-0.61003214)(3.01,-0.5575)(3.2549224,-0.50496787)(3.8823905,-0.34530285)(3.87,-0.5825)
\psbezier[linecolor=black, linewidth=0.02, arrowsize=0.05291667cm 2.0,arrowlength=1.4,arrowinset=0.0]{<-}(4.03,-0.6325)(4.51,-0.8425)(4.79,-0.8025)(4.98,-0.5825)
\rput[bl](5,-0.8){$D_x\scrL$}
\end{pspicture}
}
\end{center}
\caption{The two-surface~$D_x\scrL \subset T_x\scrM$.}
\label{figDxL}
\end{figure}

Our procedure is inspired by concepts introduced in~\cite{dgc, reghadamard} and is based on the transport
equation~\eqref{ftransport} for the regularization under the Dirac dynamics.
More precisely, as first noted in~\cite[Appendix~A]{dgc}, the transport equation~\eqref{ftransport}
means geometrically that, along a null geodesic through~$y$, the function~$f(.,y)$ simply is an affine parameter
along this geodesic with~$f(y,y)=0$. In other words, for every~$y \in \scrM$ there is a parametrized 
future-directed null curve~$\gamma(\tau)$ which satisfies the geodesic equation
\[ \ddot{\gamma}^i(\tau) - \Gamma^i_{jk}\big( \gamma(\tau) \big)\: \dot{\gamma}^j(\tau)\, \dot{\gamma}^k(\tau) \]
such that for all parameter values~$\tau$ and~$\tau'$,
\[ f \big( \gamma(\tau) , \gamma(\tau') \big) = \tau' - \tau \:. \]
With this in mind, the dynamics of the regularization can be described conveniently by a family
of null geodesics. This family must be chosen in such a way that at each spacetime point~$x \in \scrM$
and for every null direction~$v \in T_x\scrM$, it contains a unique null geodesic
with~$\gamma(0)=y$ and~$\dot{\gamma}(0) \in v \R$. This leads us to the following definition.
\begin{Def} \label{defL}
We introduce a set of parametrized null geodesics
\[ 
\scrL = \big\{ (\gamma, I) \:|\: \text{$\gamma \::\: I \rightarrow \scrM$ is a
future-directed parametrized maximal null geodesic} \big\} \]
with the following properties:
\begin{itemize}[leftmargin=2em]
\item[{\rm{(a)}}] For every~$(\gamma, I) \in \scrL$, re-parametrizing by an additive constant
\[ 
\tilde{\gamma}(\tau) := \gamma(\tau+c) \qquad \text{with} \qquad
\tau \in \tilde{I} := I - c \:. \]
gives again a geodesic in~$\scrL$.
\item[{\rm{(b)}}] For every maximal null geodesic~$\gamma(\tau)$ in~$\scrM$, there is exactly one~$\lambda>0$
such that the multiplicative re-parametrization
\beq \label{multchange}
\tilde{\gamma}(\tau) := \gamma(\lambda \tau) \qquad \text{with} \qquad
\text{$\tau \in \tilde{I} := I/\lambda$ and~$\lambda>0$} \:.
\eeq
gives a geodesic~$(\tilde{\gamma}, \tilde{I}) \in \scrL$.
\end{itemize}
\end{Def} \noindent
Next, for any spacetime point~$x \in \scrM$ we introduce the set
\beq \label{DxLdef}
D_x\scrL = \big\{ \dot{\gamma}(\tau) \:\big|\: \text{$(\gamma, I) \in \scrL$, $\tau \in I$
and $\gamma(\tau) = x$} \big\} \subset T_x\scrM \:.
\eeq
Assuming smooth dependence of the geodesics on the null direction, the
set~$D_x\scrL$ is a smooth two-dimensional surface with the topology of a sphere
(see Figure~\ref{figDxL}).
This sphere can be regarded as a representation of the {\em{celestial sphere}} at the point~$x$, where
for each spatial direction we choose a specific null vector~$\dot{\gamma}(\tau)$.
The Lorentzian metric on~$T_x\scrM$ induces a Riemannian
metric on~$D_x\scrL$. We denote the corresponding volume measure on~$D_x\scrL$ by~$d\mu_x$.

According to~\eqref{DxLdef}, the regularization at the spacetime point~$x$ is described by
a two-dimensional subset of the light cone at~$x$. 
The locally rigid regularization can be described in the same way by choosing the surface
\beq \label{Edef}
E_x\big(u(x) \big) := \big\{ \xi \in T_x\scrM \:\big|\: \text{$\xi$ lightlike and $g \big(u(x), \xi \big) =1$} \big\} \:.
\eeq
The surface~$E$ is an ellipsoid (it is even a sphere in the reference frame where~$u=(|u|,\vec{0})$).
In order to relate the Dirac dynamics to the rigid dynamics, we need to find a way to
relate the surface in~\eqref{DxLdef} to a corresponding ellipsoid~\eqref{Edef}.
This can be done in a Lorentz invariant manner
by integrating over~$D_x\scrL$ and setting
\beq \label{uxdef}
u(x) := \frac{\overline{\xi}_x}{\overline{\xi}_x^2} \qquad \text{with} \qquad
\overline{\xi}_x := \frac{1}{\mu_x \big( D_x\scrL \big)} \int_{D_x\scrL} \xi\: d\mu_x(\xi) \:.
\eeq
A short computation shows that if~$D_x \scrL$ is an ellipsoid of the form~\eqref{Edef},
then the formula~\eqref{uxdef} gives us back the vector~$u(x)$.
Therefore, the formula~\eqref{uxdef} can be understood as a change of the shape of~$D_x \scrL$
where all deviations from the ellipsoidal form are removed.

This construction leads us to the following method for deriving a dynamics of the vector field~$u$.
We begin on a Cauchy surface~$\scrN$ with a locally rigid regularization.
By choosing~$\scrL$ as generated by all parametrized null geodesics which on~$\scrN$ satisfy the identity
\[ g \Big(u \big( \gamma(\tau) \big), \dot{\gamma}(\tau) \Big) =1 \qquad \text{whenever~$\gamma(\tau) \in \scrN$}\:, \]
we can arrange that the surfaces~$D_x \scrL$ are on~$\scrN$ of the desired form~\eqref{Edef}.
Solving the geodesic equation, we obtain the corresponding Dirac dynamics of the regularization.
At some later time, we apply~\eqref{uxdef} in order to get back to the locally rigid regularization.
The only unknown in this procedure is what we mean by ``later time.'' Here we consider two
possibilities:
\begin{itemize}
\item[A.1.] {\em{The dynamics with finite relaxation time}}: We apply~\eqref{uxdef} after a finite time~$\Delta t$
has elapsed, where by time we mean the proper time as measured along the regularizing vector field~$u$. \\[-1em]
\item[A.2.] {\em{The dynamics with instantaneous relaxation time}}: We apply~\eqref{uxdef} after an infinitesimal time step.
\end{itemize}
In the next subsections, we treat these two cases after each other.

\subsection{The Dynamics with Finite Relaxation Time} \label{appA2}
We begin with the derivation of the equation for the dynamics with finite relaxation time~$\Delta t$.
To this end, given a point~$x \in \scrM$ we consider
a parametrized, future-directed null geodesic~$\gamma(\tau)$ with~$\gamma(0)=x$.
Suppose that the regularizing vector field~$u$ is defined in the past of~$x$. Then it is defined
along~$\gamma$ for all negative~$\tau$, making it possible to formulate the equation
\[ 
g\Big( \dot{\gamma}(\tau), u \big(\gamma(\tau) \big) \Big) = -\frac{\Delta t}{\tau} \qquad \text{for~$\tau<0$}\:. \]
Since the left side is strictly positive and smooth in~$\tau$, whereas the right side has a pole as~$\tau \nearrow 0$,
it is clear that for sufficiently small~$\Delta t$, this equation has a unique solution denoted by~$\tau_0<0$.
We define a corresponding null vector~$\xi_x \in T_x \scrM$ by
\beq \label{xixdef}
\xi_x(\gamma) := \frac{\dot{\gamma}(0)}{g\Big( \dot{\gamma}(\tau_0), u \big(\gamma(\tau_0) \big) \Big)} \:.
\eeq
Before going on, we point out that this definition of~$\xi_x$ does {\em{not}} depend on the
parametrization of the null geodesic. Indeed, performing a multiplicative re-pa\-ra\-metri\-za\-tion~\eqref{multchange}
and setting~$\tilde{\tau} = \tau/\lambda$, we have
\beq \label{lamscale}
\dot{\tilde{\gamma}} \big( \tilde{\tau} \big) = \frac{d}{d\tilde{\tau}} \gamma \big(\lambda \tilde{\tau} \big)
= \lambda \dot{\gamma}(\lambda \tilde{\tau})
\eeq
and thus
\begin{align*}
g\Big( \dot{\tilde{\gamma}}(\tilde{\tau}_0), u \big(\tilde{\gamma}(\tilde{\tau}_0) \big) \Big)
= \lambda\: g\Big( \dot{\gamma}(\tau_0), u \big(\gamma(\tau_0) \big) \Big)
= -\lambda\: \frac{\Delta t}{\tau_0} = \frac{\Delta t}{\tilde{\tau}_0} \:.
\end{align*}
Hence~$\gamma(\tau_0)=\tilde{\gamma}(\tilde{\tau}_0)$, showing that the point on the null geodesic
is defined independent of the parametrization. Using this fact, one concludes that
also the definition~\eqref{xixdef} is invariant under re-parametrizations, because~\eqref{lamscale}
gives rise to a factor~$\lambda$ in the numerator and in the denominator, which cancel each other.

Carrying out the above construction for all null geodesics through~$x$, we obtain a corresponding
surface
\[ D_x \scrL := \big\{ \xi_x(\gamma) \:\big|\: \text{$\gamma$ null geodesic through~$x$} \big\} \:. \]
Now we can define the regularizing vector~$u(x)$ by~\eqref{uxdef}.
In this way, we have a procedure for computing~$u(x)$ given the regularizing vector field at earlier times.
We point out that this evolution is {\em{not}} described by a differential equation, but instead 
it is computed from~$u$ at earlier times. The corresponding ``relaxation time''~$\Delta \tau$
is measured in the reference frame which is also determined by the regularizing vector field and may depend on the spatial direction. It is a desirable feature of the evolution
with finite relaxation time that no time function needs to be distinguished.

In the limit~$\Delta t \searrow 0$ when the relaxation time tends to zero, the above dynamics
reduces to the condition that the vector field~$u$ is auto-parallel, i.e.\
\[ \nabla_u u = 0 \:. \]
This evolution equation is known to suffer from the drawback that its solutions typically form
singularities. This is why taking the limit~$\Delta t \searrow 0$ does not seem the appropriate method.
For the infinitesimal dynamics, it seems a better idea to proceed time step by time step in a
distinguished foliation and take the limit of infinitesimal time steps. This method will be carried
out in the next section.

\subsection{The Dynamics with Instantaneous Relaxation} \label{appA1}
We first state the resulting dynamical equations and derive them afterward.
\begin{Thm} \label{thminfdyn}
Let~$(\scrN_t)_{t \in \R}$ be a distinguished foliation of~$\scrM$ by Cauchy surfaces.
Denoting tangential directions by Greek indices, the dynamics with instantaneous relaxation
is described by the differential equation
\[ 
\nabla_t u^i = A^{i \alpha}_j\big( \hat{u}, \nu \big)\: \nabla_\alpha u^j  \:, \]
where~$\hat{u}$ is a unit vector in the direction of~$u$,
\[ \hat{u} := \frac{u}{|u|} \:, \]
and the tensor~$A^{i \alpha}_j$ has the components
\begin{align*}
&A^{i \alpha}_j\big( \hat{u}, \nu \big) = \pi(\nu^\perp)^i_j \:\frac{2 \la \hat{u}, \nu \ra^{3}+3 J-5\la \hat{u}, \nu \ra}{2 \, \big(\la \hat{u}, \nu \ra^{2}-1\big)^{2}} \:u_\alpha \\ 
&+ \frac{\pi(\nu^\perp)^{i \alpha}}{2 \, \big(\la \hat{u}, \nu \ra^{2}-1\big)^{2}} \:\Big( \big( 2 \la \hat{u}, \nu \ra^{3}+3 J-5 \la \hat{u}, \nu \ra \big)\: \hat{u}_j
+ \big( -3 J \la \hat{u}, \nu \ra+\la \hat{u}, \nu \ra^{2}+2 \big)\: \nu_j \Big) \\
&+ \frac{\pi(\nu^\perp)^\alpha_j }{2 \, \big(\la \hat{u}, \nu \ra^{2}-1\big)^{2}} \:\Big( 
\big( 2 J \,\la \hat{u}, \nu \ra^{2}+J-3 \la \hat{u}, \nu \ra \big)\: \hat{u}^i
+ \big( -3 J \la \hat{u}, \nu \ra+\la \hat{u}, \nu \ra^{2}+2 \big)\: \nu^i \Big) \\
&- \frac{u_\alpha}{2 \,\big(\la \hat{u}, \nu \ra^6 -1\big)} \:
\bigg( \Big(-4 \la \hat{u}, \nu \ra^{5}+6 J \,\la \hat{u}, \nu \ra^{2}+12 \la \hat{u}, \nu \ra^{3}+9 J-23 \la \hat{u}, \nu \ra
\Big)\: \hat{u}^i \hat{u}_j \\
&\qquad\qquad\qquad\quad\: + \Big(-2 \la \hat{u}, \nu \ra^{4}-15 J \la \hat{u}, \nu \ra+9 \la \hat{u}, \nu \ra^{2}+8 \Big) \:\nu^i \hat{u}_j \\
&\qquad\qquad\qquad\quad\: + \Big( -6 J \,\la \hat{u}, \nu \ra^{3}-9 J \la \hat{u}, \nu \ra+11 \la \hat{u}, \nu \ra^{2}+4 \Big)\: \hat{u}^i \nu_j \\
&\qquad\qquad\qquad\quad\: + \Big( 2 \la \hat{u}, \nu \ra^{5}+15 J \,\la \hat{u}, \nu \ra^{2}-9 \la \hat{u}, \nu \ra^{3}-8 \la \hat{u}, \nu \ra \Big)\: \nu^i \nu_j \bigg) \:.
\end{align*}
Here~$J$ is the function
\beq \label{J0}
J = \frac{1}{\sqrt{\la \hat{u}, \nu \ra^2-1}}\: \log \Big( \la \hat{u}, \nu \ra + \sqrt{\la \hat{u}, \nu \ra^2-1} \Big) \:.
\eeq
\end{Thm}

Before entering the proof, we briefly explain the structure of this evolution equation.
Clearly, this equation is a system of nonlinear partial differential equation of first order.
It is not of symmetric hyperbolic form, and at present it is not known whether it can be symmetrized.
At least, in the real analytic case, one can construct solutions using the Cauchy–Kovalevskaya theorem.

%

We now enter the proof of Theorem~\ref{thminfdyn}, which will be completed at the end of this section.
It suffices to derive the dynamical equation in Min\-kowski space for a foliation by equal-time surfaces,
because the resulting formula is readily generalized to curved spacetime simply by replacing partial derivatives
by covariant derivatives.

For the computations, we work in a given reference frame~$(t, \vec{x}) \in \scrM$.
For a unit vector~$\vec{n} \in S^2 \subset \R^3$ we
let~$\zeta$ be the corresponding future-directed null vector,
\[ \zeta(\vec{n}) := (1, \vec{n}) \:. \]
Then a positive function on the sphere
\[ f_x \::\: S^2 \rightarrow \R^+ \]
defines a two-dimensional submanifold of the upper light cone centered~$x$ given by
\beq \label{DxLdefalternate}
D_x \scrL := \{ f_x(\vec{n})\: \zeta(\vec{n}) \:|\: \vec{n} \in S^2 \} \:.
\eeq
The integration measure on~$D_x \scrL$ corresponding to the Riemannian metric
(induced by the Minkowski metric) takes the form
\[ 
d\mu_x(\xi) = f_x(\vec{n})^2\: d\mu_{S^2}(\vec{n}) \]
(where~$\mu_{S^2}$ is the standard integration measure on~$S^2$ with~$\mu_{S^2}(S^2)=4 \pi$).
In this way, for every~$x$ we obtain a surface~$D_x \scrL$ contained in the upper light cone.

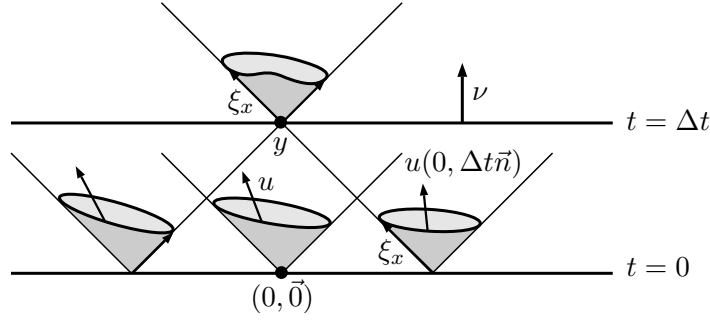
\begin{figure}
\psscalebox{1.0 1.0} 
{
\begin{pspicture}(0,27.778536)(8.02707,31.455605)
\definecolor{colour0}{rgb}{0.8,0.8,0.8}
\definecolor{colour1}{rgb}{0.9019608,0.9019608,0.9019608}
\pspolygon[linecolor=colour0, linewidth=0.02, fillstyle=solid,fillcolor=colour0](2.9520702,30.521034)(3.6170702,29.861034)(4.17707,30.401035)(4.05207,30.361034)(3.8620703,30.356035)(3.6820703,30.416035)(3.4720702,30.511036)(3.2670703,30.496035)(3.0620704,30.481035)
\pspolygon[linecolor=colour1, linewidth=0.02, fillstyle=solid,fillcolor=colour1](2.8670702,30.606035)(3.0220704,30.496035)(3.2620704,30.521034)(3.4470704,30.526035)(3.8520703,30.381035)(4.09207,30.371035)(4.21707,30.486034)(4.11707,30.596035)(3.7670703,30.716036)(3.5170703,30.756035)(3.2620704,30.781034)(3.0720704,30.781034)(2.8820703,30.741035)(2.8270702,30.676035)
\pspolygon[linecolor=colour1, linewidth=0.02, fillstyle=solid,fillcolor=colour1](0.6920703,28.793535)(0.8920703,28.668535)(1.1520703,28.568535)(1.5970703,28.438536)(2.0020704,28.378534)(2.1670704,28.403536)(2.1220703,28.483536)(1.9020703,28.608536)(1.5120703,28.753534)(1.2070704,28.828535)(0.8770703,28.873535)(0.7370703,28.878534)(0.66207033,28.848536)
\pspolygon[linecolor=colour1, linewidth=0.02, fillstyle=solid,fillcolor=colour1](4.9220705,28.576035)(5.03207,28.516035)(5.4020705,28.426035)(5.84207,28.396034)(6.1020703,28.433535)(6.24207,28.498535)(6.1520705,28.583534)(5.9020705,28.648535)(5.57707,28.678535)(5.2570705,28.683535)(4.9820704,28.643536)
\psline[linecolor=black, linewidth=0.03, arrowsize=0.05291667cm 2.0,arrowlength=1.4,arrowinset=0.0]{->}(5.61207,27.848536)(5.49207,29.033535)
\pspolygon[linecolor=colour0, linewidth=0.02, fillstyle=solid,fillcolor=colour0](4.9320703,28.546036)(5.61707,27.856035)(6.22207,28.461035)(5.96707,28.401035)(5.5620704,28.386036)(5.2470703,28.436035)(5.0470705,28.486034)
\psline[linecolor=black, linewidth=0.03, arrowsize=0.05291667cm 2.0,arrowlength=1.4,arrowinset=0.0]{->}(1.6270703,27.853535)(0.8770703,29.263535)
\pspolygon[linecolor=colour0, linewidth=0.02, fillstyle=solid,fillcolor=colour0](0.6870703,28.786036)(1.6220703,27.856035)(2.1620703,28.401035)(1.8570703,28.371035)(1.3370703,28.491035)(0.9020703,28.656034)
\pspolygon[linecolor=colour1, linewidth=0.01, fillstyle=solid,fillcolor=colour1](2.7120702,28.766035)(2.9370704,28.641035)(3.2820704,28.541035)(3.6070702,28.481035)(3.9220703,28.456036)(4.1820703,28.471035)(4.26207,28.501036)(4.17707,28.596035)(3.8370702,28.706036)(3.5620704,28.771034)(3.2520704,28.816034)(2.9070704,28.831036)(2.7820704,28.821035)
\psline[linecolor=black, linewidth=0.03, arrowsize=0.05291667cm 2.0,arrowlength=1.4,arrowinset=0.0]{->}(3.6020703,27.868536)(3.0920703,29.198536)
\psline[linecolor=colour0, linewidth=0.01, fillstyle=solid,fillcolor=colour0](2.7570703,28.718534)(3.6220703,27.848536)(4.2470703,28.488535)(4.0870705,28.448536)(3.7620704,28.453535)(3.4320703,28.503534)(3.0620704,28.588535)(2.8920703,28.658535)
\psline[linecolor=black, linewidth=0.04](0.017070312,27.838535)(8.017071,27.838535)
\psline[linecolor=black, linewidth=0.04](0.017070312,29.838535)(8.017071,29.838535)
\psline[linecolor=black, linewidth=0.02](2.0170703,29.438536)(3.6170702,27.838535)(5.21707,29.438536)
\psline[linecolor=black, linewidth=0.02](2.0170703,31.438536)(5.61707,27.838535)
\psline[linecolor=black, linewidth=0.02](5.21707,31.438536)(1.6170703,27.838535)
\pscircle[linecolor=black, linewidth=0.03, fillstyle=solid,fillcolor=black, dimen=outer](3.6020703,29.848536){0.07}
\pscircle[linecolor=black, linewidth=0.03, fillstyle=solid,fillcolor=black, dimen=outer](3.6120703,27.848536){0.07}
\psrotate(3.4870703, 28.643536){-9.880753}{\psellipse[linecolor=black, linewidth=0.04, dimen=outer](3.4870703,28.643536)(0.805,0.165)}
\psbezier[linecolor=black, linewidth=0.04](2.9045703,30.557583)(3.2463505,30.367855)(3.308284,30.592375)(3.6216784,30.46103515625)(3.9350724,30.329695)(4.0908427,30.34295)(4.1845703,30.415632)(4.278298,30.488314)(4.2083282,30.572607)(3.7671154,30.691034)(3.3259025,30.809464)(2.5809941,30.809261)(2.9045703,30.558561)
\psrotate(5.5845704, 28.543535){-3.6112623}{\psellipse[linecolor=black, linewidth=0.04, dimen=outer](5.5845704,28.543535)(0.6925,0.165)}
\psline[linecolor=black, linewidth=0.02](5.61707,27.838535)(7.21707,29.438536)
\psline[linecolor=black, linewidth=0.02](1.6170703,27.838535)(0.017070312,29.438536)
\psrotate(1.4170703, 28.633535){-15.9895315}{\psellipse[linecolor=black, linewidth=0.04, dimen=outer](1.4170703,28.633535)(0.805,0.165)}
\psline[linecolor=black, linewidth=0.03, arrowsize=0.05291667cm 2.0,arrowlength=1.4,arrowinset=0.0]{->}(5.6320705,27.843534)(4.9370704,28.528536)
\psline[linecolor=black, linewidth=0.03, arrowsize=0.05291667cm 2.0,arrowlength=1.4,arrowinset=0.0]{->}(3.5970702,29.863535)(2.9070704,30.548534)
\psline[linecolor=black, linewidth=0.03, arrowsize=0.05291667cm 2.0,arrowlength=1.4,arrowinset=0.0]{->}(1.6320703,27.838535)(2.1770704,28.383535)
\psline[linecolor=black, linewidth=0.03, arrowsize=0.05291667cm 2.0,arrowlength=1.4,arrowinset=0.0]{->}(3.6520703,29.873535)(4.19707,30.418535)
\psline[linecolor=black, linewidth=0.04, arrowsize=0.05291667cm 2.0,arrowlength=1.4,arrowinset=0.0]{->}(6.0170703,29.838535)(6.0170703,30.638535)
\rput[bl](6.15,30.15){$\nu$}
\rput[bl](3.2,27.3){$(0, \vec{0})$}
\rput[bl](8.2,29.75){$t=\Delta t$}
\rput[bl](8.2,27.8){$t=0$}
\rput[bl](5.25,29.1){$u(0, \Delta t \vec{n})$}
\rput[bl](3.3,28.95){$u$}
\rput[bl](4.9,27.95){$\xi_x$}
\rput[bl](2.9,29.95){$\xi_x$}
\rput[bl](3.5,29.4){$y$}
\end{pspicture}
}
\caption{The dynamics of the regularizing vector field.}
\label{figregdyn}
\end{figure}

Suppose that the vector field~$u$ is given
on the Cauchy surface at time~$t=0$.
For every point~$x=(0, \vec{x})$ on this Cauchy surface and for every unit vector~$\vec{n} \in S^2$,
we want to introduce a null vector~$\xi_x(\vec{n})$ of the form
\[ \xi_x(\vec{n}) = f_x(\vec{n})\: \zeta(\vec{n}) \]
with the property that
\beq \label{xicond}
\la \xi_x(\vec{n}),\, u(x) \big\ra = 1 \:.
\eeq
To this end, we choose
\[ f_x(\vec{n}) = \frac{1}{\big\la \zeta(\vec{n}),\, u(x) \big\ra} \:. \]
The corresponding two-surface~$D_x \scrL$ defined by~\eqref{DxLdefalternate}
is a hyperboloid (see Figure~\ref{figregdyn}).

Our goal is to compute~$u$ at the point~$y:=(\Delta t, \vec{0})$.
In order to implement the Dirac dynamics, we realize the locally rigid regularization at time zero
by choosing~$\scrL$ as all the parametrized null lines
\[ \scrL := \big\{ \gamma(\tau) =
(0, \vec{x}) + (\tau- \tau_0)\, \zeta(\vec{n}) \text{ with } \vec{x} \in \R^3\:, \vec{n} \in S^2, \tau_0 \in \R \big\} \:. \]
The tangent vectors of these null lines at~$y$ give rise to the function~$f_y : S^2 \rightarrow \R^+$
given by
\[ f_y(\vec{n}) := f_{\big( 0, -\Delta t\: \vec{n} \big)}(\vec{n}) \:. \]
In this way, we have implemented the transport equation~\eqref{ftransport}.
As explained in~\eqref{uxdef}, we define~$u(y)$ according to~\eqref{xicond}
by integrating over~$D_y\scrL$.

We next compute the measure~$\mu_y$ linearly in~$\Delta t$.
\begin{align}
f_y(\vec{n}) &= f_{\big( 0, -\Delta t\: \vec{n} \big)}(\vec{n})
= \frac{1}{\big\la \zeta(\vec{n}),\, u\big( 0, -\Delta t\: \vec{n} \big) \big\ra} \notag \\
&= \frac{1}{\la \zeta,\, u(0) \ra} 
+ \frac{\Delta t}{\la \zeta,\, u(0) \ra^2} \:\sum_{\alpha=1}^3 n^\alpha\: \big\la \zeta,\, \partial_\alpha u(0) \big\ra 
+ \O\big( (\Delta t)^2 \big) \notag \\
&= \frac{1}{\la \zeta,\, u(0) \ra} 
+ \frac{\Delta t}{\la \zeta,\, u(0) \ra^2} \; \zeta^i \:\big( g^j_i - \nu_i \,\nu^j \big) \big\la \zeta,\, \partial_j u(0) \big\ra 
+ \O\big( (\Delta t)^2 \big) \:, \label{fy1}
\end{align}
where~$\nu= (1, \vec{0})$ is the future-directed normal. For a more compact notation, for a
timelike vector~$v$ we denote the projections to the span of~$v$ and to its orthogonal complement by
\[ \pi(v)^i_j := \frac{v^i v_j}{v^2} \qquad \text{and} \qquad
\pi \big( v^\perp \big)^i_j := \delta^i_j -\frac{v^i v_j}{v^2} \:. \]
Then~\eqref{fy1} can be written as
\[ f_y(\vec{n}) = \frac{1}{\la \zeta,\, u(0) \ra} 
+ \frac{\Delta t}{\la \zeta,\, u(0) \ra^2} \; \zeta^i \:\pi(\nu^\perp)^j_i\: \big\la \zeta,\, \partial_j u(0) \big\ra 
+ \O\big( (\Delta t)^2 \big) \:. \]
Next, using that
\[ \la \xi_0, \nu \ra =f_0(\vec{n}) = \frac{1}{\big\la \zeta,\, u(0) \big\ra} \qquad \text{and} \qquad
\xi_0(\vec{n}) = f_0(\vec{n}) \, \zeta(\vec{n})\:, \]
we obtain
\begin{align*}
f_y(\vec{n}) &= f_0(\vec{n}) \,\bigg( 1
+ \frac{\Delta t}{\la \zeta,\, u(0) \ra} \; \zeta^i \:\pi(\nu^\perp)^j_i\: \big\la \zeta,\, \partial_j u(0) \big\ra \bigg)
+ \O\big( (\Delta t)^2 \big) \\
&= f_0(\vec{n}) \,\bigg( 1
+ \Delta t\: \la \zeta,\, u(0) \ra \; \xi_0^i \:\pi(\nu^\perp)^j_i\: \big\la \xi_0,\, \partial_j u(0) \big\ra \bigg)
+ \O\big( (\Delta t)^2 \big) \\
&= f_0(\vec{n}) \,\bigg(1 + \frac{\Delta t}{\la \xi_0, \nu \ra}\; \xi_0^i \:\pi(\nu^\perp)^j_i\:  \big\la \xi_0,\, \partial_j u(0) \big\ra \bigg)
+ \O\big( (\Delta t)^2 \big) \:.
\end{align*}
Hence
\begin{align*}
d\mu_y(\xi) &= f_y^2(\vec{n}) \:d\mu_{S^2}(\vec{n}) =
\bigg( 1 + \frac{2\:\Delta t}{\la \xi, \nu \ra} \; \xi^i \:\pi(\nu^\perp)^j_i\:  \big\la \xi,\, \partial_j u(0) \big\ra 
\bigg) \:d\mu_x + \O\big( (\Delta t)^2 \\
\xi\: d\mu_y(\xi) &= f_y^3(\vec{n})\:\zeta(\vec{n}) \:d\mu_{S^2}(\vec{n}) \\
&= \bigg( 1 + \frac{3\:\Delta t}{\la \xi, \nu \ra} \; \xi^i \:\pi(\nu^\perp)^j_i\:  \big\la \xi,\, \partial_j u(0) \big\ra 
\bigg)\: \xi\: d\mu_x(\xi) + \O\big( (\Delta t)^2 \big) \:.
\end{align*}
We thus obtain
\begin{align}
\frac{d \xi_x}{dt} &= \fint_{D_x \scrL} 
\frac{3}{\la \xi, \nu \ra} \; \xi^i \:\pi(\nu^\perp)^j_i\:  \big\la \xi,\, \partial_j u(0) \big\ra \:\xi\:
d\mu_x(\xi) \notag \\
&\quad\: - \xi_x \fint_{D_x \scrL} \frac{2}{\la \xi, \nu \ra} \; \xi^i \:\pi(\nu^\perp)^j_i\:  \big\la \xi,\, \partial_j u(0) \big\ra \: d\mu_x(\xi) \notag \\
 &= \fint_{D_x \scrL} 
\frac{1}{\la \xi, \nu \ra} \; \xi^i \:\pi(\nu^\perp)^j_i\:  \big\la \xi,\, \partial_j u(0) \big\ra \:\big( 3\xi - 2 \:\frac{u}{u^2} \big)\:
d\mu_x(\xi) \\
\frac{d u}{dt} &= - \fint_{D_x \scrL} 
\frac{1}{\la \xi, \nu \ra} \; \xi^i \:\pi(\nu^\perp)^j_i\:  \big\la \xi,\, \partial_j u(0) \big\ra \:\big( 3\,u^2\,\xi - 2 u \big)\:
d\mu_x(\xi) \:. \label{duint}
\end{align}

Our next task is to compute the integrals
\[ \fint_{D_x \scrL} \frac{1}{\la \xi, \nu \ra} \; \xi^i \,\xi^j \: d\mu_x(\xi) 
\qquad \text{and} \qquad
\fint_{D_x \scrL} \frac{1}{\la \xi, \nu \ra} \; \xi^i \,\xi^j \,\xi^k \: d\mu_x(\xi)
\:. \]
To this end, it is most conveniently to work in a reference frame where
\[ u(0) = |u|\: (1, \vec{0}) \qquad \text{and} \qquad \xi = \frac{1}{|u|}\: \zeta = \frac{1}{|u|}\: (1, \vec{n}) \:. \]
Then the integrals simplify to
\begin{align*}
\fint_{D_x \scrL} \frac{1}{\la \xi, \nu \ra} \; \xi^i \,\xi^j \: d\mu_x(\xi) &= 
\frac{1}{|u|^3} \fint_{S^2} \frac{1}{\la \zeta, \nu \ra} \; \zeta^i\, \zeta^j\: d\mu_{S^2}(\vec{n}) \\
\fint_{D_x \scrL} \frac{1}{\la \xi, \nu \ra} \; \xi^i \,\xi^j \,\xi^k \: d\mu_x(\xi) &= 
\frac{1}{|u|^4} \fint_{S^2} \frac{1}{\la \zeta, \nu \ra} \; \zeta^i\, \zeta^j\, \zeta^k \: d\mu_{S^2}(\vec{n}) \:.
\end{align*}
More generally, for~$p \in \N_0$ we set
\beq \label{Jint}
J_{i_1 \cdots i_p} = \fint_{S^2} \frac{1}{\la \nu, \zeta \ra}\:\zeta_{i_1} \cdots \zeta_{i_p}\: d\mu_{S^2}(\vec{n}) \:.
\eeq

\begin{Lemma} \label{lemmaJ}
For~$p=0,1,2$, the integrals in~\eqref{Jint} are computed by~\eqref{J0} and
\begin{align*}
J_i &= \frac{-J+\la \hat{u}, \nu \ra}{\la \hat{u}, \nu \ra^{2}-1}\: \hat{u}_i +
\frac{J \:\la \hat{u}, \nu \ra-1}{\la \hat{u}, \nu \ra^{2}-1}\: \nu_i \\
J_{ij} &= \frac{J-\la \hat{u}, \nu \ra}{2 \left(\la \hat{u}, \nu \ra^{2}-1\right)} \:g_{ij} 
+ \frac{1}{2 \big(\la \hat{u}, \nu \ra^{2}-1 \big)^{2}} \:\Big(
\big( 2 \la \hat{u}, \nu \ra^{3}+3 J-5 \la \hat{u}, \nu \ra \big) \: \hat{u}_i \hat{u}_j \\
&\quad\,+ \big( -6 J \la \hat{u}, \nu \ra+2 \la \hat{u}, \nu \ra^{2}+4 \big) \: \hat{u}_{(i} \nu_{j)}
+ \big( 2 J \,\la \hat{u}, \nu \ra^{2}+J-3 \la \hat{u}, \nu \ra \big) \: \nu_i \nu_j \Big) \\
J_{ijk} &= \frac{2 \la \hat{u}, \nu \ra^{3}+3 J-5 \la \hat{u}, \nu \ra}{2 \left(\la \hat{u}, \nu \ra^{2}-1\right)^{2}} \: \hat{u}_{(i} \: g_{jk)} 
+ \frac{-3 J \la \hat{u}, \nu \ra+\la \hat{u}, \nu \ra^{2}+2}{2 \left(\la \hat{u}, \nu \ra^{2}-1\right)^{2}}\: \nu_{(i} \: g_{jk)} \\
&\quad\:+ \frac{1}{6 \big( \la \hat{u}, \nu \ra^{2}-1 \big)^{3}} \: \Big(
\big( 8 \la \hat{u}, \nu \ra^{5}-26 \la \hat{u}, \nu \ra^{3}-15 J+33 \la \hat{u}, \nu \ra \big)\: \hat{u}_i \hat{u}_j \hat{u}_k \\
&\qquad\qquad\qquad\qquad\;\: + \big( 6 \la \hat{u}, \nu \ra^{4}+45 J \la \hat{u}, \nu \ra-27 \la \hat{u}, \nu \ra^{2}-24 \big)\: \hat{u}_{(i} \hat{u}_j \nu_{k)} \\
&\qquad\qquad\qquad\qquad\;\: + \big( -36 J \,\la \hat{u}, \nu \ra^{2}+6 \la \hat{u}, \nu \ra^{3}-9 J+39 \la \hat{u}, \nu \ra \big)\: \hat{u}_{(i} \nu_j \nu_{k)} \\
&\qquad\qquad\qquad\qquad\;\: + \big( 6 J \,\la \hat{u}, \nu \ra^{3}+9 J \la \hat{u}, \nu \ra-11 \la \hat{u}, \nu \ra^{2}-4 \big)\: \nu_i \nu_j \nu_k \Big) \:.
\end{align*}
\end{Lemma}
\Proof In the case~$p=0$, using that~$\la \nu, \zeta \ra = \nu^0 - \vec{n} \vec{\nu}$, we
obtain by direct computation
\begin{align*}
J &= \fint_{S^2} \frac{1}{\nu^0 - \vec{n}\vec{\nu}} \: d\mu_{S^2}(\vec{n}) 
= \frac{1}{2\,|\vec{\nu}|} \: \log \Big( \frac{\nu^0 + |\vec{\nu}|}{\nu^0 - |\vec{\nu}|} \Big) \\
&= \frac{1}{2\,|\vec{\nu}|} \: \log \bigg( \frac{ \big( \nu^0 + |\vec{\nu}| \big)^2}{(\nu^0)^2 - |\vec{\nu}|^2} \bigg) = 
\frac{1}{|\vec{\nu}|} \:  \log \big( \nu^0 + |\vec{\nu}| \big) \:,
\end{align*}
where in the last step we used that~$\nu$ is a unit vector. Using that~$\nu^0 = \la \hat{u}, \nu\ra$
and~$|\vec{\nu}| = \sqrt{(\nu^0)^2-1}$ gives~\eqref{J0}.

We now proceed inductively in~$p$. To this end, it
is most convenient to work in the basis~$(e_a)_{a=0,\ldots, 3}$ with
\[ e_0 = \hat{u} \:,\qquad e_1 = \nu \:, \]
and~$e_2$ and~$e_3$ two mutually orthogonal normalized spacelike vectors which 
are orthogonal to both~$e_0$ and~$e_1$. Thus the metric is
\[ g_{ab} = \la e_a, e_b \ra = \begin{pmatrix} 1 & \la \nu, \hat{u} \ra & 0 & 0 \\
\la \nu, \hat{u} \ra & 1 & 0 & 0 \\ 0 & 0 & -1 & 0 \\  0 & 0 & 0 & -1
\end{pmatrix} \:, \]
and its inverse is computed with the help of Cramer's rule to be
\[ g^{ab} = \frac{1}{1 - \la \nu, \hat{u} \ra^2} \begin{pmatrix} 1 & -\la \nu, \hat{u} \ra & 0 & 0 \\
-\la \nu, \hat{u} \ra & 1 & 0 & 0 \\ 0 & 0 & -(1 - \la \nu, \hat{u} \ra^2) & 0 \\  0 & 0 & 0 & -(1 - \la \nu, \hat{u} \ra^2)
\end{pmatrix} \:, \]
Then
\begin{align*}
J_0 &= J_i\: e^i_0 = \fint_{S^2} \frac{1}{\nu^0 - \vec{n}\vec{\nu}}\:\la \zeta, \hat{u} \ra \: d\mu_{S^2}(\vec{n}) = J \\
J_1 &= J_i\: e^i_1 = \fint_{S^2} \frac{1}{\nu^0 - \vec{n}\vec{\nu}}\:\la \zeta, \nu \ra \: d\mu_{S^2}(\vec{n}) = 1 \\
J_2 &= J_3 = 0 \:.
\end{align*}

Next,
\begin{align*}
J_{0 a} &= J_{ia}\: e^i_0 = J_a \\
J_{1 a} &= J_{ia}\: e^i_1 = \fint_{S^2} \la e_a, \zeta \ra \: d\mu_{S^2}(\vec{n}) =
\la e_a, \hat{u} \ra \\
\end{align*}
Hence, in matrix notation,
\[ J_{ab} = \begin{pmatrix} J & 1 & 0 & 0 \\
1  & \la \nu, \hat{u} \ra & 0 & 0 \\ 0 & 0 & x & 0 \\  0 & 0 & 0 & x
\end{pmatrix} \:, \]
where the parameter~$x$ is still unknown (by symmetry, the lower right $2 \times 2$-block is a multiple
of the identity matrix). In order to compute~$x$, we use that
\[ 0 = g^{ab} J_{ab} = \frac{1}{1 - \la \nu, \hat{u} \ra^2} \Big( J + \la \nu, \hat{u} \ra - 2  \la \nu, \hat{u} \ra \Big)
- 2x \:, \]
implying that
\[ x = \frac{1}{2}\: \frac{J - \la \nu, \hat{u} \ra}{1 - \la \nu, \hat{u} \ra^2} \:. \]

Finally,
\begin{align*}
J_{0 ab} &= J_{iab}\: e^i_0 = J_{ab} \\
J_{1 ab} &= J_{iab}\: e^i_1 = \fint_{S^2} \la e_a, \zeta \ra\:\la e_b, \zeta \ra \: d\mu_{S^2}(\vec{n}) =
-\frac{1}{3}\: \la e_a, e_b \ra + \frac{4}{3}\: \la e_a, \hat{u} \ra \la e_b, \hat{u} \ra  \:.
\end{align*}
This determines the tensor elements of~$J_{abc}$ except when all indices are equal to~$3$ or~$4$.
In this case, one of the indices~$3$ or~$4$ must appear an odd number of times.
As a consequence, all matrix elements vanish again by symmetry,
\[ J_{abc}=0 \qquad \text{if~$a,b,c \in \{3,4\}$} \:. \]
Rewriting these formulas in a general basis gives the result\footnote{This
computation was carried out with the help of computer algebra.
The corresponding {\textsf{Maple}} worksheet is included as an ancillary file to the arXiv submission of this paper.}.
\QED

Theorem~\ref{thminfdyn} is obtained by a straightforward computation
applying Lemma~\ref{lemmaJ} to~\eqref{duint}.

\section{Dependence on the Choice of Foliation} \label{appB}
Clearly, the Dirac dynamics modified by adiabatic projections~\eqref{Vtdef} depends on the
choice of the foliation~$(\scrN_t)_{t \in \R}$.
For the analysis of this dependence, it is useful to rewrite the modified dynamics with a
Dirac equation involving a nonlocal potential~$\B$.
First, multiplying by~$i \gamma(\nu)$ and using that~$\gamma(\nu) \big( i \partial_t - \tilde{H}) = \Dir - m$,
we obtain
\beq \label{DirV}
( \Dir - m ) \,V^{t}_{t_0}(\cdot, \y) = \gamma(\nu) \,\Big\{ i \dot{\tilde{E}}_I(t) + \tilde{E}_I(t) \,\tilde{H}(t) - \tilde{H}(t)
\Big\}\, V^t_{t_0} \:.
\eeq
The right side can be regarded as a nonlocal potential supported on the Cauchy surface~$\scrN_t$.
More specifically, it can be written in the form~$\B V^t_{t_0}$, where~$\B$ is the integral operator in
spacetime
\beq \label{Bnonloc}
(\B \psi)(x) = \int_\scrM \B(x,y)\: \psi(y)\: d\mu_\scrM(y) \:,
\eeq
and~$\B(x,y)$ is a distributional kernel supported on the hypersurface~$\scrN_t$ with~$x \in \scrN_t$.
We would like this potential to be symmetric in the sense that
\beq \label{Bsymm}
\B(x,y)^* = \B(y,x)
\eeq
(where the star is the adjoint with respect to the spin inner product), because this will make it
possible to generalize current conservation to general Cauchy surfaces (see Proposition~\ref{prpnonloc} below).
In order to arrange this to be the case, we need to rewrite the expression inside the curly brackets
in~\eqref{DirV} in such a way that it becomes a symmetric operator on the Hilbert space~$\H_m$.
Using that the scalar product~\eqref{print} involves a factor~$\gamma(\nu)$,
this will immediately imply that the resulting kernel~$\B(x,y)$ is symmetric~\eqref{Bsymm}.

In order to rewrite the curly brackets in~\eqref{DirV}, we make use of the fact that
the operator~$V^t_{t_0}$ vanishes when multiplied by the
projection operator~$\tilde{E}_{\R \setminus I}(t)$. More precisely, the last two summands
in the curly brackets can be written as
\[ \big(\tilde{E}_I(t) \,\tilde{H}(t) - \tilde{H}(t)
\big)\, V^t_{t_0} = \tilde{E}_{\R \setminus I}(t) \,\tilde{H}(t) \, V^t_{t_0}
= \big(\tilde{E}_{\R \setminus I}(t) \,\tilde{H}(t) + \tilde{H}(t)\, \tilde{E}_{\R \setminus I}(t) \big)\, V^t_{t_0} \:. \]
Next, differentiating the relation~$\tilde{E}_I(t) = \tilde{E}_I^2$ gives
\[ \dot{\tilde{E}}_I(t) = \tilde{E}_I(t)\, \dot{\tilde{E}}_I(t) + \dot{\tilde{E}}_I(t) \, \tilde{E}_I(t) \:, \]
and multiplying from the left and right by~$\tilde{E}_I(t)$ gives
\[ \tilde{E}_I(t)\, \dot{\tilde{E}}_I(t) \, \tilde{E}_I(t) = 0 \:. \]
This makes it possible to rewrite the first summand in the curly brackets as
\begin{align*}
i \dot{\tilde{E}}_I(t)\, V^t_{t_0} &= i \tilde{E}_{\R \setminus I}(t)\,\dot{\tilde{E}}_I(t)\, \tilde{E}_I(t)\, V^t_{t_0} \\
&= i \big(\tilde{E}_{\R \setminus I}(t)\,\dot{\tilde{E}}_I(t)\, \tilde{E}_I(t) -
\tilde{E}_I(t)\,\dot{\tilde{E}}_I(t)\, \tilde{E}_{\R \setminus I}(t) \big)\, V^t_{t_0} \:.
\end{align*}
Combining all the terms, we obtain the nonlocal Dirac equation
\beq \label{Dirnonlocal}
( \Dir + \B - m ) V^{\cdot}_{t_0}(\cdot, \y) = 0 \:,
\eeq
where~$\B$ is the spatial operator
\beq \label{Bex}
\begin{split}
\B &= -i \gamma(\nu) \Big( \tilde{E}_{\R \setminus I}(t)\,\dot{\tilde{E}}_I(t)\, \tilde{E}_I(t) -
\tilde{E}_I(t)\,\dot{\tilde{E}}_I(t)\, \tilde{E}_{\R \setminus I}(t) \Big) \\
&\quad\, - \gamma(\nu) \big(\tilde{E}_{\R \setminus I}(t) \,\tilde{H}(t) + \tilde{H}(t)\, \tilde{E}_{\R \setminus I}(t) \big)\:.
\end{split}
\eeq
In this way, we have rewritten the modified dynamics in terms of a Dirac equation~\eqref{Dirnonlocal}
involving a nonlocal potential~\eqref{Bnonloc} which is symmetric~\eqref{Bsymm}.

So far, we know that the modified dynamics respects current conservation in the sense
that the current integral over the surface~$\scrN_t$ in~\eqref{print}
does not depend on the choice of~$t$. In the next lemma we show that
current conservation holds even for arbitrary Cauchy surfaces, provided that the
integrand is modified by the nonlocal potential. More precisely, given a Cauchy surface~$\scrN$
and denoting its past by~$\Omega$, we modify~\eqref{print} to
\begin{align}
(\psi | \phi)_\scrN &:= \int_\scrN \Sl \psi \,|\, \gamma(\nu)\, \phi \Sr_x\: d\mu_\scrN(x) \label{c1} \\
&\quad\;\; -i \int_\Omega d\mu_\scrM(x) \int_{\scrM \setminus \Omega} d\mu_\scrM(y)\;
\Sl \psi(x) \,|\, \B(x,y)\, \psi(y) \Sr_x \label{c2} \\
&\quad\;\; +i \int_{\scrM \setminus \Omega} d\mu_\scrM(x) \int_\Omega d\mu_\scrM(y)\;
\Sl \psi(x) \,|\, \B(x,y)\, \psi(y) \Sr_x \:. \label{c3}
\end{align}

\begin{figure}
\psscalebox{1.0 1.0} 
{
\begin{pspicture}(0,26.035076)(9.977692,28.510231)
\definecolor{colour0}{rgb}{0.7019608,0.7019608,0.7019608}
\definecolor{colour1}{rgb}{0.9019608,0.9019608,0.9019608}
\pspolygon[linecolor=colour0, linewidth=0.02, fillstyle=solid,fillcolor=colour0](0.6950537,28.045076)(1.5300537,28.055077)(2.5725536,28.140078)(3.6125536,28.285076)(4.4875536,28.430077)(4.9275537,28.490076)(5.2975535,28.450077)(6.0975537,28.250076)(6.6775537,28.092577)(7.370054,27.937576)(7.995054,27.847576)(8.700054,27.830076)(9.205053,27.760077)(7.687554,27.570076)(6.725054,27.385077)(6.0350537,27.225077)(5.560054,27.115076)(4.9000535,26.990076)(4.440054,26.945076)(4.1350536,26.965076)(3.3200538,27.180077)(2.6550536,27.445076)(1.6600537,27.802576)(1.2200537,27.932577)
\pspolygon[linecolor=colour1, linewidth=0.02, fillstyle=solid,fillcolor=colour1](0.02005371,28.080076)(0.2850537,28.065077)(0.8350537,28.000076)(1.6500537,27.790077)(2.5550537,27.450077)(3.0925536,27.245077)(3.6025536,27.075077)(4.0525537,26.955076)(4.3175535,26.935078)(4.9275537,26.995077)(5.6175537,27.130077)(6.9425535,27.397577)(7.937554,27.602577)(9.052554,27.762577)(9.955053,27.902576)(9.957554,26.055077)(0.03005371,26.055077)
\psbezier[linecolor=black, linewidth=0.04](0.013773764,28.08411)(2.215283,27.92751)(3.8738291,28.367542)(4.862216,28.472765292936675)(5.850603,28.57799)(7.2029333,27.529495)(9.965054,27.89699)
\psbezier[linecolor=black, linewidth=0.04](0.025053712,28.077578)(1.5393314,28.129198)(3.345186,26.953835)(4.3450537,26.937576904296876)(5.3449216,26.921318)(7.061592,27.586056)(9.965054,27.897577)
\rput[bl](7,28.2){$\scrN$}
\rput[bl](7,27){$\scrN'$}
\rput[bl](4.2,27.5){$\Omega \setminus \Omega'$}
\rput[bl](4.4,26.35){$\Omega'$}
\end{pspicture}
}
\caption{Current conservation in the presence of a nonlocal potential.}
\label{fignonloc}
\end{figure}
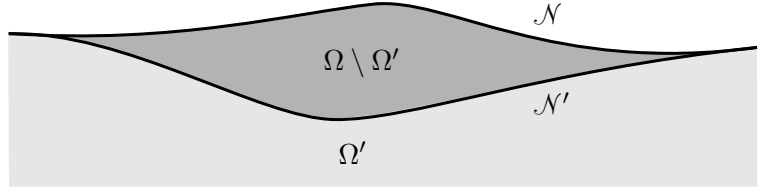

\begin{Prp} \label{prpnonloc} Let~$\psi, \phi$ be smooth solutions of the Dirac equation~\eqref{Dirnonlocal}
involving a symmetric nonlocal potential~\eqref{Bsymm}. Let~$\scrN$ and~$\scrN'$ be two
Cauchy surfaces for which the integrals in~\eqref{c1}--\eqref{c3} are well-defined and finite,
and for which the sets~$\Omega \setminus \Omega'$ and~$\Omega' \setminus \Omega$ are both
relatively compact (see Figure~\ref{fignonloc}). Then the current integrals coincide,
\[ (\psi | \phi)_\scrN = (\psi | \phi)_{\scrN'} \:. \]
\end{Prp}
\Proof We choose a Cauchy surface~$\scrN''$ which lies to the past of~$\scrN$ and~$\scrN'$ with the property
that the sets~$\Omega \setminus \Omega''$ and~$\Omega' \setminus \Omega''$ are both relatively compact.
Then, by applying the above proposition in two steps for~$\scrN$ and~$\scrN''$
and subsequently for~$\scrN''$ and~$\scrN$, one sees that it suffices to consider
the case that~$\Omega' \subset \Omega$ (this case is shown in Figure~\ref{fignonloc}).

In this case, by rearranging the integration domains, one obtains
\begin{align*}
(\psi | \phi)_\scrN - (\psi | \phi)_{\scrN'} &= \int_\scrN \Sl \psi \,|\, \gamma(\nu)\, \phi \Sr_x\: d\mu_\scrN(x) 
- \int_{\scrN'} \Sl \psi \,|\, \gamma(\nu)\, \phi \Sr_x\: d\mu_{\scrN'}(x) \\
&\quad\: -i \int_{\Omega \setminus \Omega'} d\mu_\scrM(x) \int_\scrM d\mu_\scrM(y)\;
\Sl \psi(x) \,|\, \B(x,y)\, \psi(y) \Sr_x \\
&\quad\: +i \int_\scrM d\mu_\scrM(x) \int_{\Omega \setminus \Omega'} d\mu_\scrM(y)\;
\Sl \B(y,x)\, \psi(x) \,|\, \psi(y) \Sr_x \:,
\end{align*}
where in the last line we used the symmetry of the kernel~\eqref{Bsymm}.
Applying the Green's formula, the two surface integrals can be recovered as boundary terms of
a spacetime integral. We thus obtain
\begin{align*}
&(\psi | \phi)_\scrN - (\psi | \phi)_{\scrN'} = -i \int_{\Omega \setminus \Omega'}
\Big( \Sl \psi \,|\, (\Dir - m) \, \phi \Sr_x - \Sl (\Dir - m) \, \psi \,|\, \phi \Sr_x \Big)\: d\mu_\scrM(x) \\
&\quad\, -i \int_{\Omega \setminus \Omega'} 
\Big(  \Sl \psi(x) \,|\, (\B \psi)(x) \Sr_x - \Sl (\B \psi)(x) \,|\, \psi(x) \Sr_x \Big) \:d\mu_\scrM(x) \\
&=-i \int_{\Omega \setminus \Omega'}
\Big( \Sl \psi \,|\, (\Dir +\B - m) \, \phi \Sr_x - \Sl (\Dir +\B - m) \, \psi \,|\, \phi \Sr_x \Big)\: d\mu_\scrM(x) = 0 \:,
\end{align*}
giving the result.
\QED
We remark that the form of the nonlocal modification of this conservation law is inspired by surface
layer integrals for causal fermion systems; in particular the commutator inner product in~\cite{dirac}.
We also point out that if~$\B$ is a local potential (i.e.\ a multiplication operator), then
the additional terms~\eqref{c2} and~\eqref{c3} vanish, giving us back the usual form of the
scalar product.

After the above preparations, we are in the position to analyze how our results on
fermiogenesis depend on the choice of the foliation~$(\scrN_t)_{t \in \R}$.
To this end, we consider the nonlocal potential~\eqref{Bex}.
As already mentioned above, it is supported on the Cauchy surface in the
sense that~$\B(x,y)$ vanishes unless~$x$ and~$y$ lie on the same Cauchy surface~$\scrN_t$.
This shows in particular that changing the foliation also changes the potential~$\B$.
However, this does not necessarily mean that also the dynamics changes, because
what counts is the action of the potential on~$V^t_{t_0}$. More precisely, two potentials~$\B$
and~$\B'$ describe the same dynamics if
\beq \label{Bhomrep}
\B V^{\cdot}_{t_0} = \B' \,V^{\cdot}_{t_0} \:.
\eeq
If the system is homogeneous, then both~$\B$ and~$\B'$ can be represented by multiplication
operators in momentum space. This means that~\eqref{Bhomrep} holds, even if~$\B$ and~$\B'$
are represented by different integral kernels in spacetime.
This consideration shows that, if we go over to the general non-homogeneous situation,
the dependence of the modified dynamics on the choice of foliation
involves the length scale~$\ell_\text{macro}$ of macroscopic physics
as the length-scale on which spatial homogeneity is violated.
Next, one should keep in mind that, describing a cutoff in momentum space on the scale~$\varepsilon^{-1}$,
the kernel~$\B(x,y)$ decays on the scale~$\varepsilon$ (and is highly oscillatory).
Putting these scalings together, we conclude that
changing the foliation affects fermiogenesis as worked out in this
paper only by
\beq \label{errorscale}
\text{corrections of higher order in} \qquad \frac{\varepsilon}{\ell_\text{macro}} \:.
\eeq
Such corrections seems negligible in most situation of physical interest.

\section{Failure of a Naive Particle Detection} \label{appgauge}
In this appendix, we explain why the operator~$\eta_\Lambda$ used in~\eqref{fermiogenesis}
for the detection of fermiogenesis must be adjusted carefully to the local geometry.
Indeed, the basic problem can be understood already in Minkowski space from
the necessity of gauge invariance, as is illustrated in the following simple example.

\begin{Example} {\bf{(Time-dependent gauge transformations)}} {\em{
Consider the Dirac equation in the Minkowski vacuum after a time-dependent gauge transformation, i.e.\
\beq \label{tgauge}
\Dir = e^{-i \theta(t,\vec{x})} \,i \Pdd \,e^{i \theta(t,\vec{x})} 
= i \Pdd + \big( \Pdd_x \theta(x) \big) \:.
\eeq
The Hamiltonian in~\eqref{tilHdef} and the spatial operator in~\eqref{Atildef} take the form
\begin{align}
\tilde{H} &= -i \gamma^0 \sum_{\alpha=1}^3 \gamma^\alpha \partial_\alpha 
- \gamma^0 \sum_{\alpha=1}^3 \gamma^\alpha (\partial_\alpha \theta) + (\partial_t \theta)
+ \gamma^0 m \\
\tilde{A}(t) &= -i \gamma^0 \sum_{\alpha=1}^3 \gamma^\alpha \partial_\alpha 
- \gamma^0 \sum_{\alpha=1}^3 \gamma^\alpha (\partial_\alpha \theta) \:. \label{Atdef}
\end{align}

The simplest candidate for the operator~$\eta_\Lambda$ is to choose a
non-negative test function~$\eta \in C^\infty_0(\R^3, \R)$,  with
\[ \eta(\x) = \eta(-\x) \qquad \text{and} \qquad \eta(0) =1 \]
and to set
\beq \label{etaLam}
\eta_\Lambda(\x,\y) := \eta \Big( \frac{\y-\x}{\Lambda} \Big) \:,
\eeq
Then~\eqref{trLam} can be written formally as a double integral,
\[ \tr \big( \eta_\Lambda\: \tilde{\Pi}(t) \big) = -\int_{\R^3} d^3x \int_{\R^3} d^3y \:
\Tr_{\C^4} \Big( \eta_\Lambda(\x,\y)\:   \tilde{P}^\varepsilon\big( (t,\y), (t,\x) \big)\; \gamma^0  \Big) \:. \]
Using that the interaction consist merely of the local phase transformations in~\eqref{tgauge},
we conclude that
\[ \tr \big( \eta_\Lambda\: \tilde{\Pi}(t) \big) = -\int_{\R^3} d^3x \int_{\R^3} d^3y \;
e^{-i \theta(t,\vec{x}) + i \theta(t,\vec{y})} 
\Tr_{\C^4} \Big( \eta_\Lambda(\x,\y)\: P^\varepsilon\big( (t,\y), (t,\x) \big)\; \gamma^0 \Big) \:, \]
where~$P^\varepsilon(x,y)$ is the regularized kernel in the Minkowski vacuum.
Here the integrals typically diverges as a direct consequence of the fact that we are in infinite spatial volume.
However, differentiating formally with respect to~$t$, we obtain a finite and well-defined expression,
provided that the potential~$\theta$ decays sufficiently fast at spatial infinity.
Likewise, fermiogenesis in a finite time interval~$[t_0, t_1]$ is well-defined as the difference
\begin{align}
\int_{t_0}^{t_1} &B(t)\: dt = \tr \big( \eta_\Lambda\: \tilde{\Pi}(t) \big) \Big|_{t_0}^{t_1} \notag \\
&= -\int_{\R^3} d^3x \int_{\R^3} d^3y \;
\Big( e^{-i \theta(t_1,\vec{x}) + i \theta(t_1,\vec{y})} - e^{-i \theta(t_0,\vec{x}) + i \theta(t_0,\vec{y})}
\Big) \notag \\
&\qquad\qquad\qquad\qquad\qquad \times \: \Tr_{\C^4} \Big( \eta_\Lambda(\x,\y)\: P^\varepsilon\big( (t,\y), (t,\x) \big)\; \gamma^0 \Big) \:. \label{gaugeP}
\end{align}

The appearance of the gauge phases in~\eqref{gaugeP} suggests that this expression
may not be gauge invariance. In order to see this in detail, we consider the situation that~$\theta$
vanishes at time~$t_0$ and expand in orders of~$\theta(t_1, .)$. Linearly in this potential, we obtain the
contribution
\[ i\tr \big( \theta \,\eta\, \Pi(t_1) - \eta\, \theta\, \Pi(t_1) \big) 
= i\tr \big( \eta\, [\Pi(t_1), \eta] \big) = 0 \:,  \]
because the operators~$\Pi(t_1)$ and~$\eta$ commute due to homogeneity and the fact that~$\eta$ is scalar.
Quadratically in~$\theta$, we obtain
\[ \frac{1}{2}\:\tr \Big( \theta^2 \,\eta\, \Pi(t_1) + \eta\, \theta^2\, \Pi(t_1) -2 \theta\, \eta\, \theta\, \Pi(t_1) \Big) 
= -\frac{1}{2}\:\tr \Big( \big[\eta, \theta \big]\, \big[\Pi(t_1), \theta \big] \Big) \:. \]
By choosing~$\eta$ appropriately, one can arrange that this trace is non-zero.
This shows that defining fermiogenesis via~\eqref{fermiogenesis} with~$\eta_\Lambda$ according to~\eqref{etaLam}
is not a sensible concept, because gauge invariance is violated.

In order to avoid this problem, one might want to replace the operator~$\eta_\Lambda$ in~\eqref{etaLam}
by an operator derived from the spatial Dirac operator~$\tilde{A}(t)$ in~\eqref{Atdef}. For example, one could
choose a non-negative test function~$\theta \in C^\infty_0(\R)$ and introduce~$\eta_\Lambda$ via the
functional calculus for the self-adjoint operator~$\tilde{A}(t)$,
\[ \eta_\Lambda := \theta\Big( \frac{\tilde{A}(t)}{\Lambda} \Big) \in \Lin(\H) \:. \]
However, this method does not resolve the problem of gauge dependence. The reason is that
the electric potential~$A_0=\partial_t \theta$ does not enter the operator~$\tilde{A}(t)$ (see~\eqref{Atdef}),
but it clearly changes the phases in~\eqref{gaugeP}.
}}\QEDrem
\end{Example}

In curved spacetime, the problem discussed in this example for gauge transformations
appears similarly for infinitesimal spatial diffeomorphism which depend on time.
In other words, the trace~\eqref{trLam} depends in a non-trivial way on the choice of the
shift function. Moreover, the time derivative in~\eqref{fermiogenesis} depends on the
choice of the lapse function. This explains why the operator~$\eta_\Lambda$ cannot be
computed intrinsically on the Cauchy surface~$\scrN$, but why we need to use the form of the
metric (or equivalently the Dirac operator or the form of the Hadamard expansion) in a neighborhood
of the Cauchy surface.

\Thanks{{{\em{Acknowledgments:}}
M.J.\ gratefully acknowledges support by the Studienstiftung des deutschen Volkes
and the Hanns-Seidel-Stiftung. We would like to thank Marco van den Beld Serrano
and the referees for valuable comments.

\providecommand{\bysame}{\leavevmode\hbox to3em{\hrulefill}\thinspace}
\providecommand{\MR}{\relax\ifhmode\unskip\space\fi MR }
\providecommand{\MRhref}[2]{%
  \href{http://www.ams.org/mathscinet-getitem?mr=#1}{#2}
}
\providecommand{\href}[2]{#2}

\end{document}